\documentclass[12pt]{nostro}

\usepackage{epsfig}\parskip 5pt plus 1pt
\newcommand{\nn}{\nonumber}

\newcommand{\be}{\begin{equation}}
\newcommand{\ee}{\end{equation}}
\newcommand{\bea}{\begin{eqnarray}}
\newcommand{\eea}{\end{eqnarray}}
\newcommand{\tc}{ \theta_{13} }
\newcommand{\tatm}{ \theta_{23} }

\def\simge{\mathrel{%
   \rlap{\raise 0.511ex \hbox{$>$}}{\lower 0.511ex \hbox{$\sim$}}}}
\def\simle{\mathrel{
   \rlap{\raise 0.511ex \hbox{$<$}}{\lower 0.511ex \hbox{$\sim$}}}}

\begin{document}
\begin{frontmatter}
\thispagestyle{empty}
\begin{flushright}
%{\tt hep-ph/ }\\
{IFT-UAM/CSIC-05-48}\\
{ROMA-1419-05}
\end{flushright}
\title{$\nu_\mu$ disappearance at the SPL, T2K-I, NO$\nu$A and the Neutrino Factory} 
\author[Madrid]{A. Donini}
\author[Madrid]{E. Fern\'andez-Mart\'{\i}nez}
\author[Roma]{D. Meloni}
\author[Madrid]{S. Rigolin}
\address[Madrid]{I.F.T. and Dep. F\'{\i}sica Te\'{o}rica, 
Univ. Autonoma Madrid., E-28049, Madrid, Spain}
\address[Roma]{INFN, Sezione di Roma e Dip. di Fisica, 
Univ. di Roma ``La Sapienza'', P.le A. Moro 2, I-00185 Roma, Italy} 
\vspace{.3cm}
\begin{abstract}
We study the measurement of the atmospheric neutrino oscillation parameters, $\theta_{23}$ and $\Delta m^2_{23}$,
at the $\nu_\mu$ disappearance channel at three conventional beam facilities, the SPL, T2K-phase I and NO$\nu$A. 
These two parameters have been shown to be of crucial importance in the measurement of two of the unknowns of the PMNS mixing matrix, $\theta_{13}$
and the leptonic CP-violating phase $\delta$. In our analyis, the effect of the two discrete
ambiguities, ${\rm sign}(\Delta m^2_{23})$ and ${\rm sign}(\tan 2 \theta_{23})$, is explicitly
taken into account. We analyse also the $\nu_\mu$ disappearance channel at the Neutrino Factory, 
and combine it with the ``golden'' $\nu_e \to \nu_\mu$ and ``silver'' $\nu_e \to \nu_\tau$ appearance channels to study its impact
on the measurement of $\theta_{13}$ and $\delta$.
Eventually, we present the sensitivity of the four facilities to different observables: $\theta_{13}$, $\delta$, maximal $\theta_{23}$,
the sign of the atmospheric mass difference, $s_{atm}$, and the $\theta_{23}$-octant, $s_{oct}$.
\end{abstract}

\vspace*{\stretch{2}}
\begin{flushleft}
% insert here the PACS number
\end{flushleft}
%\begin{center}%
\end{frontmatter}

%
%%%%%%%%%%%%%%%%%%%%%%%%%%%%%%%%%%%%%%%%%%%%%%%%%%%%%%%%%%%%%%%%%%%%%%
%
\newpage
%
%%%%%%%%%%%%%%%%%%%%%%%%%%%%%%%%%%%%%%%%%%%%%%%%%%%%%%%%%%%%%%%%%%%%%%
%
\section{Introduction}
\label{sec:intro}

The results of atmospheric, solar, accelerator and reactor neutrino experiments \cite{exp} show 
that flavour mixing occurs not only in the hadronic sector, as it has been known for long, but 
in the leptonic sector as well. The experimental results point to two very distinct mass 
differences\footnote{A third mass difference, $\Delta m^2_{LSND} \sim 1$ eV$^2$, suggested by 
the LSND experiment \cite{lsnd}, has not being confirmed yet \cite{boone} and will not be 
considered in this paper.}, $\Delta m^2_{sol} \approx 8.2 \times 10^{-5}$ eV$^2$ and 
$|\Delta m^2_{atm}| \approx 2.5 \times 10^{-3}$ eV$^2$.
Only two out of the four parameters of the three-family leptonic mixing matrix $U_{PMNS}$ 
\cite{neutrino_osc} are known: $\theta_{12} \approx 34^\circ$ and $\theta_{23}\approx 45^\circ$. 
The other two parameters, $\theta_{13}$ and $\delta$, are still unknown: for the mixing angle
$\theta_{13}$ direct searches at reactors \cite{chooz} and three-family global analysis of the 
experimental data \cite{Fogli:2005cq} give the upper bound $\theta_{13} 
\leq 11.5^\circ$, whereas for the leptonic CP-violating phase $\delta$ we have no information 
whatsoever. Two additional discrete unknowns are the sign of the atmospheric mass difference and 
the $\theta_{23}$-octant (if $\theta_{23} \neq 45^\circ$). 

The full understanding of the leptonic mixing matrix constitutes, together with the discrimination 
of the Dirac/Majorana character of neutrinos and with the measurement of their absolute mass scale, the main 
neutrino-physics goal for the next decade. 
In the recent past, most of the experimental breakthroughs in neutrino physics have been achieved
by exploiting the so-called ``disappearance channels'': by observing a deficit in the neutrinos that reach 
the detector with respect to those expected to be emitted from the source, a positive and eventually unambiguous 
signal of neutrino oscillations has been established. 
The SK detector has gathered indirect evidence of $\nu_\mu \to \nu_\tau$ conversion of atmospheric neutrinos. 
However, no direct observation of $\tau$'s is possible at this detector. To observe directly $\mu \to \tau$ conversion, 
two detectors are under construction at the Gran Sasso Laboratory \cite{OPERA,ICARUS}.
The SNO experiment \cite{SNO} has shown that a fraction of the $\nu_e$'s emitted by the Sun core reach the
Earth converted into $\nu_\mu$'s and $\nu_\tau$'s (and not into unobservable sterile neutrinos). 
However, the SNO detector is not able to distinguish between $\tau$'s and $\mu$'s and thus measure 
the subleading oscillations $\nu_e \to \nu_\tau,\nu_\mu$ directly.
For this reason, new-generation experiments have been proposed to look for the fleeting 
and intimately related parameters $\theta_{13}$ and $\delta$ through the more promising 
``appearance channels'' such as $\nu_e \to \nu_\mu$ or $\nu_\mu \to \nu_e$ (the ``golden channel'') 
and $\nu_e \to \nu_\tau$ (the ``silver channel''). However, strong correlations between $\theta_{13}$ and 
$\delta$ \cite{Cervera:2000kp} and the presence of parametric degeneracies in the 
($\theta_{13},\delta$) parameter space, \cite{Burguet-Castell:2001ez}-\cite{Barger:2001yr}, make 
the simultaneous measurement of the two variables extremely difficult. 
A further problem arises from our present imprecise knowledge of atmospheric parameters, whose uncertainties
are far too large to be neglected when looking for such tiny signals as those expected in appearance 
experiments \cite{Donini:2005rn}.
Most of proposed solutions to these problems suggests the combination of different experiments and facilities, such as
Super-Beam's (of which T2K \cite{Itow:2001ee} is the first approved one), $\beta$-Beam's \cite{Zucchelli:sa} 
or the Neutrino Factory \cite{Geer:1997iz,Apollonio:2002en}.

Clearly, the $\nu_\mu$ disappearance channel is the best place to reduce further the uncertainties
on atmospheric parameters. It must be reminded, however, that this channel is afflicted by a four-fold degeneracy 
\cite{Donini:2004iv}: the sign of the atmospheric mass difference and the $\theta_{23}$-octant
are poorly measured through this channel, indeed\footnote{This is why we have to deal with these ambiguities when looking 
to appearance signals, ultimately.}. It has been proposed to face this problem using atmospheric neutrinos 
at T2K-II, \cite{Huber:2005ep,Gandhi:2005wa}, or at a magnetized iron calorimeter, \cite{Gandhi:2004bj,Choubey:2005zy}.

The study of the disappearance $\nu_\mu \to \nu_\mu$ channel at several proposed facilities is the main goal 
of this paper (partially presented in Ref.~\cite{Donini:2005gy}). How the combination of appearance and disappearance channels 
at different facilities can be used to soften the degeneracy problem in the measurement of $\theta_{13}$ and $\delta$ 
is also analysed. We first study in detail the disappearance channel at three proposed SuperBeam
facilities, T2K-I \cite{Itow:2001ee} (that is under construction at the Jaeri site), 
NO$\nu$A \cite{Ayres:2004js} (that passed first scrutiny and will rely upon the existing NuMI facility at FermiLab) 
and the SPL \cite{Gomez-Cadenas:2001eu}. We will show how energy resolution is crucial to reduce significantly atmospheric 
parameter uncertainties and how our present ignorance of $\theta_{13}$ and $\delta$ affects the results in the 
disappearance channel. 

We then investigate the $\nu_\mu \to \nu_\mu$ disappearance channel at the Neutrino Factory.
This channel, although already considered in the literature (see Refs.~\cite{Barger:1999jj,Bueno:2000fg}, for example), has never
gained the center of the stage on its own, being overshadowed by appearance channels such as the ``golden''
$\nu_e \to \nu_\mu$ and the ``silver'' \cite{Donini:2002rm} $\nu_e \to \nu_\tau$ transitions.
This channel is able to reduce atmospheric parameter uncertainties to an unprecedented level and, using the 
intertwine with appearance channels, to solve many of the discrete ambiguities for $\theta_{13} \geq 3^\circ - 4^\circ$.
In this range of $\theta_{13}$, combination of appearance and disappearance channels is rather effective and 
$\theta_{13}$ and $\delta$ can also be measured unambiguously. 
Unfortunately, appearance and disappearance signals are optimized with different baselines: degeneracies are
solved most effectively with a $L = 3000$ km baseline in the appearance channel and with a $L = 7000$ km baseline
in the disappearance channel\footnote{This was known for long (see Refs.~\cite{Donini:1999jc} 
and \cite{Huber:2003ak} at this regard).}.

Eventually, we present a comparison of the sensitivity to $\theta_{13}$, to the sign of the atmospheric mass
difference, to the $\theta_{23}$-octant and to maximal $\theta_{23}$ of the different facilities. Their
CP-discovery potential \cite{Donini:2004iv} is also shown. 
As expected, the Neutrino Factory outperforms the considered SuperBeams in every aspect, with the notable exception 
of the CP-discovery potential that seems to be larger for T2K-II. The Neutrino Factory is in this case limited by
the ``accidental'' flow of degeneracies towards $|\delta| = 0^\circ$ or $180^\circ$ for very small values
of $\theta_{13}$, see Refs.~\cite{Burguet-Castell:2002qx,Donini:2003vz}. 

The paper is organized as follows: in Sect.~\ref{sec:setup} we shortly introduce the four facilities and
the neutrino-nucleon cross-section; in Sect.~\ref{sec:sb} we compare the potential of the three conventional beams
in the measurement of the atmospheric parameters $\theta_{23}$ and $\Delta m^2_{23}$; 
in Sect.~\ref{sec:nf} we study the potential of the Neutrino Factory in the measurement of the atmospheric
parameters and combine the $\nu_\mu$ disappearance channel with ``golden'' and ``silver'' appearance channels;
in Sect.~\ref{sec:CPdis} we show the sensitivity to several unknowns of the considered facilities 
combining appearance and disappearance channels; in Sect.~\ref{sec:concl} we eventually draw our conclusions. 

%
%%%%%%%%%%%%%%%%%%%%%%%%%%%%%%%%%%%%%%%%%%%%%%%%%%%%%%%%%%%%%%%%%%%%%%
%
\section{The experimental setup} 
\label{sec:setup}
%
%%%%%%%%%%%%%%%%%%%%%%%%%%%%%%%%%%%%%%%%%%%%%%%%%%%%%%%%%%%%%%%%%%%%%%
%

In this section we describe, briefly, the four facilities that we will use in 
the following and we remind the neutrino-nucleon cross-section used throughout the paper.

\subsection{The T2K Beam}
\label{sec:setup:t2k}

A Super-Beam is a conventional neutrino beam with a proton intensity higher than that of existing 
(or under construction) beams such as K2K \cite{Nakaya:2005gz}, NuMI \cite{Kopp:2004sc} and the 
CNGS \cite{Kodama:2004db}. 
With respect to the $\beta$-Beam \cite{Zucchelli:sa} and the Neutrino Factory \cite{Geer:1997iz}, 
neutrino beams of a new design, 
it has the advantage of a well known technology. On the other hand, the flux composition 
(with $\nu_\mu$ as the main component for a $\pi^+$ focusing, plus a small but unavoidable admixture 
of $\bar{\nu}_\mu$, $\nu_e$ and $\bar{\nu}_e$) limits its sensitivity to $\nu_\mu \to \nu_e$ oscillations.

\begin{figure}[t!]
\vspace{-0.5cm}
\begin{center}
\begin{tabular}{cc}
\hspace{-0.3cm} \epsfxsize7.5cm\epsffile{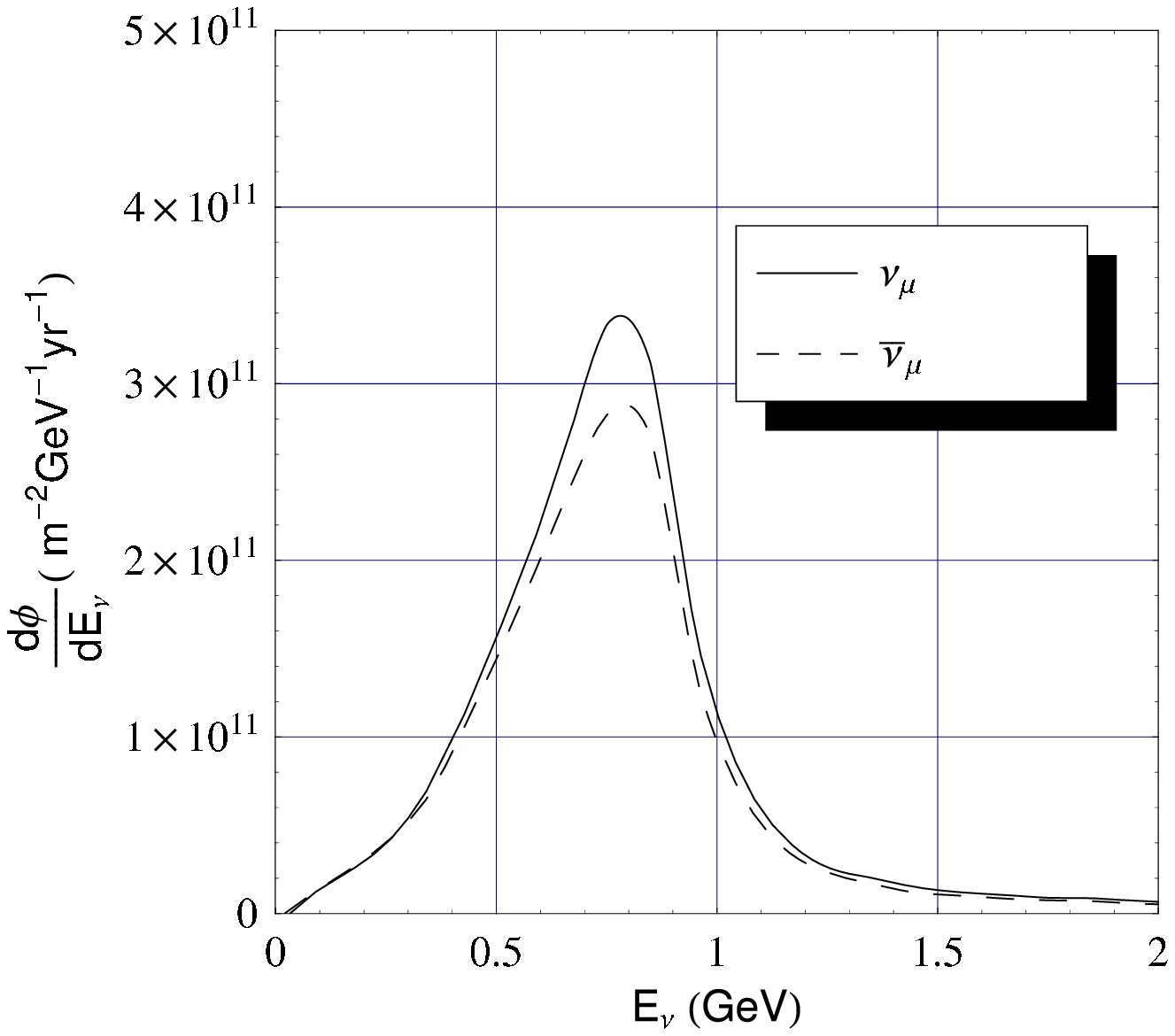}&
\hspace{-0.3cm} \epsfxsize7.5cm\epsffile{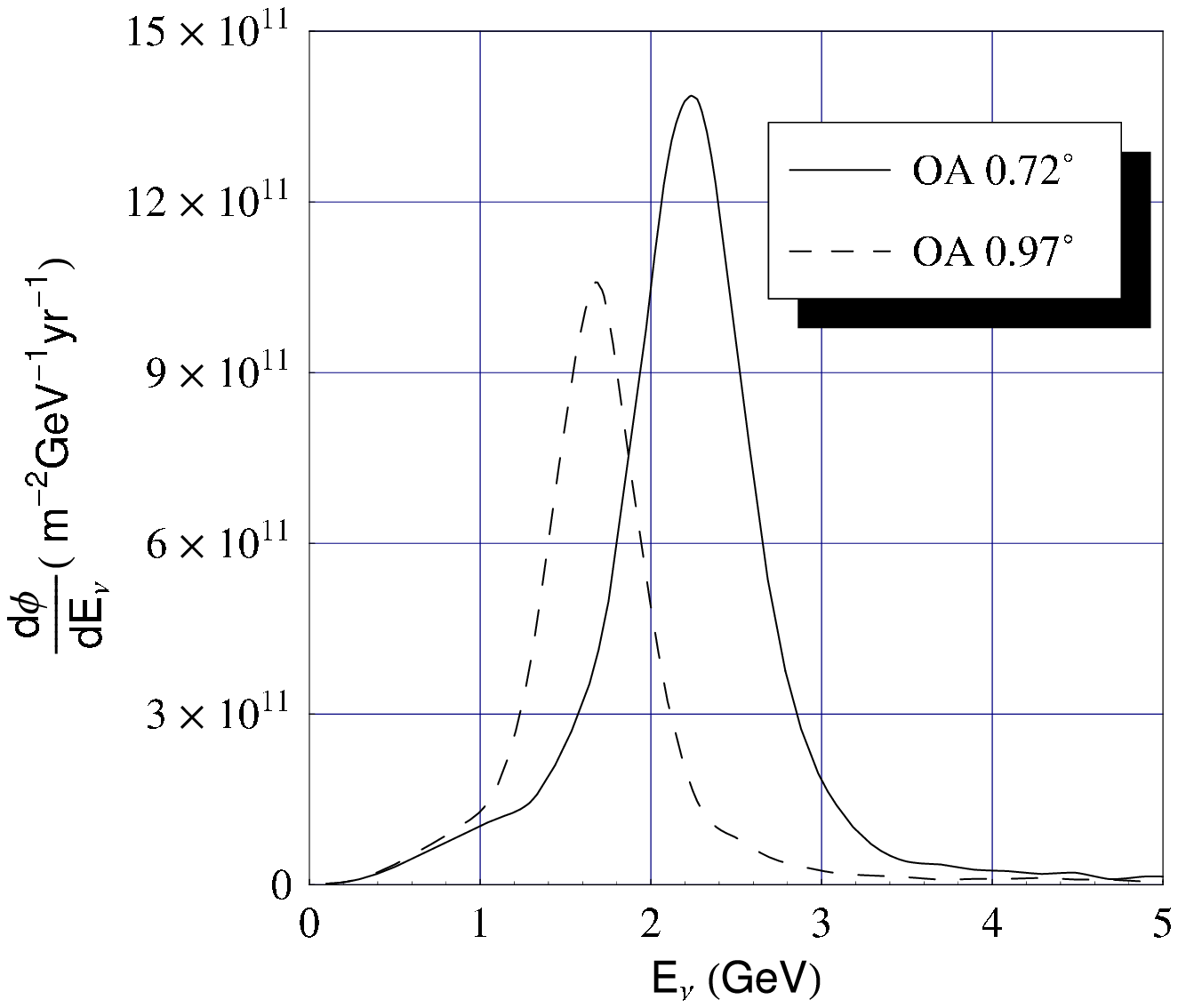} \\
\end{tabular}
\caption{\it 
Left: T2K-I fluxes at the Kamioka location (295 km baseline), \cite{Itow:2001ee}; 
Right: NO$\nu$A fluxes at the far location (812 km baseline), \cite{Huber:2004ka}.}
\label{fig:fluxes2}
\end{center}
\end{figure}

The T2K-I facility \cite{Itow:2001ee} has been approved and it is under construction. It consists of a conventional beam 
with 0.75 MWatt power 50 GeV protons produced at the J-Parc site in Tokai aiming at the Super-Kamiokande
water \v Cerenkov detector with a baseline $L = 295$ km.
The characteristic feature of the T2K beam is (apart from its power, that makes this beam a first-generation Super-Beam) 
that the detector is not on the longitudinal axis of the beam. This feature has a threefold advantage: first, it allows 
that neutrinos produced through protons whose energy is fixed by the requirement that they may be used for different 
purposes have a $\nu_\mu \to \nu_e$ oscillation probability peaked at $L = 295$ km, where the detector is located; 
second, the off-axis beam is narrower than the on-axis one, thus improving the matching of the $L/E$ ratio with the 
first peak of the oscillation; third, it allows a reduction of the beam-driven background.
Notice that the off-axis angle has not yet been chosen, to arrange the $L/E$ ratio to slight variations in the 
measured value of $\Delta m^2_{23}$. The technical design of the tunnel is such that the off-axis angle can be modified
ranging between $2^\circ$ and $3^\circ$.
The T2K fluxes for a $2^\circ$ off-axis angle as they are expected at the Kamiokande site are shown in 
Fig.~\ref{fig:fluxes2}(left).
Following the Letter of Intent \cite{Itow:2001ee}, we have considered four 200 MeV bins with $E_{min} = 400$ MeV. 
The muon and electron identification efficiencies per bin are presented in Fig.~\ref{fig:effT2K}.
The average neutrino energy is $<E_\nu> = 0.75$ GeV with 5 years of $\pi^+$ running, only.

The dominant background is made of pions from neutral currents. The detector is extremely well understood, 
being the SuperKamiokande water \v Cerenkov. As for the SPL, the dominant sources of systematic error 
(apart for the neutrino-nucleon cross-sections) are the definition of the fiducial volume in the detector and the 
overall normalization of the flux. We have introduced a ``pessimistic'' 5 \% global systematic error.

A second phase has been proposed for the T2K experiment: if a positive signal in $\nu_\mu \to \nu_e$ conversion 
is observed, the beam power is planned to be increased up to 4 MWatt and the detector mass up to 1 Mton (500 Kt fiducial mass).
A 5 years $\pi^+$ and 5 years $\pi^-$ running are envisaged. Such scenario will be called T2K-II in Sect.~\ref{sec:CPdis}.

\subsection{The NO$\nu$A Beam}
\label{sec:setup:nova}

The NO$\nu$A facility Letter of Intent \cite{Ayres:2004js} has passed the first scrutiny. It consists of a conventional
beam with 0.7 MWatt power 120 GeV protons produced at the NuMI in FermiLab aiming at a site located at $L = 812$ km 
from the source. It is still not decided the typology of the detector that will be located there, with two options
considered so far: a low-$Z$ calorimeter; or, an Icarus-like liquid argon bubble chamber.
As for T2K, the NO$\nu$A detector will be placed at a non-zero off-axis angle with respect to the longitudinal 
beam direction, to reduce backgrounds and to have a narrow beam to match the $L/E$ oscillation peak ratio.
A reduced version of the detector will be placed at the FermiLab site to extrapolate the expected rates and their energy spectrum 
in the far detector in the absence of oscillation.
The NO$\nu$A fluxes for a $0.72^\circ$ off-axis angle as they are expected at the far site are shown in 
Fig.~\ref{fig:fluxes2}(right).

In the following, we consider a totally active 30 kton detector made of 
liquid scintillator and PVC \cite{MenaRequejo:2005hn} with $6.5 \times 10^{20}$ pot/y (no proton driver is considered, \cite{Albrow:2005kw}). 
The average neutrino energy is $<E_\nu> = 2.22$ GeV, with 5 years of $\pi^+$ run only \cite{Ayres:2004js}.
Events are grouped in three $\sim$ 650 MeV bins with $E_{min} = 1000$ MeV, 
with a constant muon identification efficiency $\epsilon_\mu = 0.9$ \cite{Huber:2002rs} and a constant electron identification 
efficiency $\epsilon_e = 0.24$ \cite{MenaRequejo:2005hn}. 

The dominant background for the disappearance channel are $\nu_\mu \to \nu_x$
neutral currents with a rejection factor at the level of $5 \times 10^{-3}$
(for the appearance channel the background is typically from electron
neutrinos in the beam and from neutral current events faking electron neutrinos, 
with a rejection factor approximately of $2 \times 10^{-3}$).

The main source of systematic errors are the incomplete knowledge of the neutrino flux
and the extrapolation of the flux and of the backgrounds from the near to the far detector. 
The near detector will be used to improve the present knowledge of low energy neutrino cross-sections 
and to complement the results of the MINER$\nu$A experiment \cite{Budd:2005ga} (operating before the bulk of NO$\nu$A data is collected).
We have introduced a ``pessimistic'' 5\% global systematic error.

\subsection{The SPL}
\label{sec:setup:sb}

\begin{figure}[t!]
\vspace{-0.5cm}
\begin{center}
\begin{tabular}{cc}
\hspace{-0.3cm} \epsfxsize7.5cm\epsffile{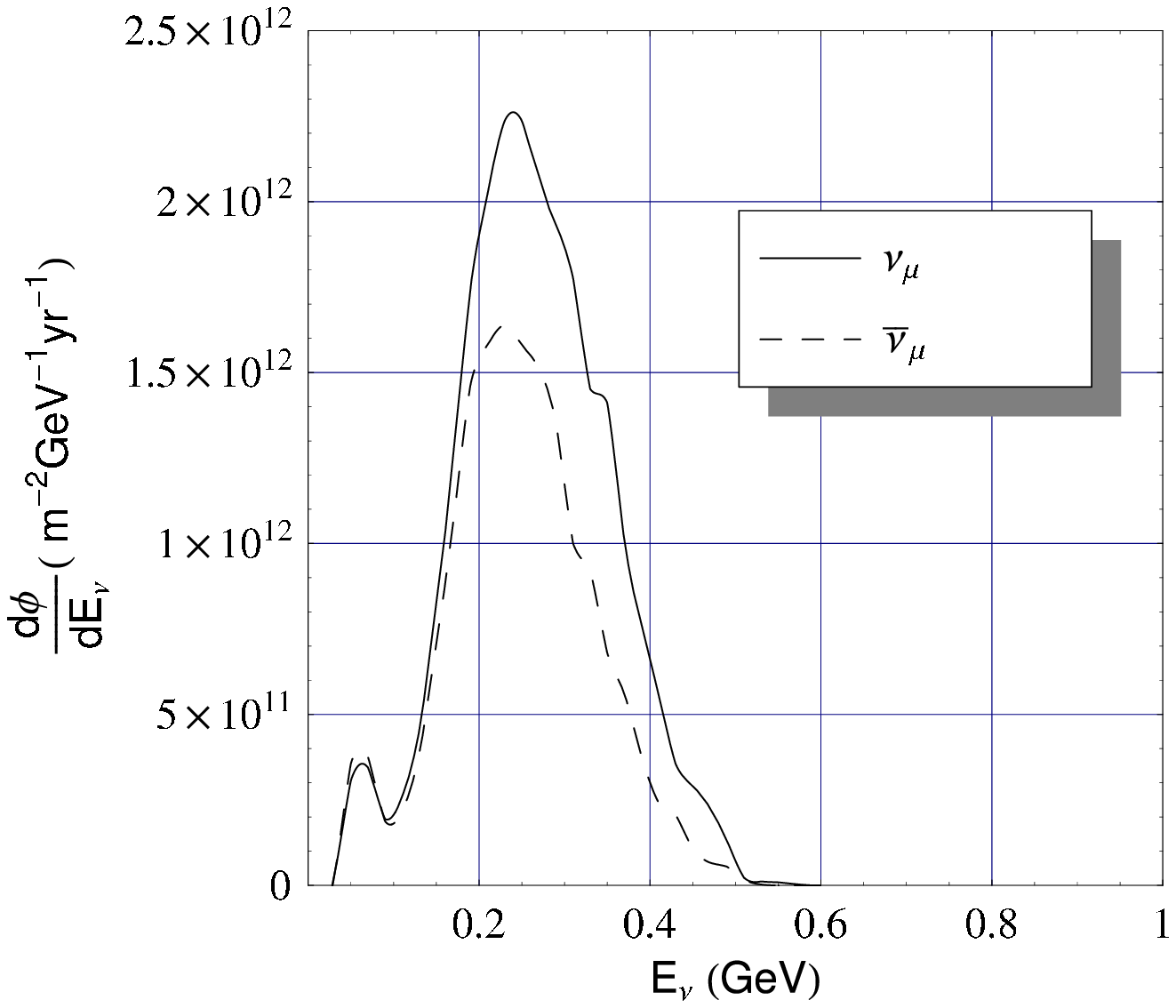}&
\hspace{-0.3cm} \epsfxsize7.5cm\epsffile{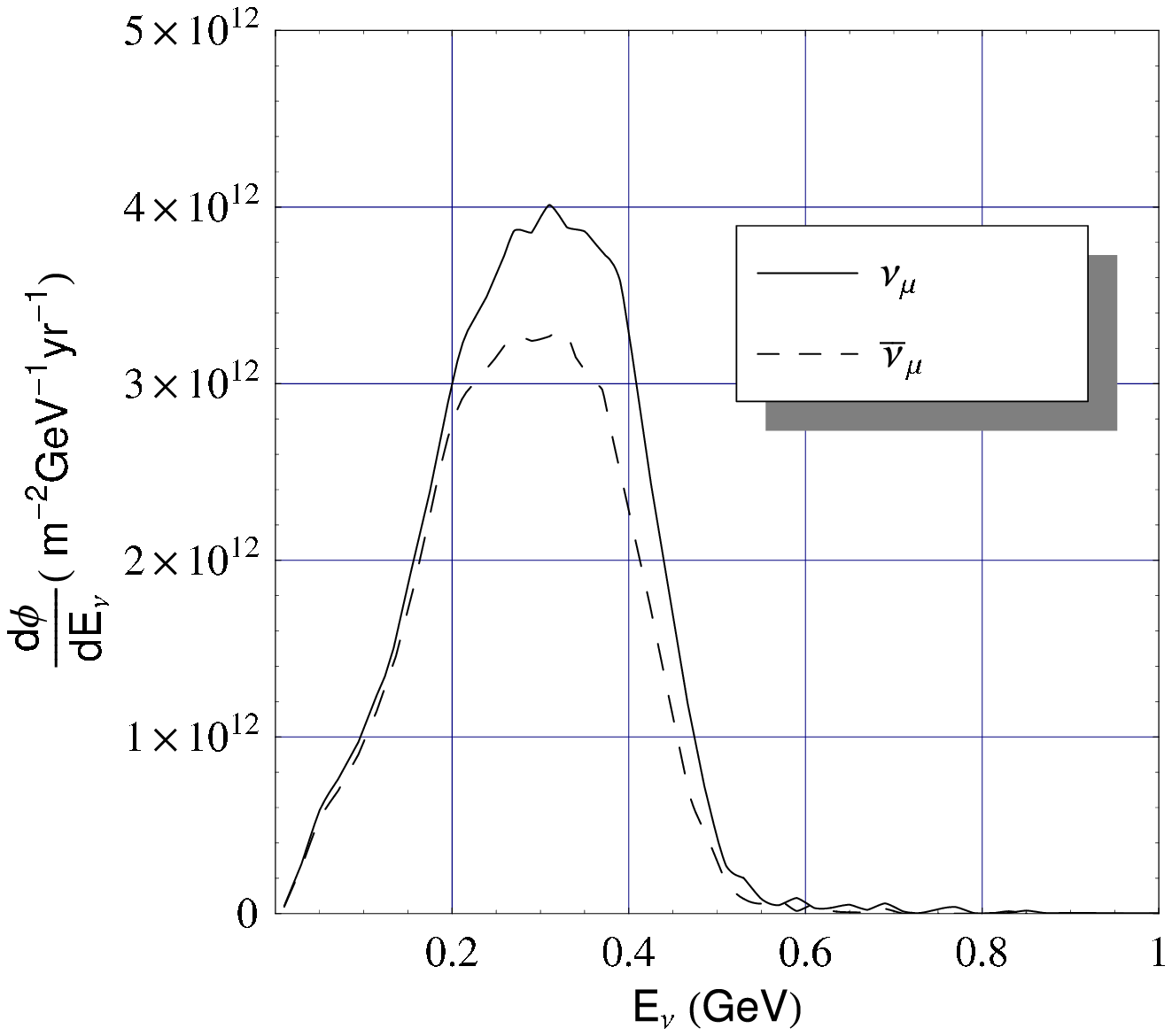} \\
\end{tabular}
\caption{\it 
Left: Standard SPL Super-Beam fluxes at the Fr{\'{e}}jus location (130 km baseline), \cite{gilardoni}; 
Right: Optimized SPL Super-Beam fluxes at the Fr{\'{e}}jus location (same baseline) \cite{Campagne:2004wt}.}
\label{fig:fluxes}
\end{center}
\end{figure}

The SPL is a proposal for a high intensity conventional beam with a 4 MWatt 2.2 GeV proton driver
located at CERN, Ref.~\cite{Gomez-Cadenas:2001eu}. The neutrino fluxes have first been computed in a full simulation of 
the beamline in Ref.~\cite{gilardoni}, assuming a decay tunnel length of 60~m. The corresponding fluxes 
are shown in Fig.~\ref{fig:fluxes}(left). However, the original beam design was conceived as
the first stage of a Neutrino Factory, and it was not optimized as a facility to look for $\nu_\mu \to \nu_e$ 
oscillations on its own. Such an optimization has been presented in Ref.~\cite{Campagne:2004wt}, and the modified
fluxes are shown in Fig.~\ref{fig:fluxes}(right). We will consider both the ``old'' and ``new'' neutrino fluxes
to study the potential of the $\nu_\mu$ disappearance channel at this facility to measure the atmospheric
parameters $\theta_{23}$ and $\Delta m^2_{23}$. 

For both fluxes, a UNO-like \cite{Jung:1999jq} 1 Mton water \v Cerenkov detector (with a 440 kton fiducial volume)
located inside the Fr\'ejus tunnel with a baseline $L = 130$ km is considered. The average $\nu_\mu (\bar \nu_\mu)$ 
(anti)neutrino energy for both the ``old'' and the ``new'' fluxes is $E_\nu \in [250 - 300]$ MeV, such as 
to match the first oscillation peak. 2 years of $\pi^+$ running and 8 years of $\pi^-$ running are considered.
The detector characteristics when exposed to the SPL beam were studied 
in detail in Ref.~\cite{Burguet-Castell:2003vv}, where an almost constant $\epsilon_e = 70$ \% efficiency
to identify electrons was found. Reasonably enough, due to the low energy of the neutrino flux, no binning 
was considered (being the neutrino energy comparable with the Fermi motion). 
Following Ref.~\cite{Blondel:2004cx}, however, we try to consider a very coarse binning grouping events in two bins 
with energies $E_1 \in [0,250]$ MeV and $E_2 \in [250,600]$ MeV. We have considered a constant bin independent 
muon identification efficiency, $\epsilon_{\mu 1} = \epsilon_{\mu 2} = 0.7$. 
The electron identification efficiency is $\epsilon_{e^-} = 0.707, \epsilon_{e^+} = 0.671$ \cite{Casper:2002sd}.
Notice that in the appearance channel no binning is considered, since in the energy range $E_1$ practically no event
is observed.

The background is made dominantly of pions from neutral currents and it is negligible in the range of energy $E_2$.
The dominant sources of systematic error (apart for the neutrino-nucleon cross-sections)
are the definition of the fiducial volume in the detector and the overall normalization of the flux. 
We have introduced a ``pessimistic'' 5 \% global systematic error.

\subsection{The Neutrino Factory}
\label{sec:setup:nf}

\begin{figure}[t!]
\vspace{-0.5cm}
\begin{center}
\begin{tabular}{cc}
\hspace{-0.3cm} \epsfxsize7.5cm\epsffile{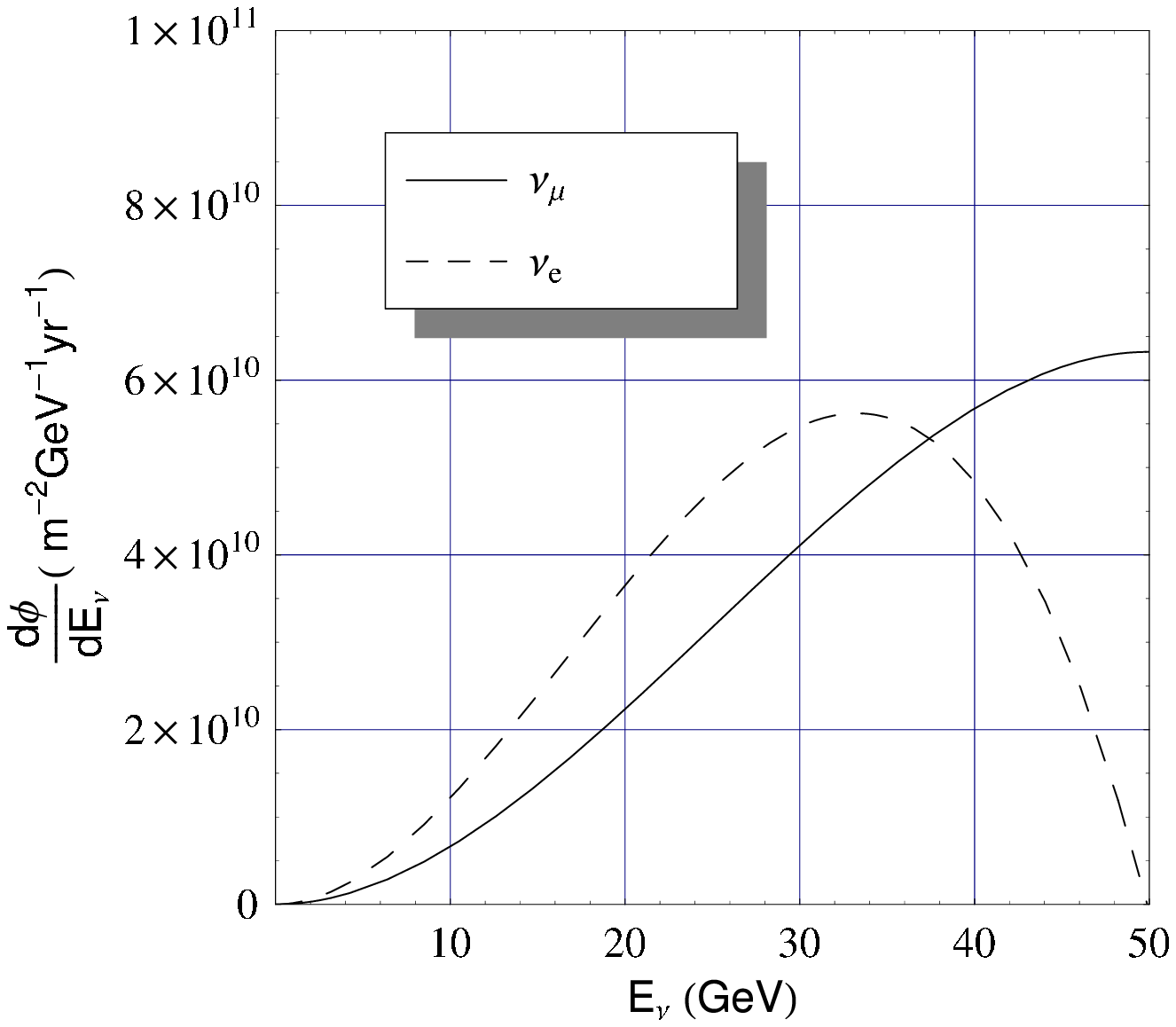} &
\hspace{-0.3cm} \epsfxsize7.5cm\epsffile{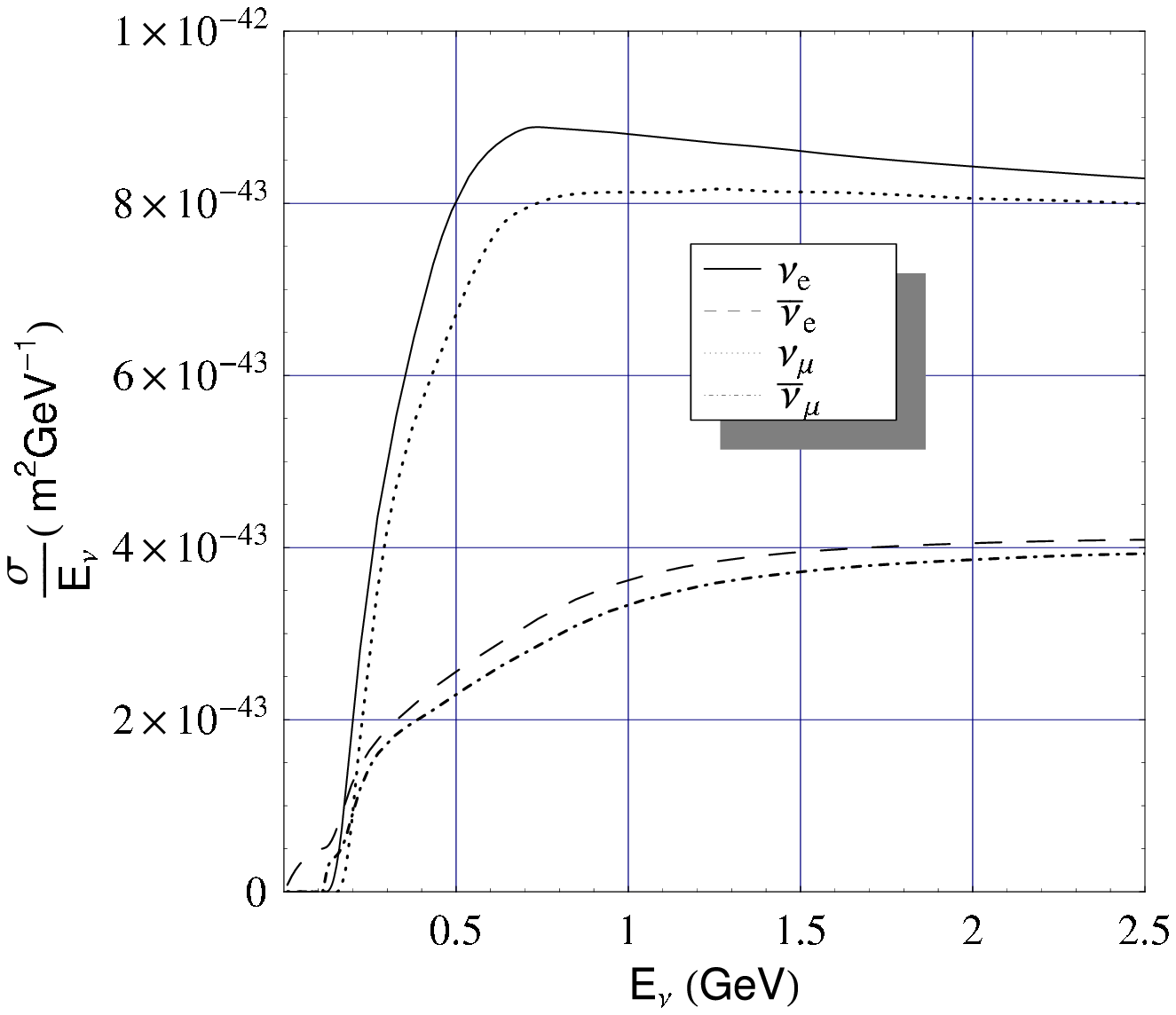}
\end{tabular}
\caption{\it 
Left: 50 GeV Neutrino Factory fluxes at 3000 km \cite{Cervera:2000kp};
Right: the $ \nu N$ and $\bar \nu N$ cross-sections on water \cite{lipari}.}
\label{fig:fluxes3}
\end{center}
\end{figure}

The Neutrino Factory proposal considered here is a 50 GeV muon storage ring fuelled through a 4 MWatt SPL-like
Super-Beam \cite{Blondel:2000gj}, with $2\times 10^{20}$ muons decaying in each straight
section of the storage ring per year. Five years of data taking for each muon polarity are envisaged.
Two detectors of different technology are considered: a 40 kton Magnetized Iron Detector (MID) located at 
$L = 3000$ km or $L = 7000$ km; and a 10 kton Emulsion Cloud Chamber (ECC) at $L= 732$ km.
This proposal corresponds to the design of a possible CERN-based Neutrino Factory Complex,
with detectors located at the Gran Sasso Laboratory as the shortest baseline 
and at a second site to be defined as the medium and long baseline. 
These detectors have been especially optimized to look for a particular signal: 
the MID for the ``golden'' channel $\nu_e \to \nu_\mu$, and the ECC for the ``silver'' channel $\nu_e \to \nu_\tau$. 
The corresponding neutrino fluxes are shown in Fig.~\ref{fig:fluxes3}(left).

The detectors background, muon identification efficiency and systematics for the golden and silver channels 
at this specific facility have been studied in Refs.~\cite{Cervera:2000vy} (the Magnetized Iron Detector) and 
\cite{Autiero:2003fu} (the Emulsion Cloud Chamber). 
It has been noticed that the MID simulation must be updated due to the growing 
evidence for the relevance of low energy bins that were sacrified at first to reduce backgrounds in 
Ref.~\cite{Cervera:2000kp}. To take advantage of the low energy bins, in this paper we will consider for the disappearance 
channel 4 GeV bins when the MID is located at the $L = 3000$ km baseline and 5 GeV bins when the MID is located at the $L = 7000$ km 
baseline, instead of the 10 GeV bins considered in the appearance channels. The disappearance signal is much different from 
the golden and silver appearance signals. In particular, not such strong cuts as for the appearance channels signals must be 
applied to reduce the background. A reasonable constant muon identification efficiency $\epsilon_\mu = 0.7$ has been applied when 
the disappearance channel is considered throughout the paper. A global 2 \% systematic error has been included. 

\subsection{The neutrino cross-section}
\label{sec:setup:xs}

An important source of systematic error is our present poor knowledge of the $\nu N$ and $\bar \nu N$ cross-sections
for energies below 1~GeV~\cite{Zeller:2003ey}: either there are very few data (the case of neutrinos) or there are no 
data at all (the case of antineutrinos). On top of that, the few available data have generally not been taken on 
the target used in the experiments (either water, iron, lead or plastics), and the extrapolation from different nuclei 
is complicated by nuclear effects that at the considered energies can play an important role. 

We use for each detector a different cross-section following Ref.~\cite{lipari}. For example, 
we show in Fig.~\ref{fig:fluxes3}(right) the cross-sections on water used for the water \v Cerenkov 
detectors throughout the paper. Notice the difference between the $\nu_e N$ and $\bar \nu_e N$ cross-sections: 
the former, being an interaction between the $\nu_e$ and a neutron inside the oxygen nucleus, is affected by 
nuclear effects and thus shows a threshold energy. The latter is mainly a $\bar \nu_e$ interaction with the
protons of the two hydrogens, approximately free. This effect, although less pronounced, is visible also for 
$\nu_\mu$ and for $\bar\nu_\mu$. This feature is quite relevant for neutrino/antineutrino of hundreds of MeV energy, 
region where different cross-sections can easily differ by a factor 2. Be aware that there are other nuclear effects
(see Ref.~\cite{Serreau:2004kx} and references therein) not included yet in any of the available calculations that could 
play an important effect at the cross-section threshold energy.

In the rest of the paper, we will make use of cross-sections on iron, lead and scintillator for the MID, the ECC 
and the NOVA detectors, respectively.

%
%%%%%%%%%%%%%%%%%%%%%%%%%%%%%%%%%%%%%%%%%%%%%%%%%%%%%%%%%%%%%%%%%%%%%%
%
\section{The measurement of $\theta_{23}$ and $\Delta m^2_{23}$ at SPL, T2K and NO$\nu$A.}
\label{sec:sb}
%
%%%%%%%%%%%%%%%%%%%%%%%%%%%%%%%%%%%%%%%%%%%%%%%%%%%%%%%%%%%%%%%%%%%%%%
%

\subsection{Correlations and degeneracies in $\nu_\mu$ disappearance}
\label{sec:sb:deg}

A conventional beam with $\nu_\mu$ neutrinos of moderate energy can perform an independent measurement 
of the atmospheric parameters $\theta_{23}$ and $\Delta m^2_{23}$ via the $\nu_\mu$ disappearance channel: 
it is expected that this kind of facilities will reduce the error on the atmospheric mass difference 
to less than 10 \% in a few years for $\Delta m^2_{23} \geq 2.2 \times 10^{-3}$ eV$^2$ \cite{Petyt:1998zd}. 
The expected error on the atmospheric angle depends on the value of $\theta_{23}$ itself, 
the smallest error for large but non maximal mixing (as it has been shown in Ref.~\cite{Minakata:2004pg}). 
It is interesting to study in detail the parameter correlations and degeneracies that affect this 
measurement and that can induce large errors. We remind the vacuum oscillation probability expanded to the second order 
in the small parameters $\tc$ and $(\Delta_{12}L/E)$ \cite{Akhmedov:2004ny}:
\bea
P(\nu_\mu \to \nu_\mu) & = & 1-  \left [ \sin^2 2 \theta_{23} -s^2_{23} \sin^2 2 \theta_{13} \cos
    2\theta_{23} \right ]\, \sin^2\left(\frac{\Delta_{23} L}{2}\right) \cr
& - & \left(\frac{\Delta_{12} L}{2}\right) [s^2_{12} \sin^2 2 \theta_{23} + \tilde{J} 
s^2_{23} \cos \delta] \, \sin(\Delta_{23} L) \cr
& - & \left(\frac{\Delta_{12} L}{2}\right)^2 [c^4_{23} \sin^2 2\theta_{12}+
s^2_{12} \sin^2 2\theta_{23} \cos(\Delta_{23} L)]
\label{eq:probdismu}
\eea
where $\tilde{J}=\cos \theta_{13} \sin 2\theta_{12}\sin 2\theta_{13}\sin 2\theta_{23}$
and $\Delta_{23}=\Delta m^2_{23}/2 E$, $\Delta_{12}=\Delta m^2_{12}/2 E$.
The first term in the first parenthesis is the dominant one and is symmetric under $\tatm \to \pi/2-\tatm$.
This is indeed the source of our present ignorance on the $\theta_{23}$-octant \cite{Fogli:1996pv}, parametrized
by $s_{oct} = {\rm sign}(\tan 2 \theta_{23})$. This symmetry is lifted by the other terms, 
that introduce a mild CP-conserving $\delta$-dependence also, albeit through subleading effects very difficult 
to isolate. 
%We present our results for the $\nu_\mu$ disappearance channel in the ($\theta_{23},\Delta m^2_{23}$) plane: 
%as a consequence, we do not need to specify the $\theta_{23}$-octant, 
%since the interval $\theta_{23} \in [35^\circ,55^\circ]$ is spanned explicitly. 

Considering that the sign of the atmospheric mass difference, $s_{atm} = {\rm sign}(\Delta m^2_{23})$, is unknown
at present, for any experimental input ($\bar \theta_{23},\Delta m^2_{atm}$) we must solve two systems of equations: 
\be
\label{eq:enedismu}
N^{\pm}_{\mu\mu}( \bar \theta_{23}, \Delta m^2_{atm}; \bar s_{atm})= 
N^{\pm}_{\mu\mu} ( \theta_{23}, |\Delta m^2_{23}|; \bar s_{atm}) \, , 
\ee
\be
\label{eq:enedismusign}
N^{\pm}_{\mu\mu}( \bar \theta_{23}, \Delta m^2_{atm}; \bar s_{atm})=
N^{\pm}_{\mu\mu} ( \theta_{23}, |\Delta m^2_{23}|; -\bar s_{atm}) \, , 
\ee
where $\bar s_{atm}$ is the physical mass hierarchy.

For non maximal $\bar \theta_{23}$ four different solutions are found: 
for $| \Delta m^2_{23}| \sim \Delta m^2_{atm}$ we get two solutions from eq.~(\ref{eq:enedismu}), 
i.e. the input value $\theta_{23} = \bar \theta_{23}$ and $\theta_{23} \simeq \pi/2 - \bar \theta_{23}$, 
being the second solution not exactly at $\theta_{23} = \pi/2 - \bar \theta_{23}$ due to the small $\theta_{23}$-octant asymmetry; 
and two more solutions from eq.~(\ref{eq:enedismusign}) at a different value of $|\Delta m^2_{23}|$ \cite{Donini:2004iv}. 
In eq.~(\ref{eq:probdismu}) we can see that changing sign to $\Delta m^2_{23}$ the second term becomes positive: 
a change that must be compensated with an increase in $|\Delta m^2_{23}|$ to give 
$P^\pm_{\mu\mu}(\Delta m^2_{atm}; \bar s_{atm}) = P^\pm_{\mu\mu}(|\Delta m^2_{23}|; - \bar s_{atm})$.

\begin{figure}[t!]
\vspace{-0.5cm}
\begin{center}
\begin{tabular}{cc}
\hspace{-1.0cm} \epsfxsize8.25cm\epsffile{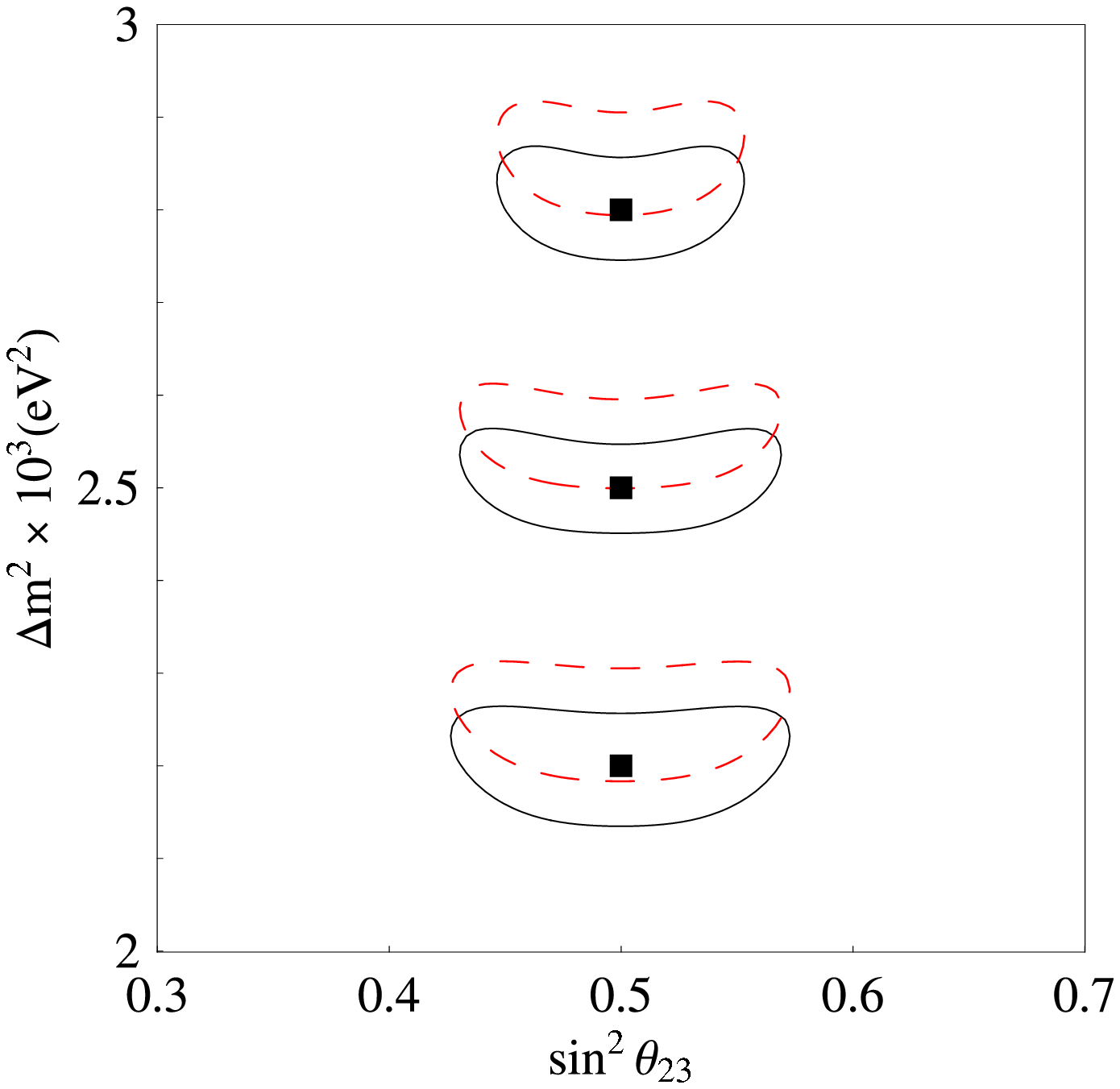} &
\hspace{-0.5cm} \epsfxsize8.25cm\epsffile{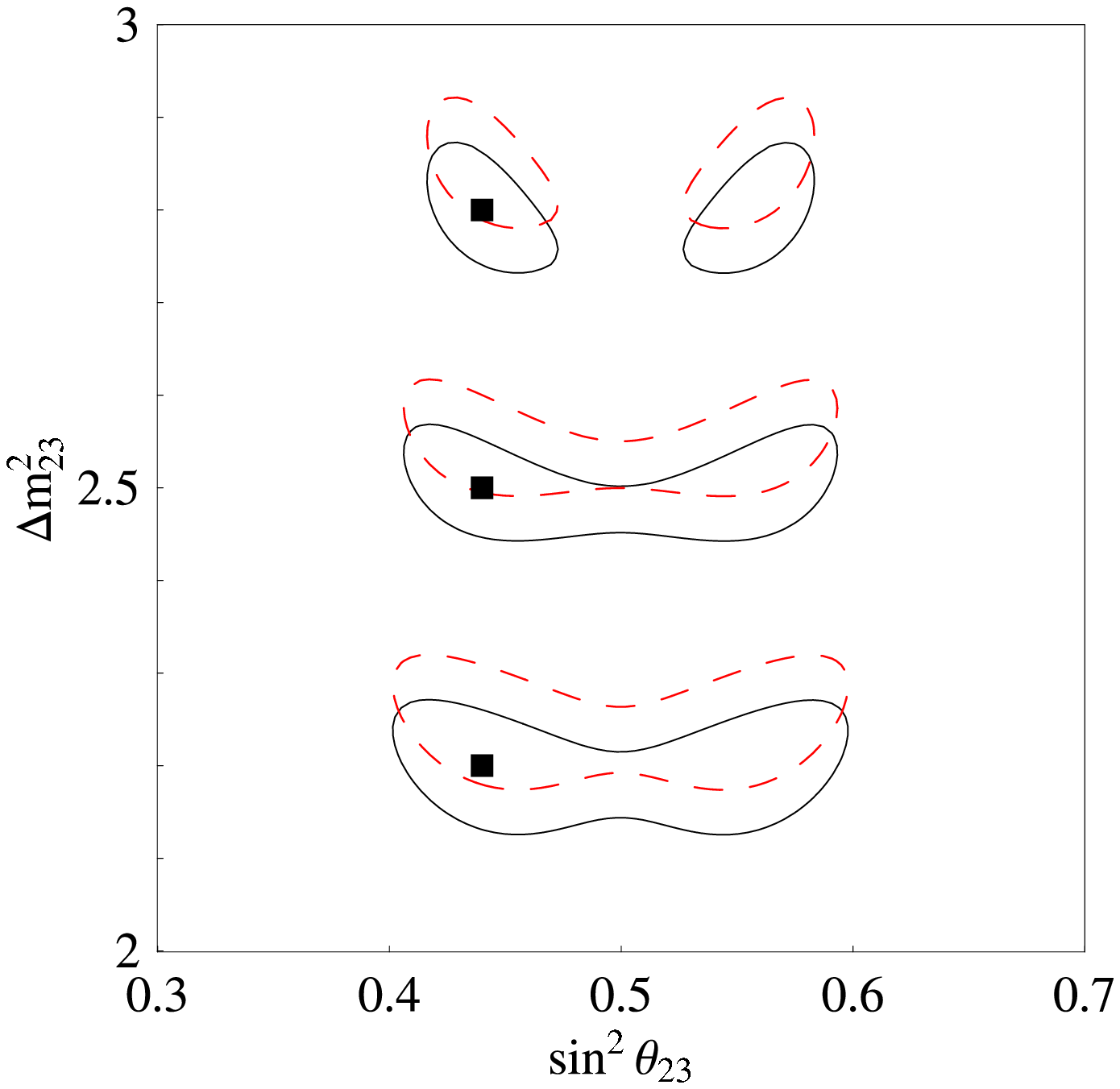} \\
\end{tabular}
\caption{\it The sign degeneracy at T2K-I; left: $\theta_{23} = 45^\circ$; right: $\theta_{23} = 41.5^\circ$.} 
\label{fig:degen}
\end{center}
\end{figure}

The result of a fit to the disappearance channel data at the T2K-phase I experiment is depicted in Fig.~\ref{fig:degen} 
for three different values of the atmospheric mass difference: $\Delta m^2_{23} = (2.2, 2.5, 2.8) \times 10^{-3}$ eV$^2$.
Fixed values of the solar parameters have been used, $\Delta m^2_{12} = 8.2 \times 10^{-5}$ eV$^2$; $\theta_{12} = 33^\circ$. 
For maximal atmospheric mixing $\theta_{23}=45^\circ$, Fig.~\ref{fig:degen} (left), two solutions
are found at 90 \% CL when both choices of $s_{atm}$ are considered. On the other hand, using a non maximal input atmospheric
angle $\theta_{23} = 41.5^\circ$ ($\sin^2 \theta_{23} = 0.44$) \cite{Fogli:2005cq} four degenerate solutions are found, 
Fig.~\ref{fig:degen}(right). In general, we must therefore speak of a two-fold or four-fold degeneracy in the disappearance channel,
as it was pointed out in Ref.~\cite{Donini:2004iv}.

Notice how the {\it disappearance sign clones} appears at a value of $|\Delta m^2_{23}|$ higher than the input value, as it 
was expected from eq.~(\ref{eq:probdismu}). The shift in the vertical axis is a function of $\theta_{13}$ and $\delta$, 
that in this particular case have been kept fixed to $\theta_{13} = 0^\circ = \delta$. The degeneracy can be softened or solved 
by using detectors with large baselines that can exploit matter effects, as it will be shown in Sect.~\ref{sec:nf}. 
Notice, eventually, that the uncertainty on the atmospheric parameters can be enhanced once we take into account that $\theta_{13}$ and $\delta$ 
are unknown \cite{Minakata:2004pg}. This will be studied in Sect.~\ref{sec:sb:t13del}.

\subsection{A matter of conventions}
\label{sec:sb:conve}

It is useful to open here a short parenthesis to address a problem that arised recently concerning the ``physical'' 
meaning of the variable used to fit the ``atmospheric'' mass difference, $\Delta m^2_{atm}$.
Notice, first of all, that the experimentally measured solar mass difference $\Delta m^2_{sol}$ can 
be unambiguously identified with the three-family parameter $\Delta m^2_{12} = m^2_2 - m^2_1$.
This is not true for the experimentally measured atmospheric mass difference $\Delta m^2_{atm}$. 
Since the subleading solar effects are, at present, barely seen in atmospheric neutrino experiments, we can define 
in different ways the three-family parameter to be used in the fits: for example, using 
$\Delta m^2_{23} = m^2_3 - m^2_2$ (the choice adopted throughout this paper), or $\Delta m^2_{13} = m^2_3 - m^2_1$, 
or even $\Delta m^2 = \left ( \Delta m^2_{23} + \Delta m^2_{13} \right )/2$, \cite{Fogli:2001wi}, we will get 
in general the same results using present data. When future experiments aiming to the measurement of 
the atmospheric mass difference at the level of $10^{-4}$ eV$^2$ will be running, however, different choices of 
the fitting parameter will give different results. 

\begin{figure}[t!]
\vspace{-0.5cm}
\begin{center}
\begin{tabular}{ccc}
\hspace{-1.0cm} \epsfxsize5cm\epsffile{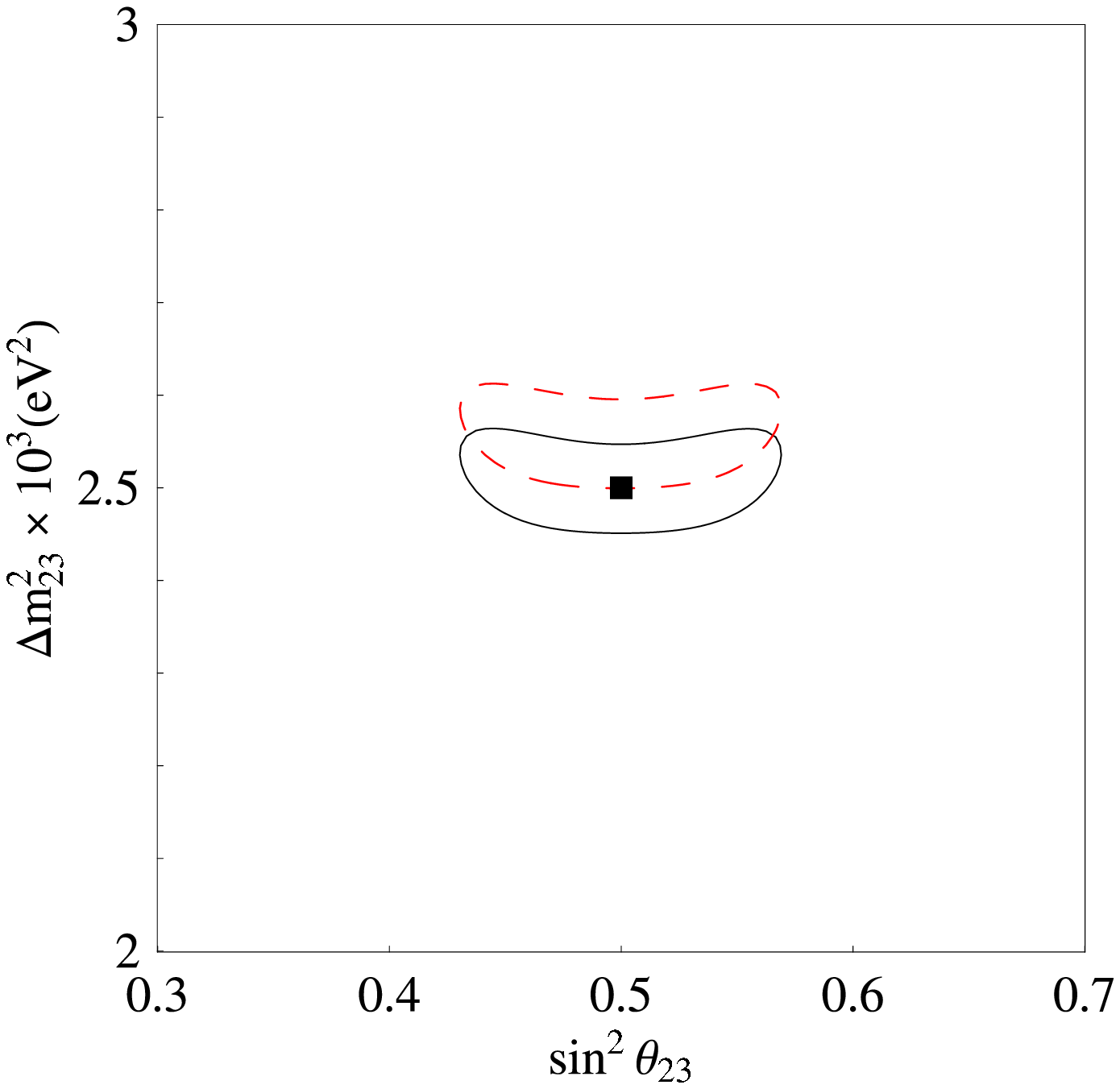} &
\hspace{-0.5cm} \epsfxsize5cm\epsffile{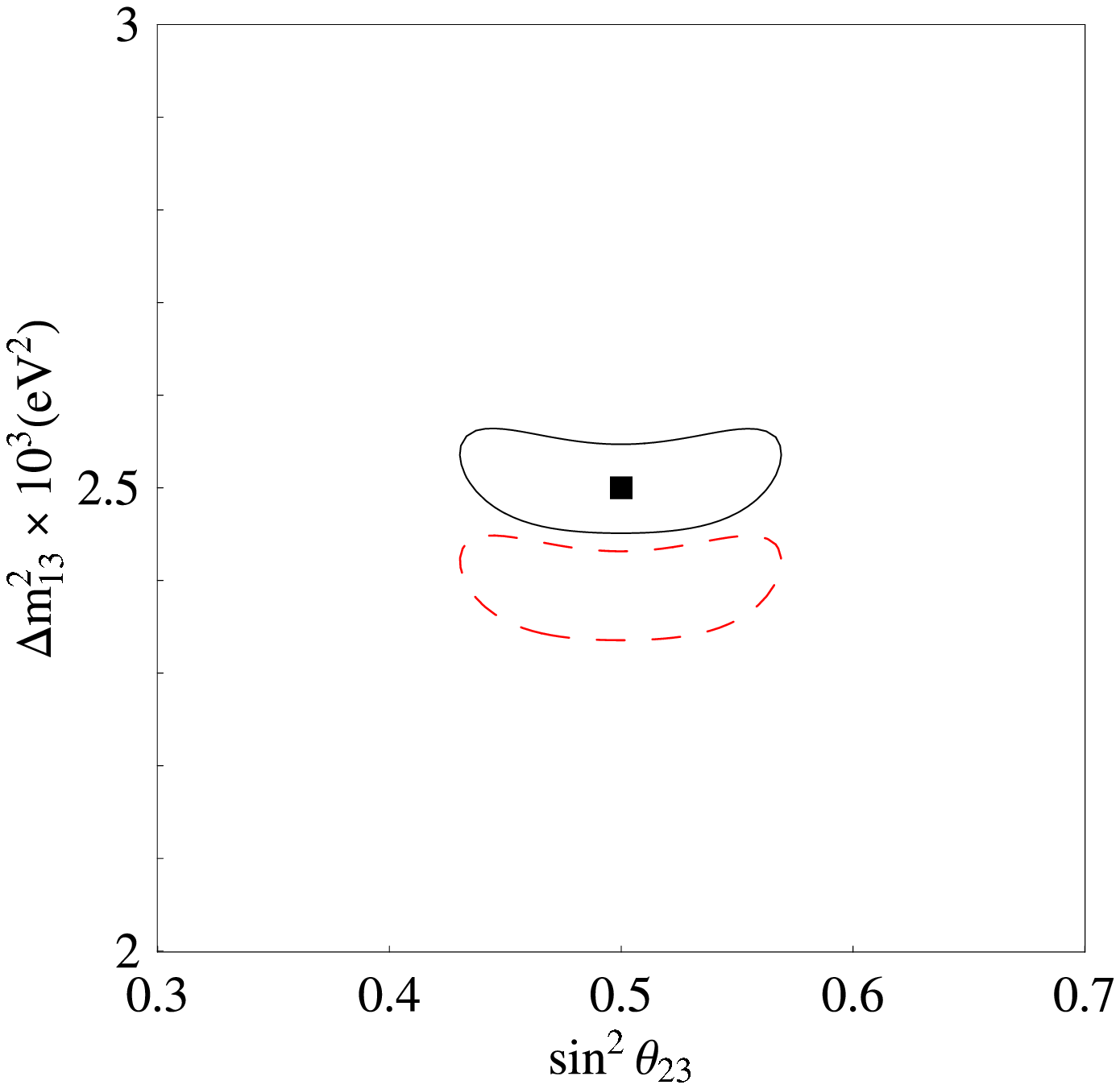} &
\hspace{-0.5cm} \epsfxsize5cm\epsffile{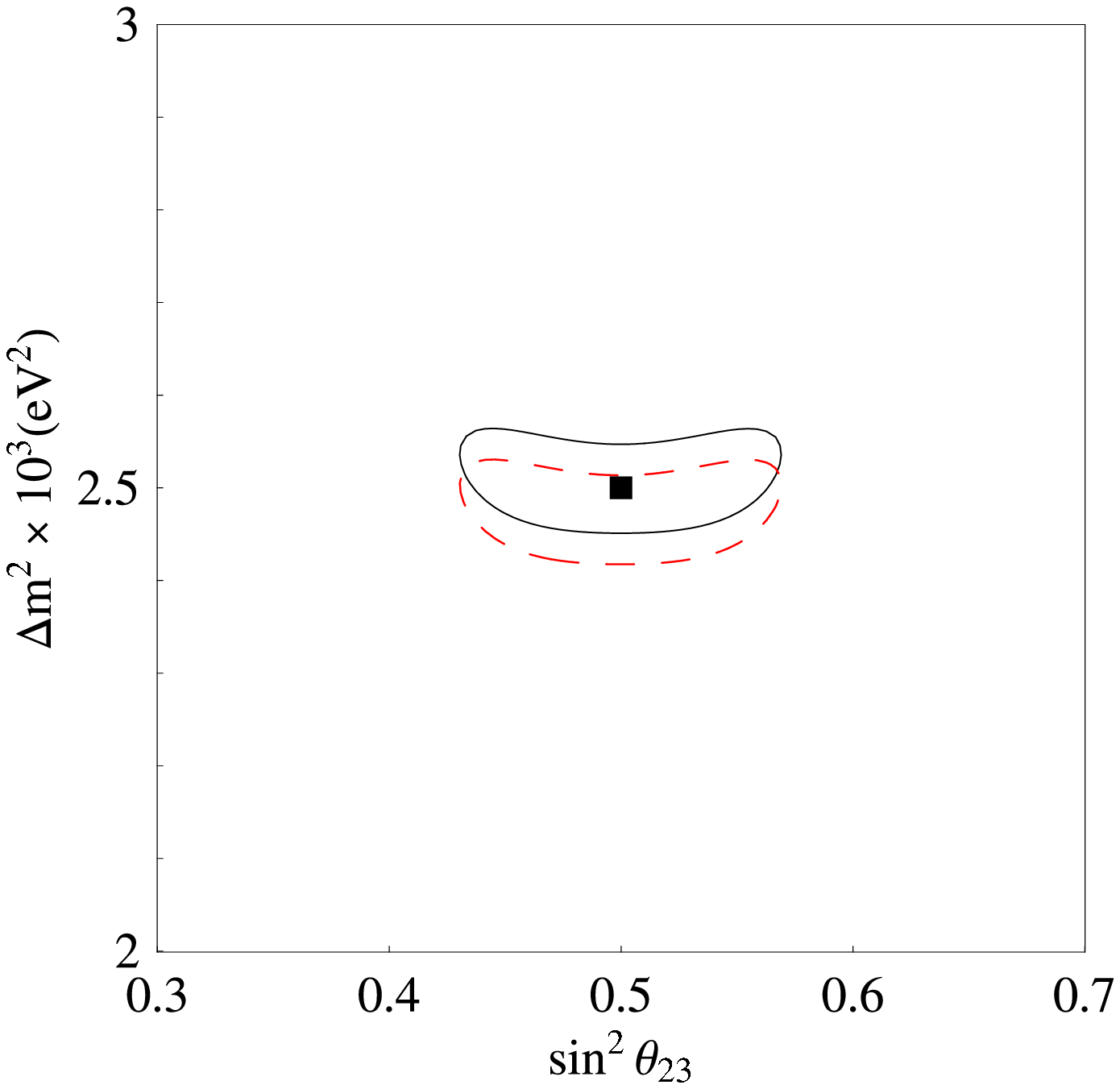} \\
\end{tabular}
\caption{\it  Different choices of the three-family ``atmospheric'' mass difference; 
left: $\Delta m^2_{23}$; middle: $\Delta m^2_{13}$; right: $\Delta m^2$, \cite{Fogli:2001wi}.} 
\label{fig:conve}
\end{center}
\end{figure}

This can be observed in Fig.~\ref{fig:conve}, where the three choices introduced above are compared. 
We plot in the three panels the 90 \% CL contours resulting from a fit to the experimental data corresponding to
the input value, $\Delta m^2_{atm} = 2.5 \times 10^{-3}$, in normal hierarchy, but fitted in turn
in $\Delta m^2_{23}$ (left panel), $\Delta m^2_{13}$ (middle panel) and $\Delta m^2$ (right panel).
As it can be seen in the figure, the contour corresponding to the normal hierarchy, $s_{atm} = \bar s_{atm}$,
is always located around the input value (in each plot the input value corresponds to a different fitting 
variable, though).
On the other hand, the contour obtained for the inverted hierarchy is located above, below and on top of the input value,
respectively, depending on the fitting variable. This is a clear consequence of the fact that the difference between 
each of the possible choices is $O (\Delta m^2_{12})$ and it is reflected in a different form of 
eq.~(\ref{eq:probdismu}). 

Up to here, it is perfectly clear what happens whenever we use a certain three-family variable to fit the 
results that at present are given under the label $\Delta m^2_{atm}$. 
A philosophical discussion, however, arised around the ``physical'' meaning of the 
different choices reported above. The idea is that the ``physically meaningful'' quantities to be measured in 
oscillation experiments are the ``oscillation frequencies''. For three-family mixing, three ``frequencies'' can 
be defined, the shortest being the solar oscillation frequency (as we said, unambiguously related to the mass
difference $\Delta m^2_{12}$). In normal hierarchy, the middle frequency is related to $\Delta m^2_{23}$ and the
longest one to $\Delta m^2_{13}$. In inverted hierarchy these two frequencies are interchanged and
the middle frequency will be related to $\Delta m^2_{31}$ and not to $\Delta m^2_{32}$. For this reason, 
it has been suggested that plots in different hierarchies should be presented using different variables where
maintaining the ordering of the oscillation frequencies. For example, if we choose to identify $\Delta m^2_{atm}$ 
with $\Delta m^2_{23}$
in normal hierarchy (i.e., with the middle frequency), we should identify it with $\Delta m^2_{31}$ in inverted 
hierarchy (i.e., again with the middle frequency). As a consequence, we could not plot contours for the two hierarchies
on the same figure (the vertical axis corresponds to different variables depending on the choice of $s_{atm}$).
This is a drawback of giving to the oscillation frequencies a deeper physical meaning than to the mass
differences. 
A new variable to be identified with $\Delta m^2_{atm}$ has been introduced in Ref.~\cite{Fogli:2001wi}: 
$\Delta m^2 = \left ( \Delta m^2_{23} + \Delta m^2_{13} \right )/2$. When changing hierarchy, this variable
just flip its sign (since $\Delta m^2_{23} \to - \Delta m^2_{13}$ and viceversa). It is thus possible 
to present both hierarchies in the same figure with $|\Delta m^2|$ in the vertical axis. 
The oscillation frequency related to this variable is neither the longest nor the middle one, 
but it maintains its role of next-to-longest or a-bit-longer-than-the-middle-one in both hierarchies\footnote{
It should be stressed that even fitting in inverted hierarchy using $\Delta m^2$ as the fitting variable, the 
allowed region is shifted with respect to that corresponding to the normal hierarchy. This can be observed in 
Fig.~\ref{fig:conve}(right), and it implies that $|\Delta m^2|_{NH} \neq | \Delta m^2 |_{IH}$.}. 

It seems to us that the ``physical meaning'' of a frequency is not deeper than that of a mass difference, and that
it is perfectly acceptable to use any variable to fit the experimental data when considering a full three-family 
analysis.
What is really important is to be consistent with the adopted choice, in particular when adding appearance and 
disappearance data, something that we will do in Sect.~\ref{sec:sb:t13del}. 
In the rest of the paper we adopt $\Delta m^2_{23}$ as fitting variable.

\subsection{The importance of energy resolution}
\label{sec:sb:bin}

It is extremely important that an experiment whose goal is to improve significantly the present uncertainties on the 
atmospheric parameters may be able to use energy dependence. A counting experiment is certainly limited, as it has 
been shown in Refs.~\cite{Donini:2004hu,Donini:2004iv}.

\begin{figure}[t!]
\vspace{-0.5cm}
\begin{center}
\begin{tabular}{cc}
\hspace{-1.0cm} \epsfxsize8cm\epsffile{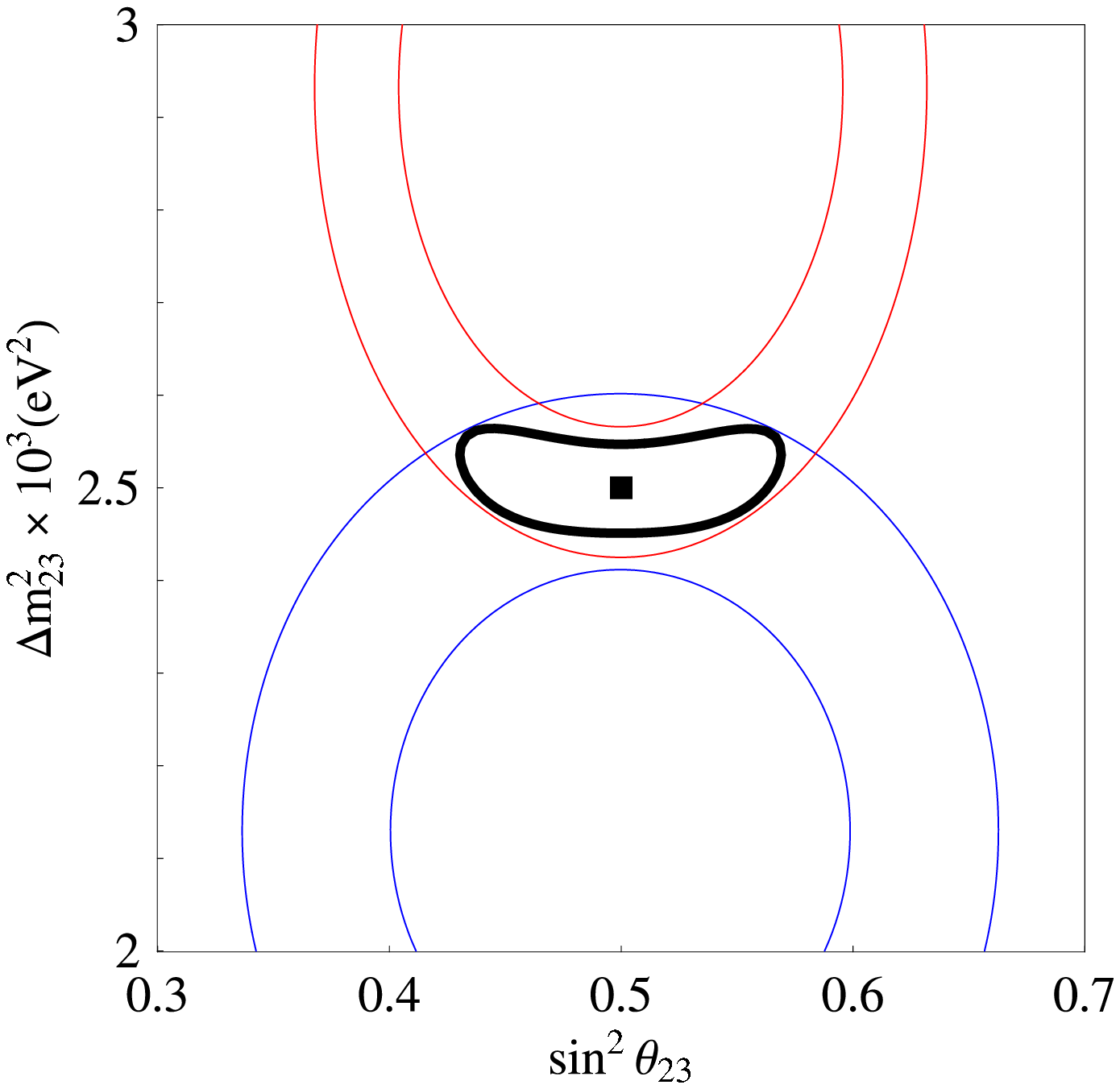} &
\hspace{-0.5cm} \epsfxsize8cm\epsffile{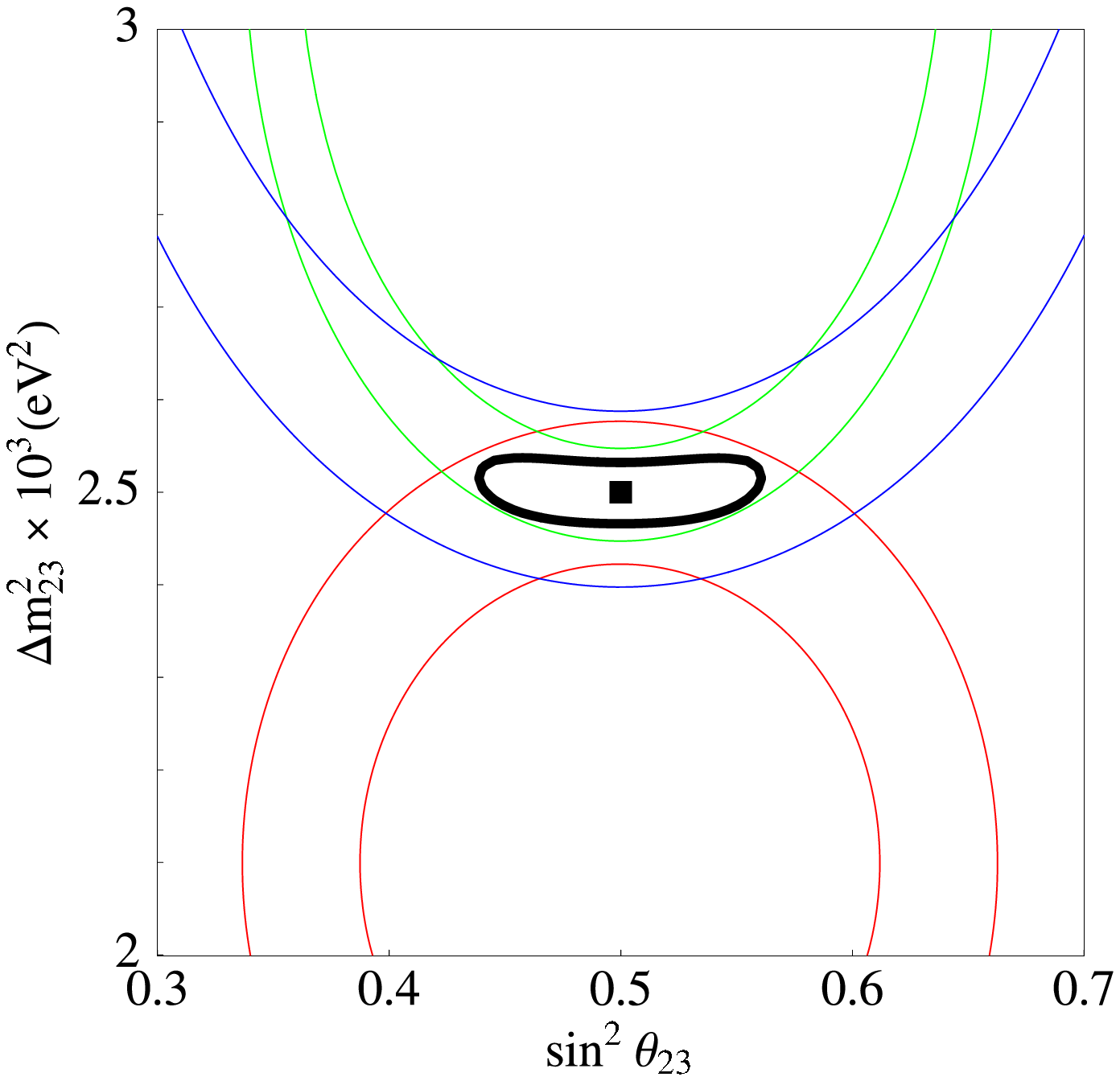} \\
\hspace{-1.0cm} \epsfxsize8cm\epsffile{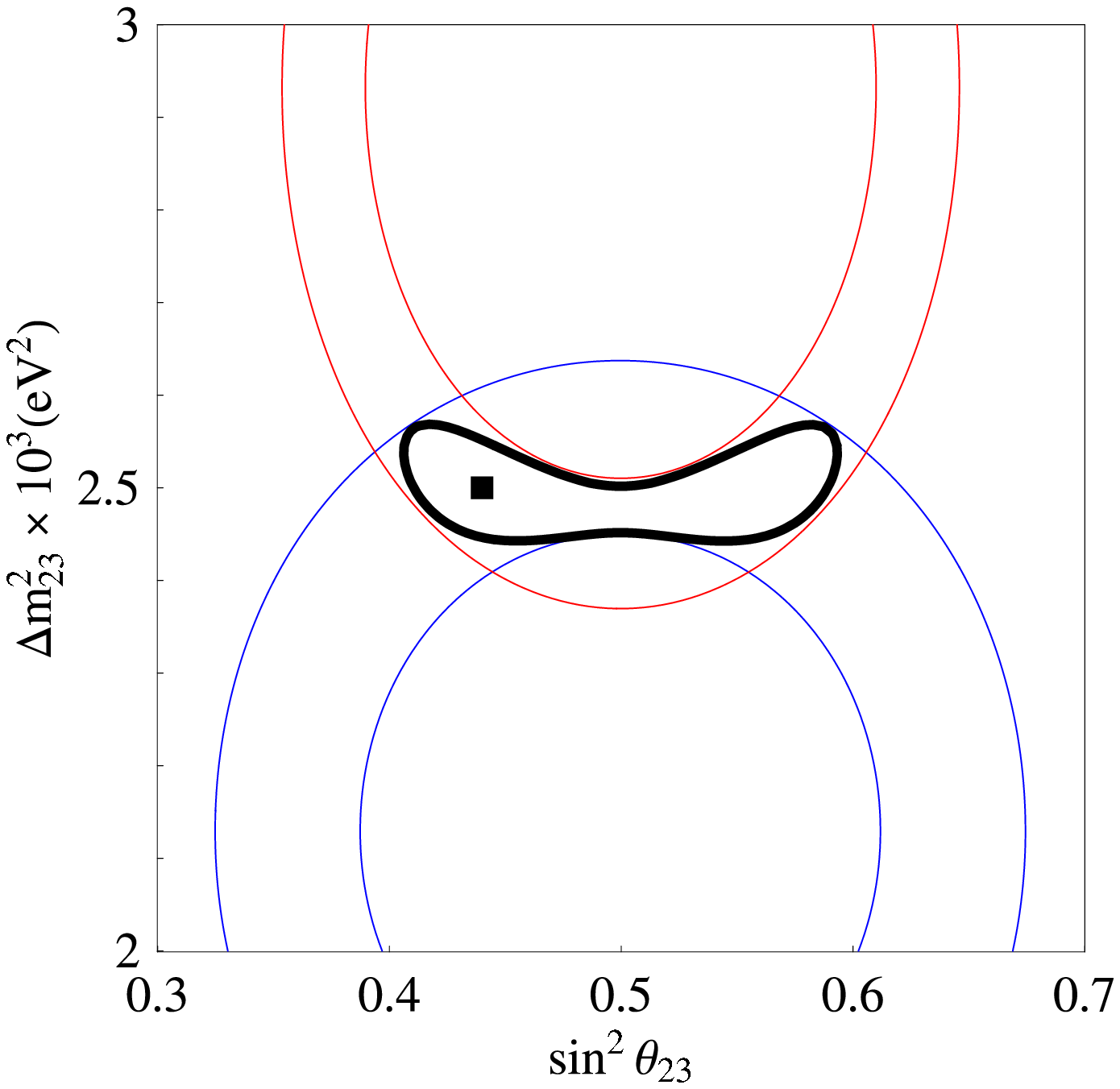} &
\hspace{-0.5cm} \epsfxsize8cm\epsffile{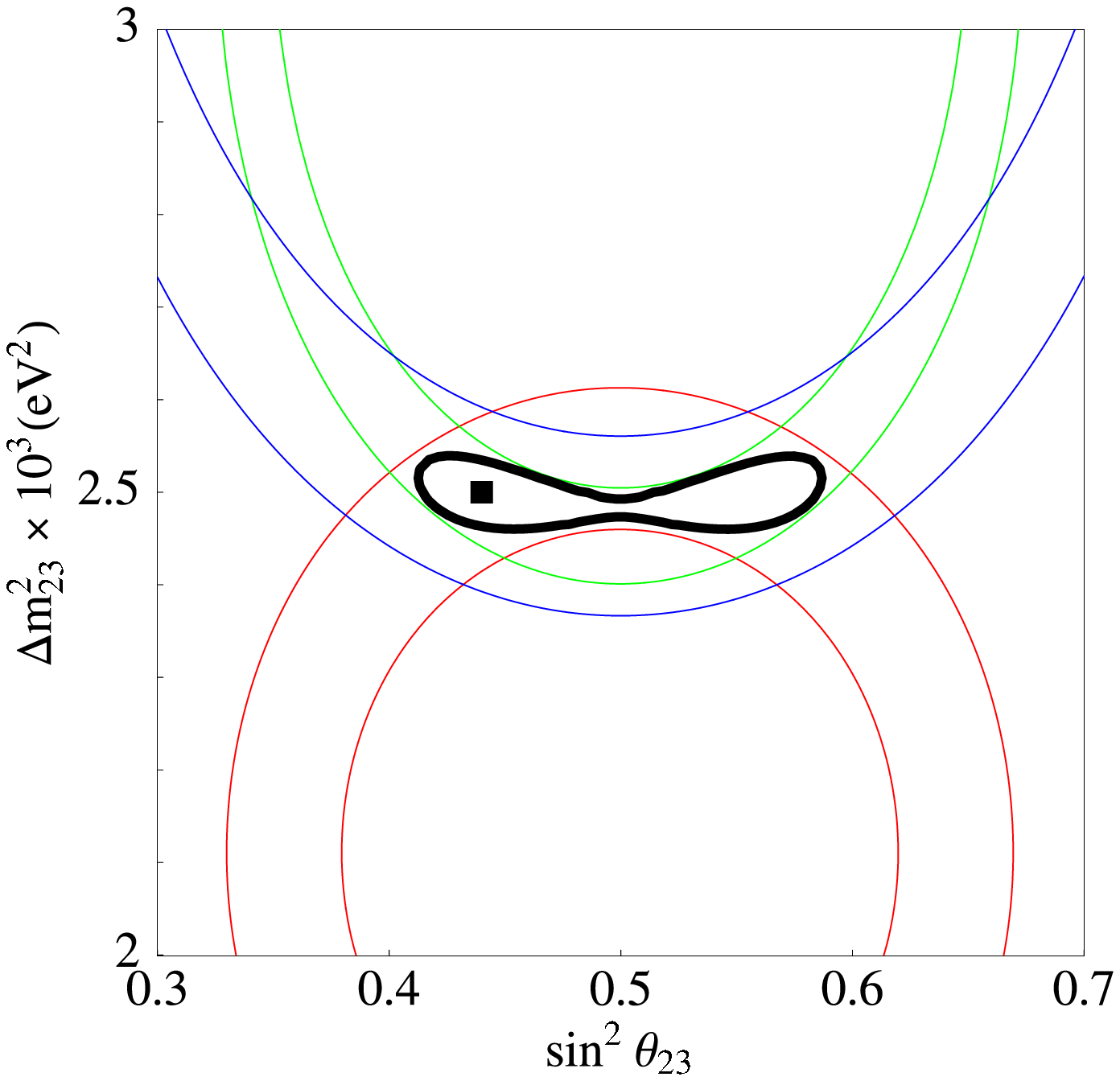} \\
\end{tabular}
\caption{\it Binning at T2K-I (left) and NO$\nu$A (right); top: $\theta_{23} = 45^\circ$; bottom: $\theta_{23} = 41.5^\circ$.}
\label{fig:T2Kbins}
\end{center}
\end{figure}

In Fig.~\ref{fig:T2Kbins} we present a comparison of the disappearance channel at T2K-phase I (left panels) 
and NO$\nu$A (right panels). Both maximal mixing, $\theta_{23} = 45^\circ$ (top panels), 
and non maximal mixing, $\theta_{23} = 41.5^\circ$ (bottom panels), are used as input values. 
In each plot we present the 90\% CL contours in the $(\theta_{23},\Delta m^2_{23})$
plane for different energy bins (only the bins corresponding to neutrinos 
with energy just below and above the peak energy are reported for T2K, all bins for NO$\nu$A). 
Again, solar parameters are kept fixed to their present best fit values, $\Delta m^2_{12} = 8.2 \times 10^{-5}$ eV$^2$, 
$\theta_{12} = 33^\circ$, and the unknowns are fixed to $\theta_{13} = \delta = 0^\circ$.
Both for T2K-phase I and NO$\nu$A 5 years of $\pi^+$ are considered.
Notice that the contours corresponding to neutrinos with an energy below 
and above the oscillation peak have a different shape. As a consequence, the combination of different bins 
significantly increases the $\theta_{23}$ resolution of the experiment with respect to a counting experiment. 
%(as it was observed in Ref.~\cite{Donini:2004iv}). 
It can be seen that T2K-I is able to measure $\Delta m^2_{23}$ with a precision of less than $10^{-4}$ eV$^2$ for
$\Delta m^2_{23} = 2.5 \times 10^{-3}$ eV$^2$. However, it is not able to exclude maximal mixing at 90 \% CL
for $\theta_{23} = 41.5^\circ$. The same happens for NO$\nu$A, see Fig.~\ref{fig:T2Kbins}(right).

\begin{figure}[t!]
\vspace{-0.5cm}
\begin{center}
\begin{tabular}{cc}
\hspace{-1.0cm} \epsfxsize8cm\epsffile{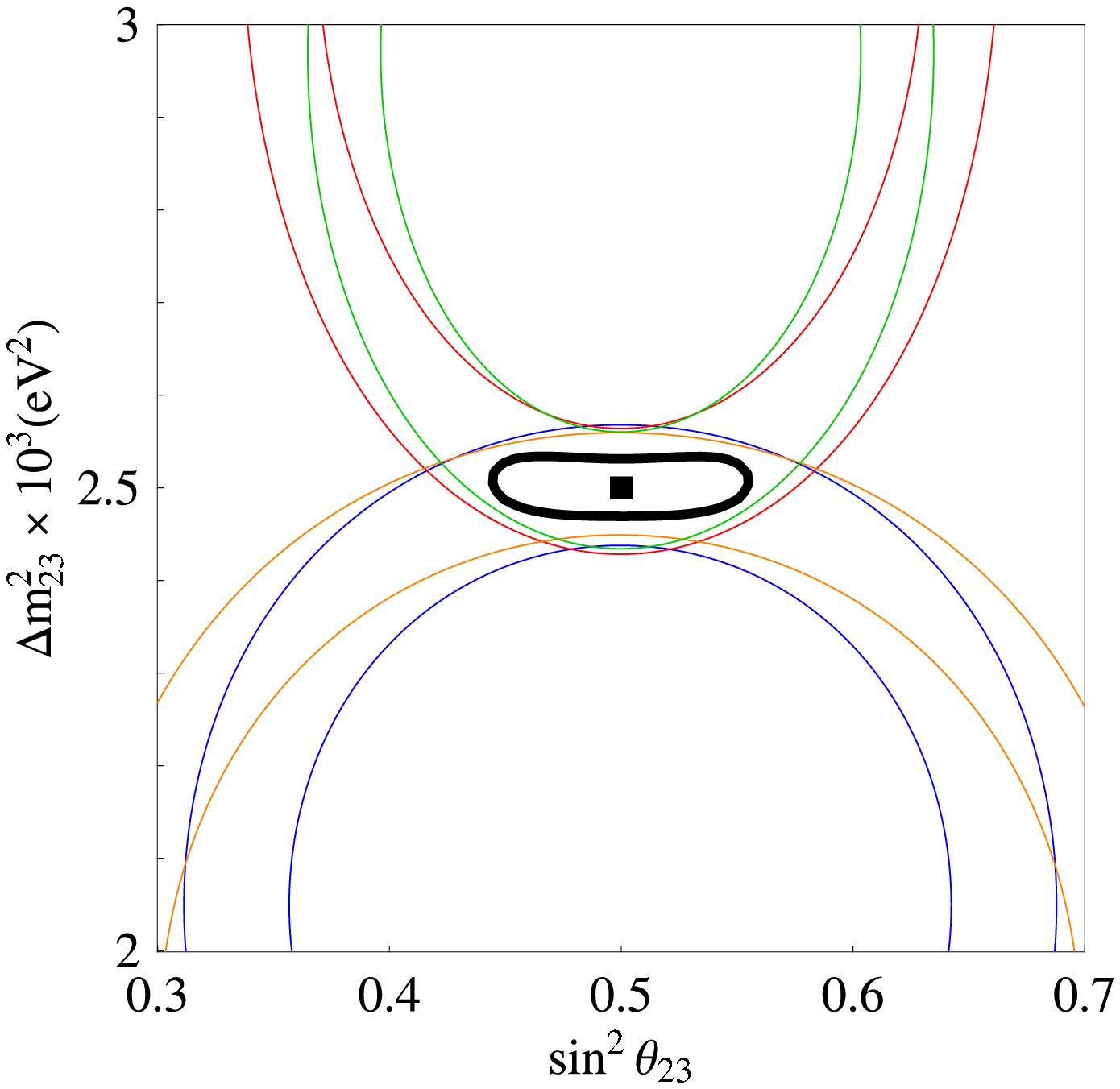} &
\hspace{-0.5cm} \epsfxsize8cm\epsffile{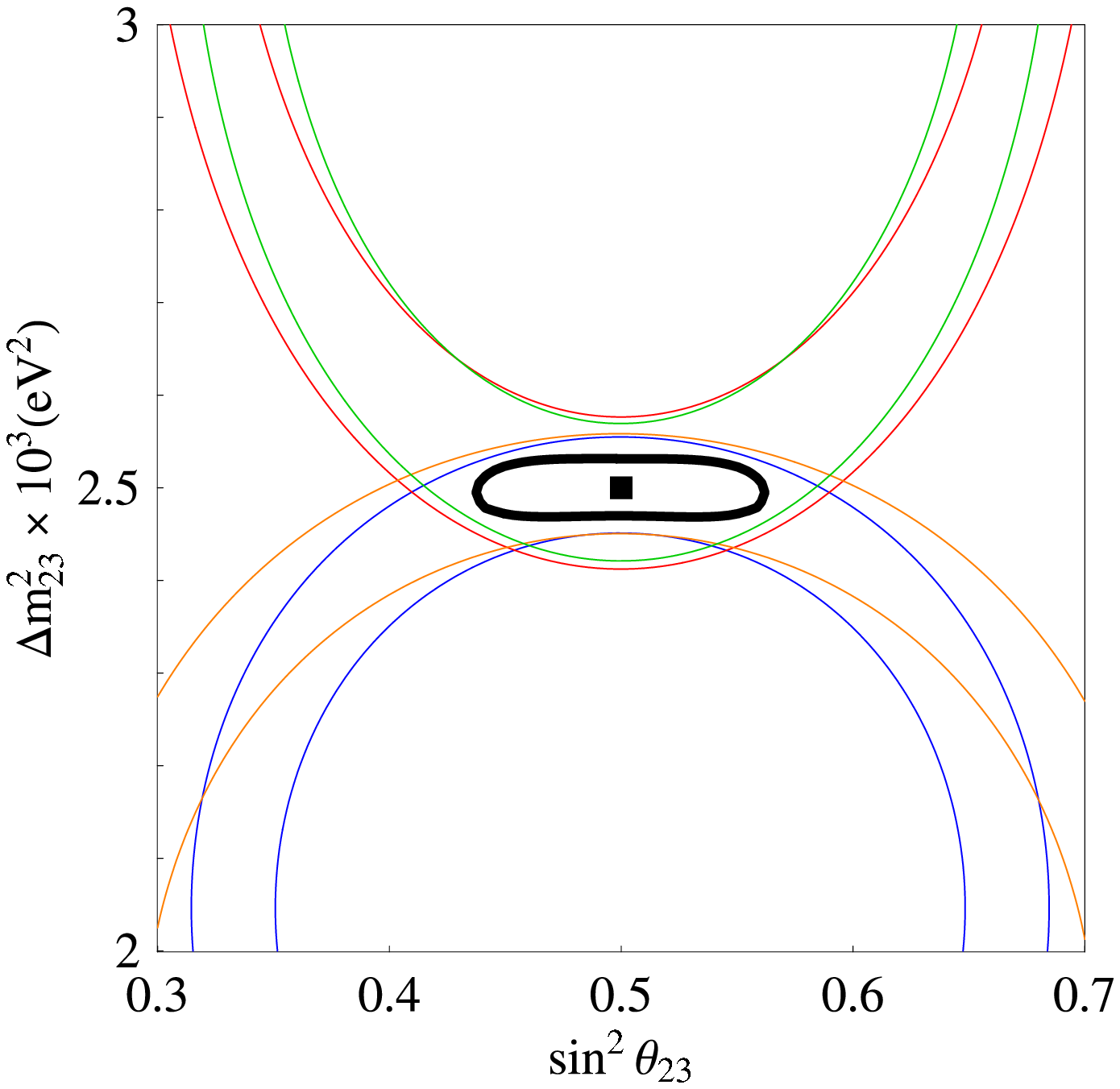} \\
\hspace{-1.0cm} \epsfxsize8cm\epsffile{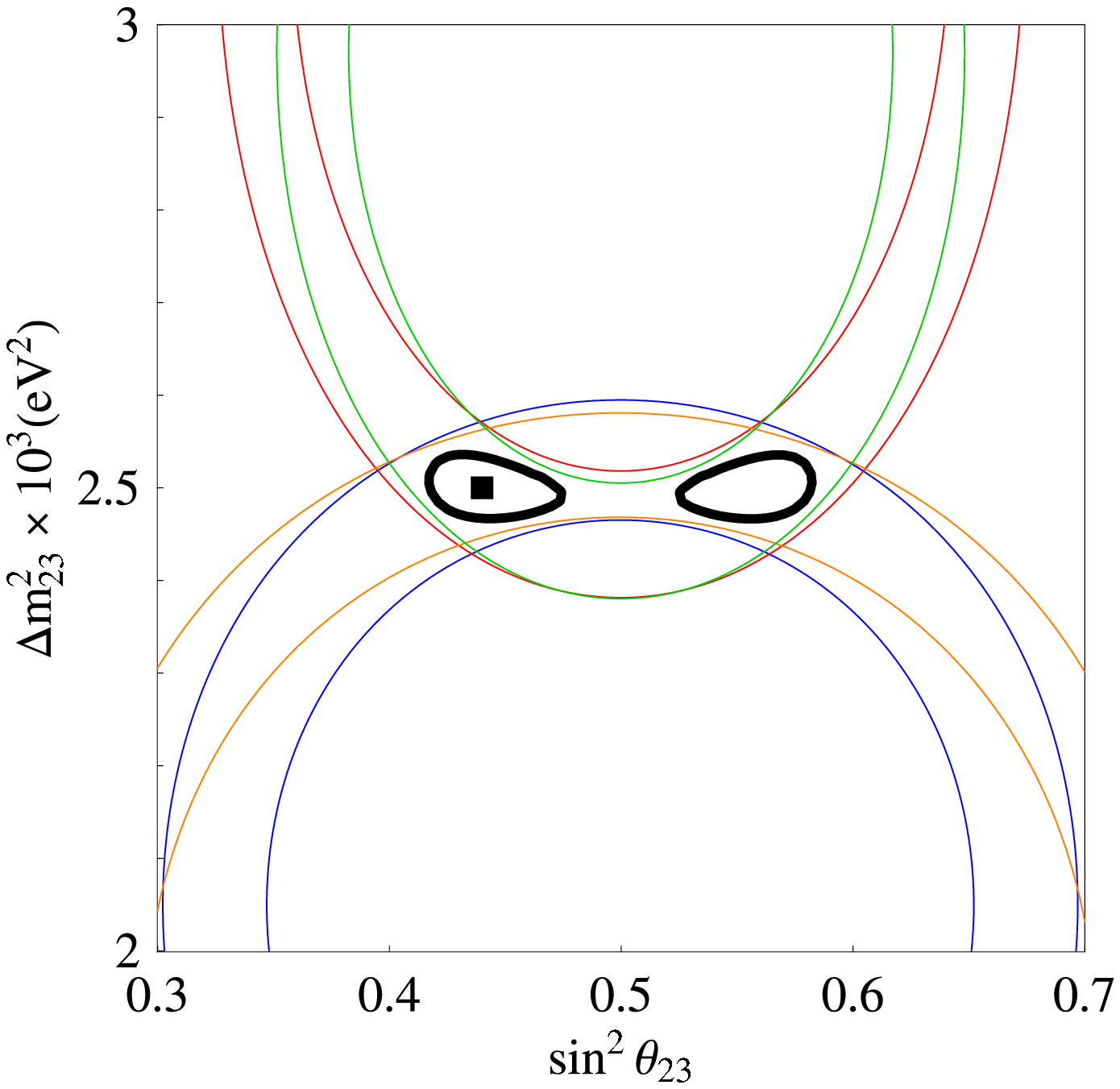} &
\hspace{-0.5cm} \epsfxsize8cm\epsffile{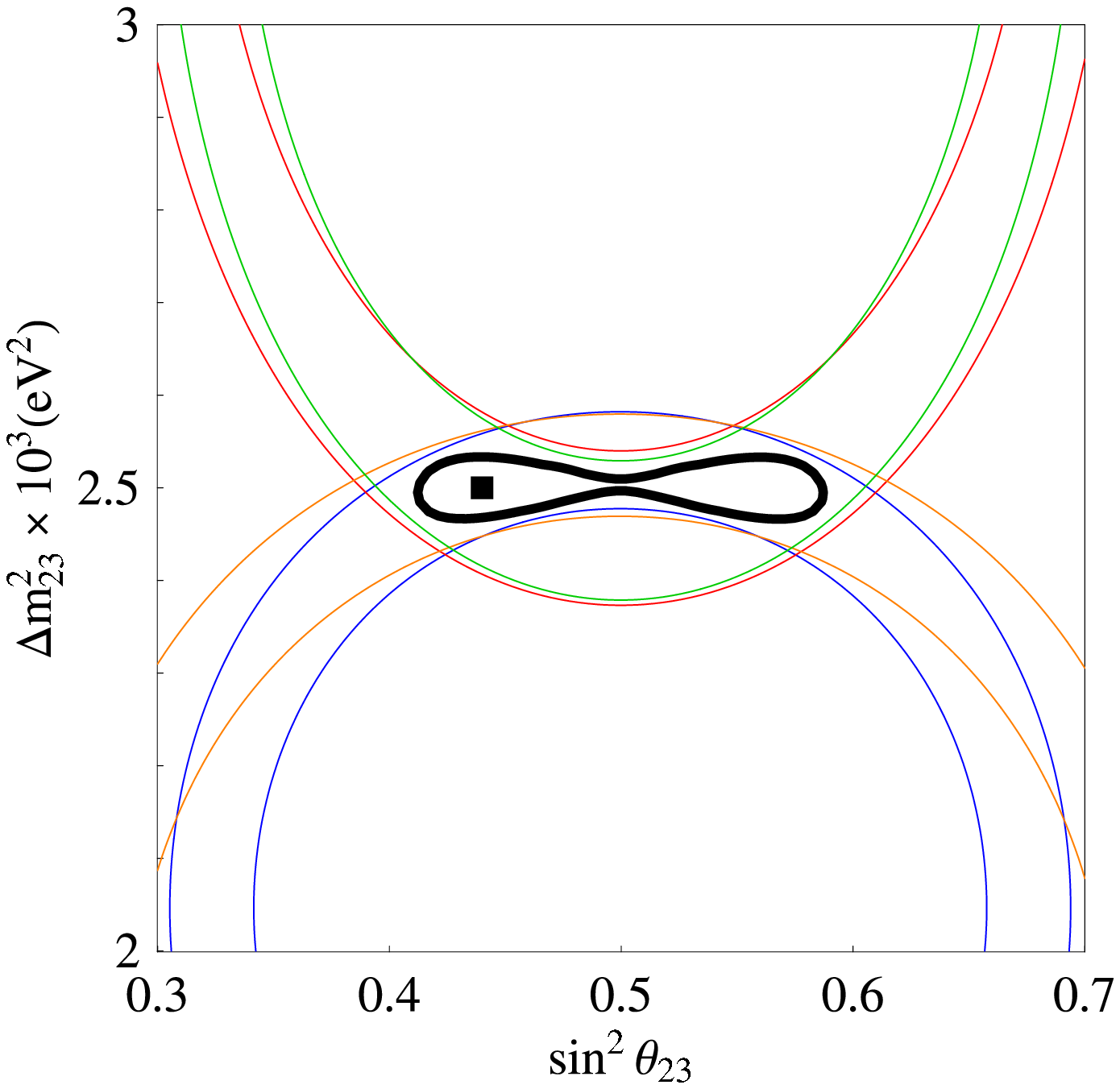} \\
\end{tabular}
\caption{\it Binning at the standard (left) and optimized SPL (right); top: $\theta_{23} = 45^\circ$; bottom: $\theta_{23} = 41.5^\circ$.}
\label{fig:SPLbins}
\end{center}
\end{figure}

In Fig.~\ref{fig:SPLbins} we present a comparison of the disappearance
channel at the SPL using the neutrino fluxes of Ref.~\cite{gilardoni}, Fig.~\ref{fig:fluxes}(left panels), 
or the neutrino fluxes of Ref.~\cite{Campagne:2004wt}, Fig.~\ref{fig:fluxes}(right panels),
again for both maximal mixing, $\theta_{23} = 45^\circ$ (top panels), and non maximal mixing, 
$\theta_{23} = 41.5^\circ$ (bottom panels). The 90 \% CL contours corresponding to both SPL energy bins are drawn.
In this case, both neutrinos and antineutrinos fluxes are produced by $\pi^+$ and $\pi^-$ decays with 2 and 8 years
of data taking, respectively. Notice that, however, being the neutrino and antineutrino 
average energies extremely similar in this setup, the neutrino (antineutrino) contours are almost superimposed. 
As a consequence, experimental information from the neutrino and the antineutrino fluxes is not complementary 
and we observe just an increase in the statistics. For this reason, although in this setup a 1 Mton detector 
is considered, the resolution in $\theta_{23}$ is not astonishing. 
A big improvement with respect to the results presented for this facility in 
Ref.~\cite{Donini:2004iv,Donini:2005rn} is represented by the spectral information: 
in the new analysis two energy bins are considered for both setups. 
Thanks to this, maximal mixing can be excluded at 90 \% CL for $\theta_{23} = 41.5^\circ$.
As a final comment, notice that the new neutrino fluxes of 
Ref.~\cite{Campagne:2004wt} do not improve the $\theta_{23}$ resolution with respect to old fluxes from Ref.~\cite{gilardoni}. 
This is because the new fluxes have been optimized to look for the $\nu_\mu \to \nu_e$ appearance signal and not to the 
$\nu_\mu \to \nu_\mu$ disappearance one. In particular, the average neutrino and antineutrino energies are identical
(see Fig.~\ref{fig:fluxes}) and thus the small complementarity of the two fluxes reduces.
In the rest of the paper we will consider the SPL with old fluxes, only. 

It is remarkable how a relatively ``small'' experiment such as T2K-I, with only 5 years of data taking
in one polarity and with a 22.5 Kton water \v Cerenkov detector, has a resolution in $\Delta m^2_{23}$ and $\theta_{23}$
not much worse than the SPL facility, with a much larger 1 Mton water \v Cerenkov and 10 years of data taking in 
both pion polarities (compare the left panels of Figs.~\ref{fig:T2Kbins} and \ref{fig:SPLbins}).
One of the reasons is that the SPL has both neutrino and antineutrino beams with an average energy corresponding to the 
oscillation peak for the $L = 130$ baseline. As a consequence, information coming from the two beams just add statistically 
but it is not complementary (in the absence of matter effects). 
It would be better to run the SPL with neutrinos again but at a different energy (not at the oscillation peak) after the first
two years with $\pi^+$. 

\subsection{The impact of $\theta_{13}$ and $\delta$}
\label{sec:sb:t13del}

Up to this moment we have kept the unknown parameters $\theta_{13}$ and $\delta$ as external fixed quantities, 
$\theta_{13} = \delta = 0^\circ$, following what we have done for the solar parameters $\Delta m^2_{12}$ and $\theta_{12}$.
However, the two sets of parameters should be treated differently. Indeed, we do have a good measure of solar 
parameters and it has been shown in the literature that the impact of solar parameter uncertainties in the measurement
of atmospheric parameters is negligible \cite{Donini:2005rn}. 
The main effect is the shift in the atmospheric mass difference fitting variable, 
as it has been discussed in Sect.~\ref{sec:sb:conve}. This is not the case for the unknown parameters ($\theta_{13},\delta$).
Being both parameters unknown, we must fit the atmospheric parameters introducing them as free variables to be 
reconstructed at the same time with ($\theta_{23},\Delta m^2_{23}$). 
A recent comprehensive three-families analysis of present solar, atmoshperic and reactor data can be found, for 
example, in Ref.~\cite{Fogli:2005cq}.

\begin{figure}[t!]
\vspace{-0.5cm}
\begin{center}
\begin{tabular}{cc}
\hspace{-1.0cm} \epsfxsize8.25cm\epsffile{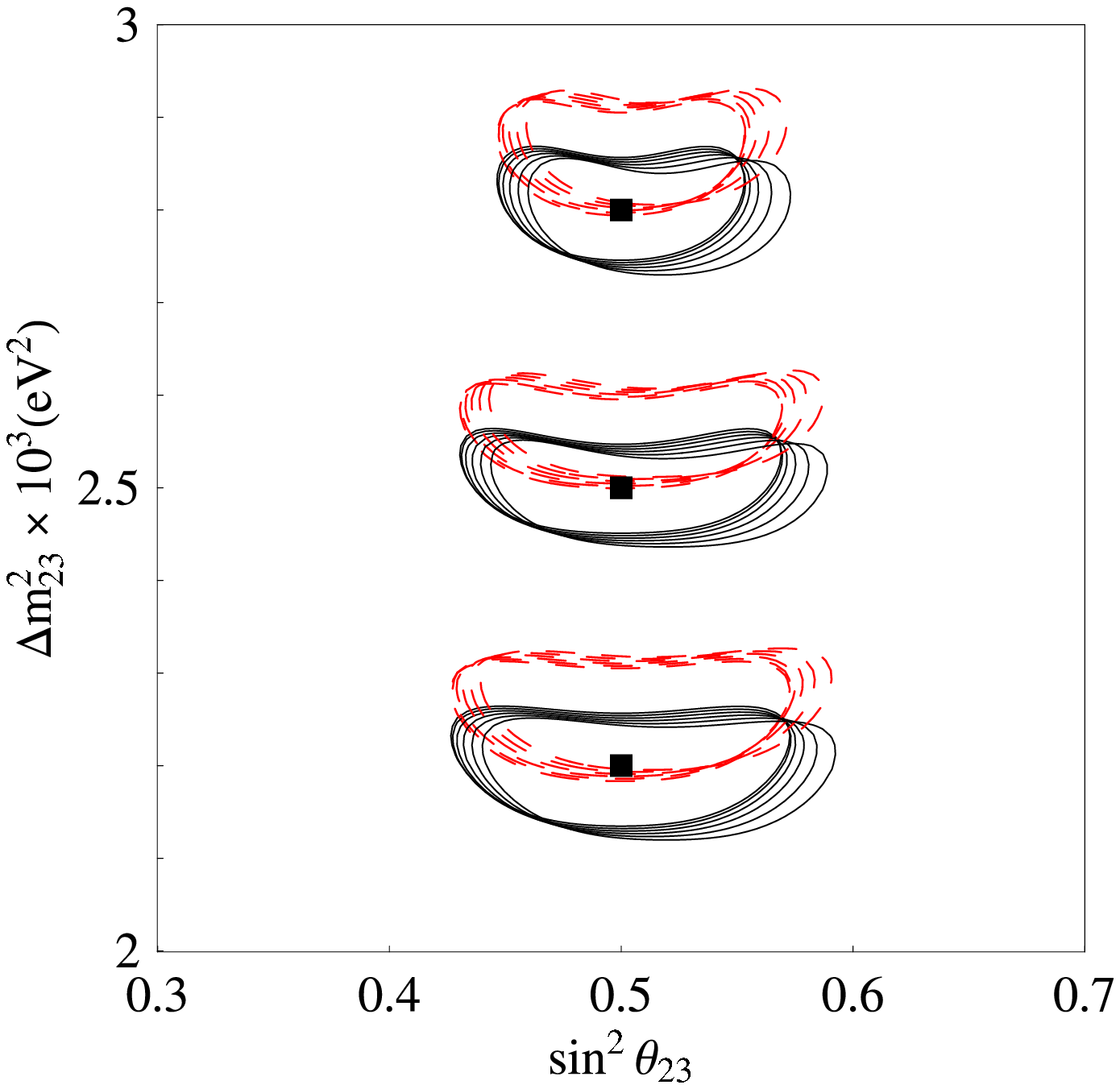} & \hspace{-0.5cm} \epsfxsize8.25cm\epsffile{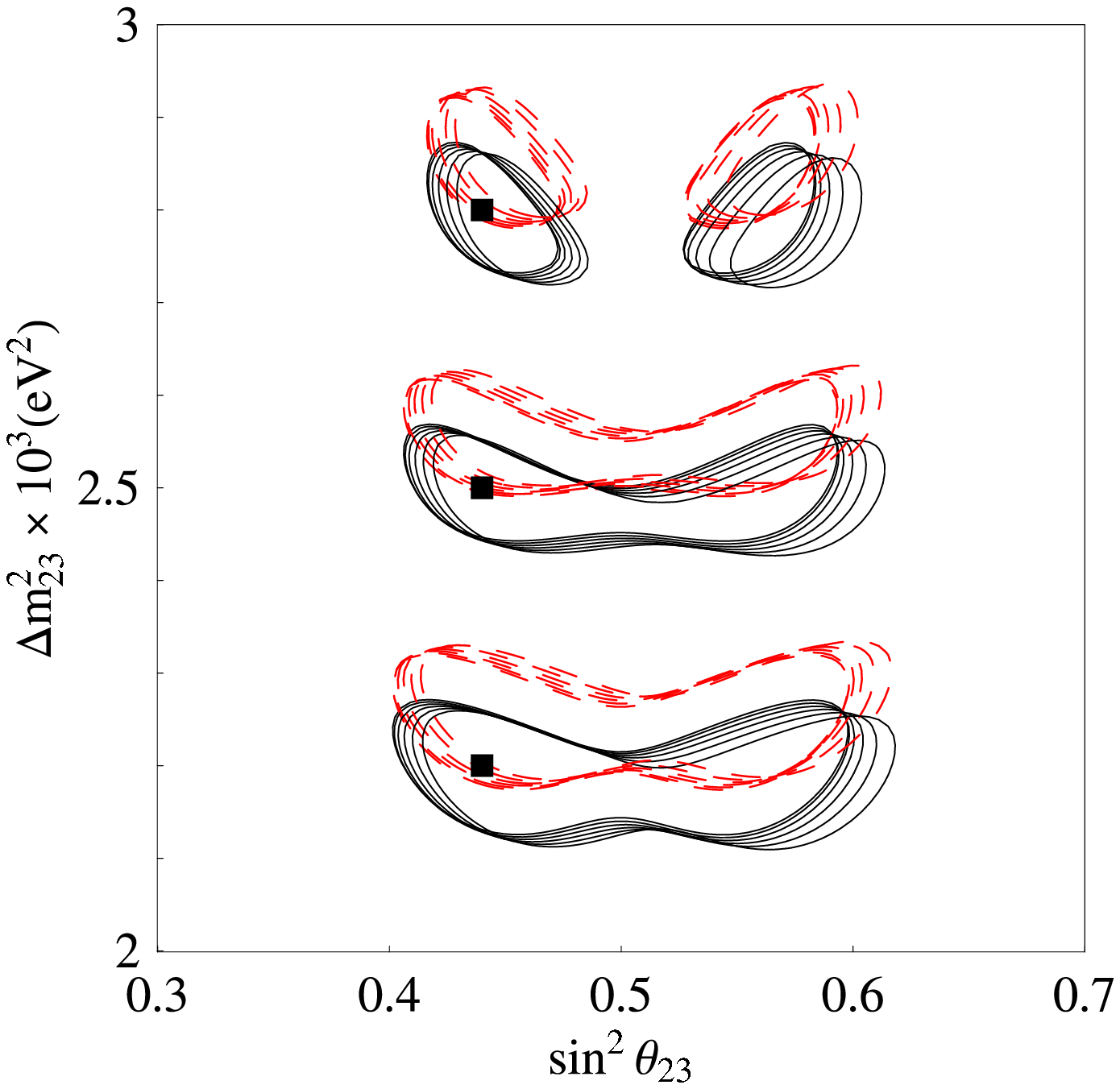}
\end{tabular}
\caption{\it The impact of $\theta_{13}$ at T2K-I; left: $\theta_{23} = 45^\circ$; right: $\theta_{23} = 41.5^\circ$.} 
\label{fig:th13}
\end{center}
\end{figure}

In Fig.~\ref{fig:th13} we show the effect of a varying $\theta_{13}$ (but fixed $\delta$) on the previous fits at T2K-I
for $\theta_{23} = 45^\circ$ (left panel) and $\theta_{23} = 41.5^\circ$ (right panel) and three values of the atmospheric mass
difference, $\Delta m^2_{23} = (2.2,2.5,2.8) \times 10^{-3}$ eV$^2$. Both choices of $s_{atm}$ are shown on the same plot.
The input values of the unknowns are: $\bar \theta_{13} = 0^\circ$, $\bar \delta = 0^\circ$. In the fit, $\theta_{13}$
is free to vary in the range $\theta_{13} \in [0^\circ, 10^\circ]$.
Notice that the input values can be fitted with increasing values of $\theta_{13}$ if $\theta_{23}$ is also increased, 
resulting in a shift of the 90 \% CL contours to the right. This is a consequence of the $\theta_{23}$-asymmetric second term 
in the first parenthesis of eq.~(\ref{eq:probdismu}).

Such a naive treatment of the experimental data is, however, not correct. The considered Super-Beam facilities
have indeed been proposed to look for the appearance $\nu_\mu \to \nu_e$ channel, the oscillation probability 
in vacuum expanded to the second order in the small parameters $\tc$ and $(\Delta_{12}L/E)$ 
\cite{Cervera:2000kp,Akhmedov:2004ny} for which is:
\bea
\label{eq:appnue}
P^\pm (\nu_\mu \to \nu_e ) &=& 
s_{23}^2 \sin^2 (2 \theta_{13}) \sin^2 \left ( \frac{ \Delta_{23} L}{ 2 } \right ) \\
&+& \sin ( 2 \theta_{12} ) \sin ( 2 \theta_{23} ) \sin (2 \theta_{13} )
      \cos \left ( \mp \delta - \frac{\Delta_{23} L }{2} \right ) 
\sin \left ( \frac{ \Delta_{12} L }{ 2 } \right )
\sin \left ( \frac{ \Delta_{23} L }{ 2 } \right ) \nn \\
&+& c_{23}^2 \sin^2 (2 \theta_{12}) \sin^2 \left ( \frac{ \Delta_{12} L }{ 2 } \right ) \, , \nn
\eea
where $\pm$ refers to neutrinos and antineutrinos, respectively.
It is clear that to take properly into account the effect of ($\theta_{13},\delta$) we should combine 
(whenever possible) appearance and disappearance signals at a given facility.

\begin{figure}[t!]
\vspace{-0.5cm}
\begin{center}
\begin{tabular}{cc}
\hspace{-1.0cm} \epsfxsize8.25cm\epsffile{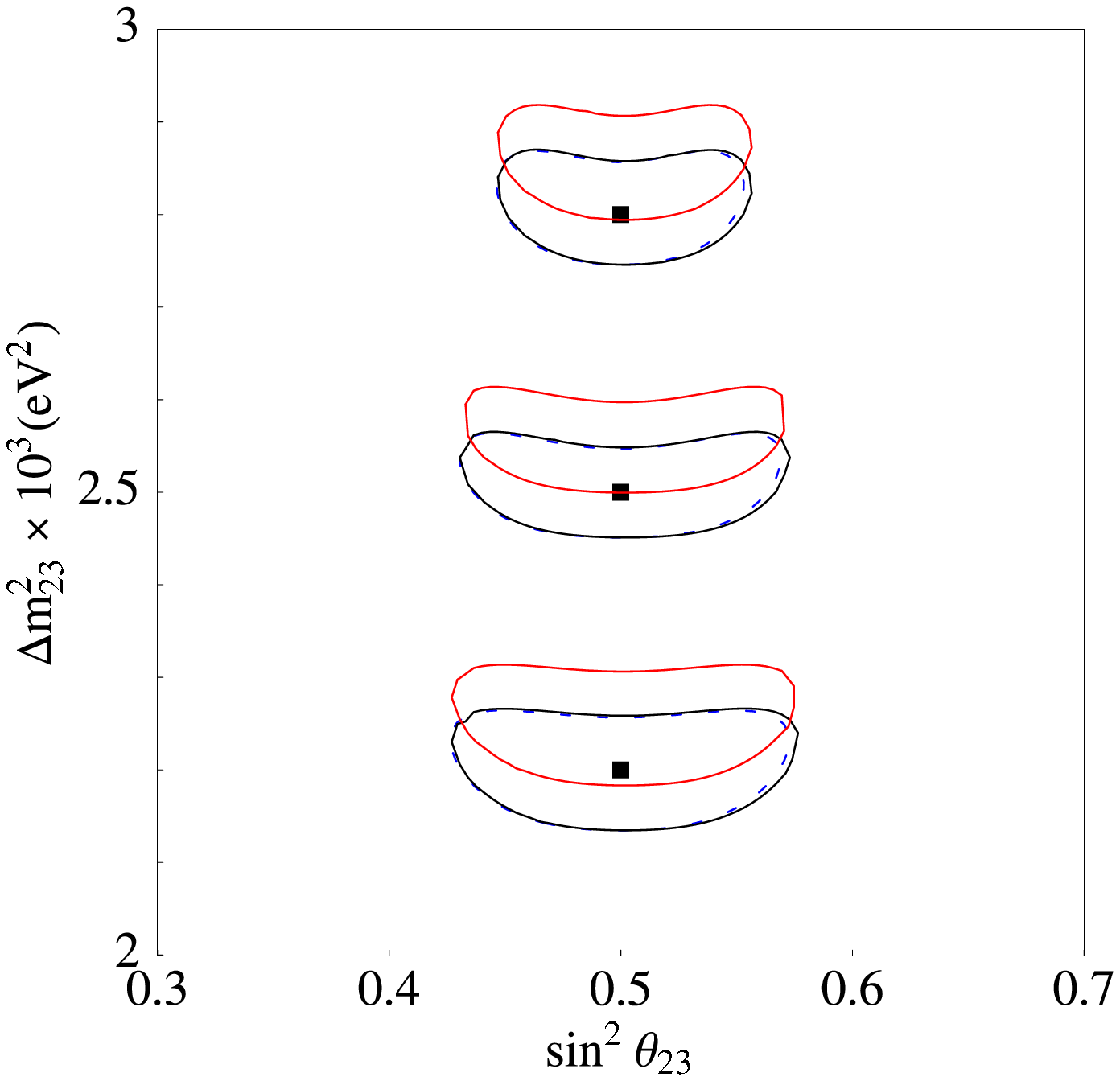} & \hspace{-0.5cm} \epsfxsize8.25cm\epsffile{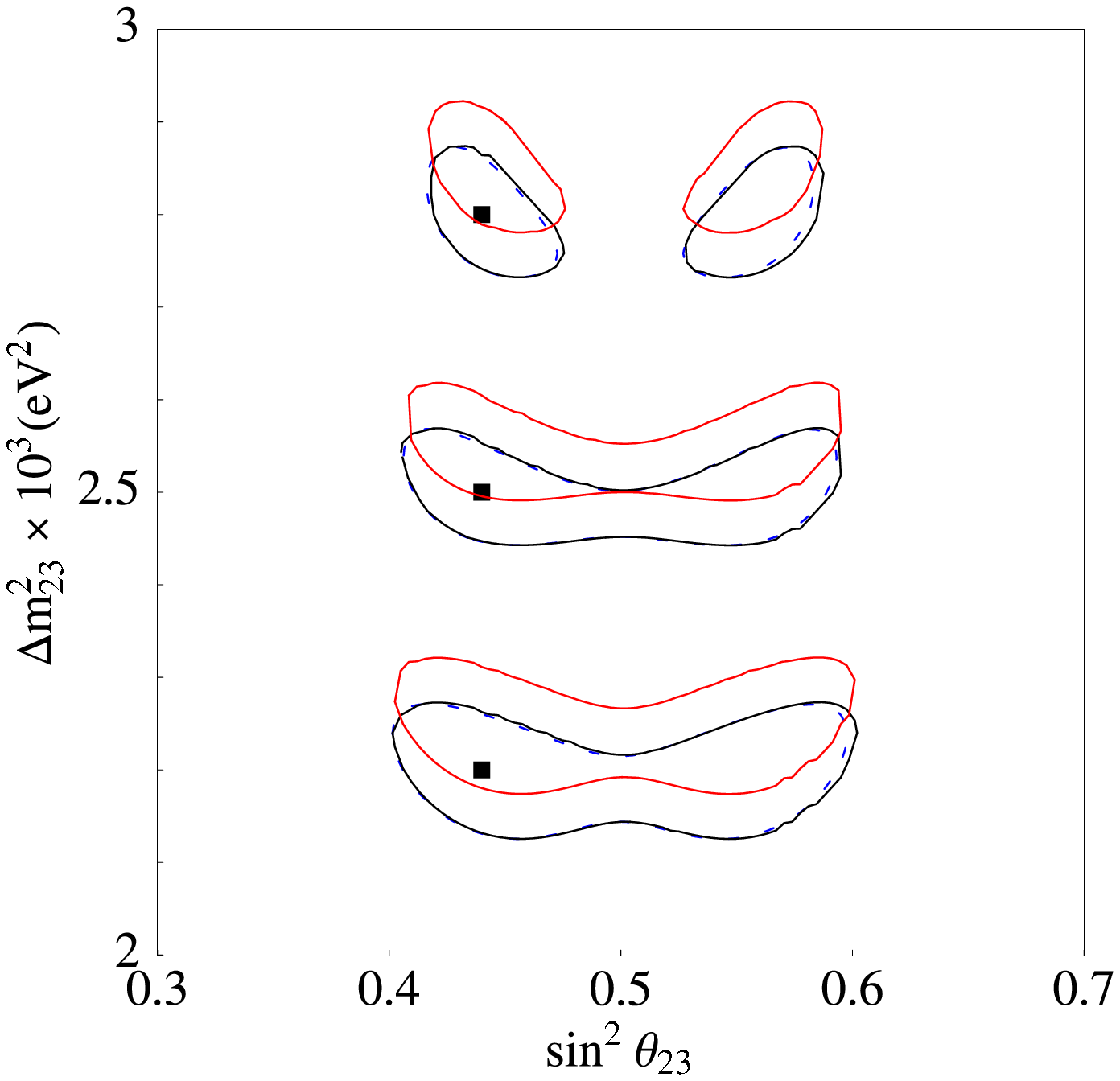} \\
\hspace{-1.0cm} \epsfxsize8.25cm\epsffile{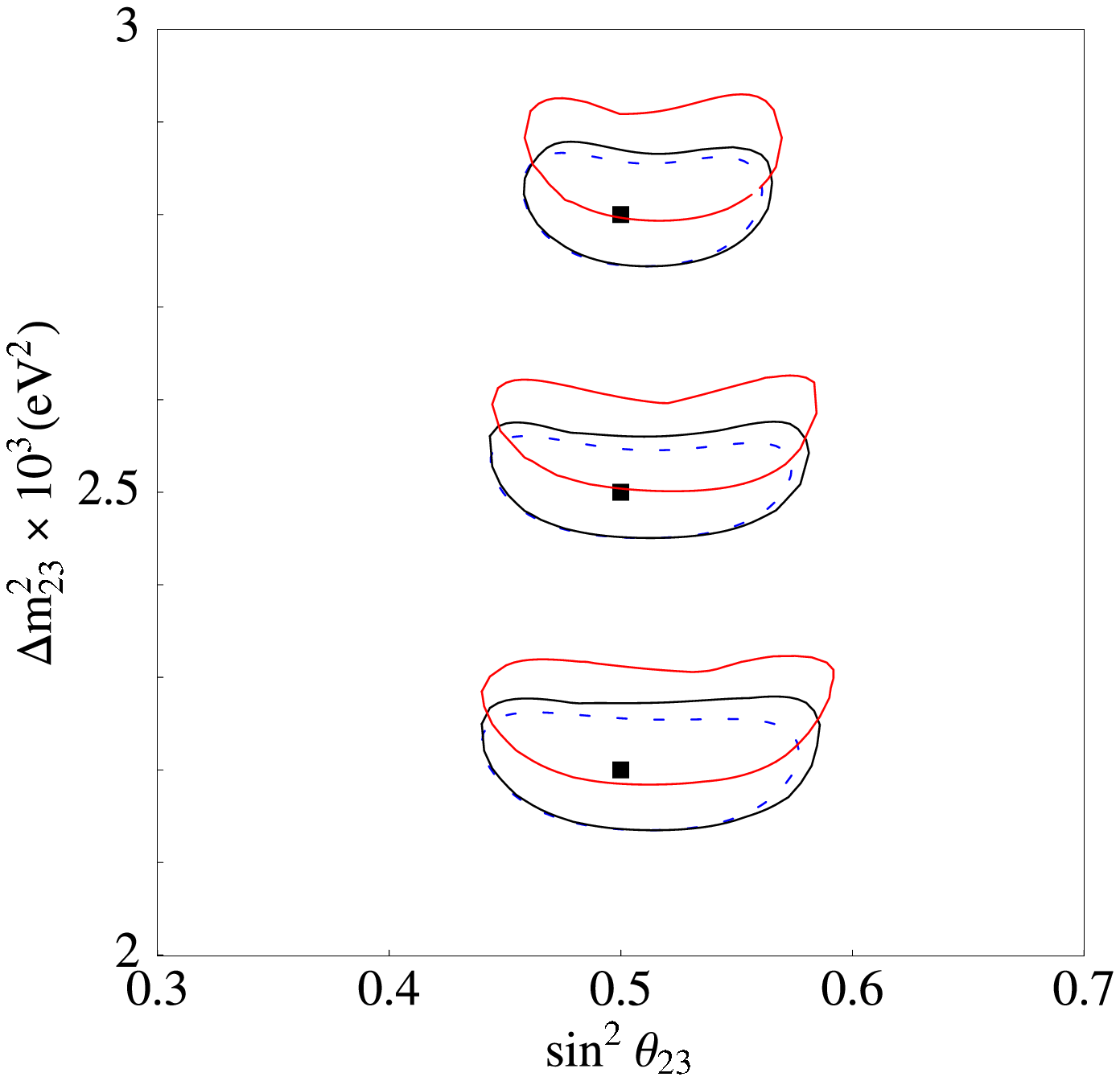} & \hspace{-0.5cm} \epsfxsize8.25cm\epsffile{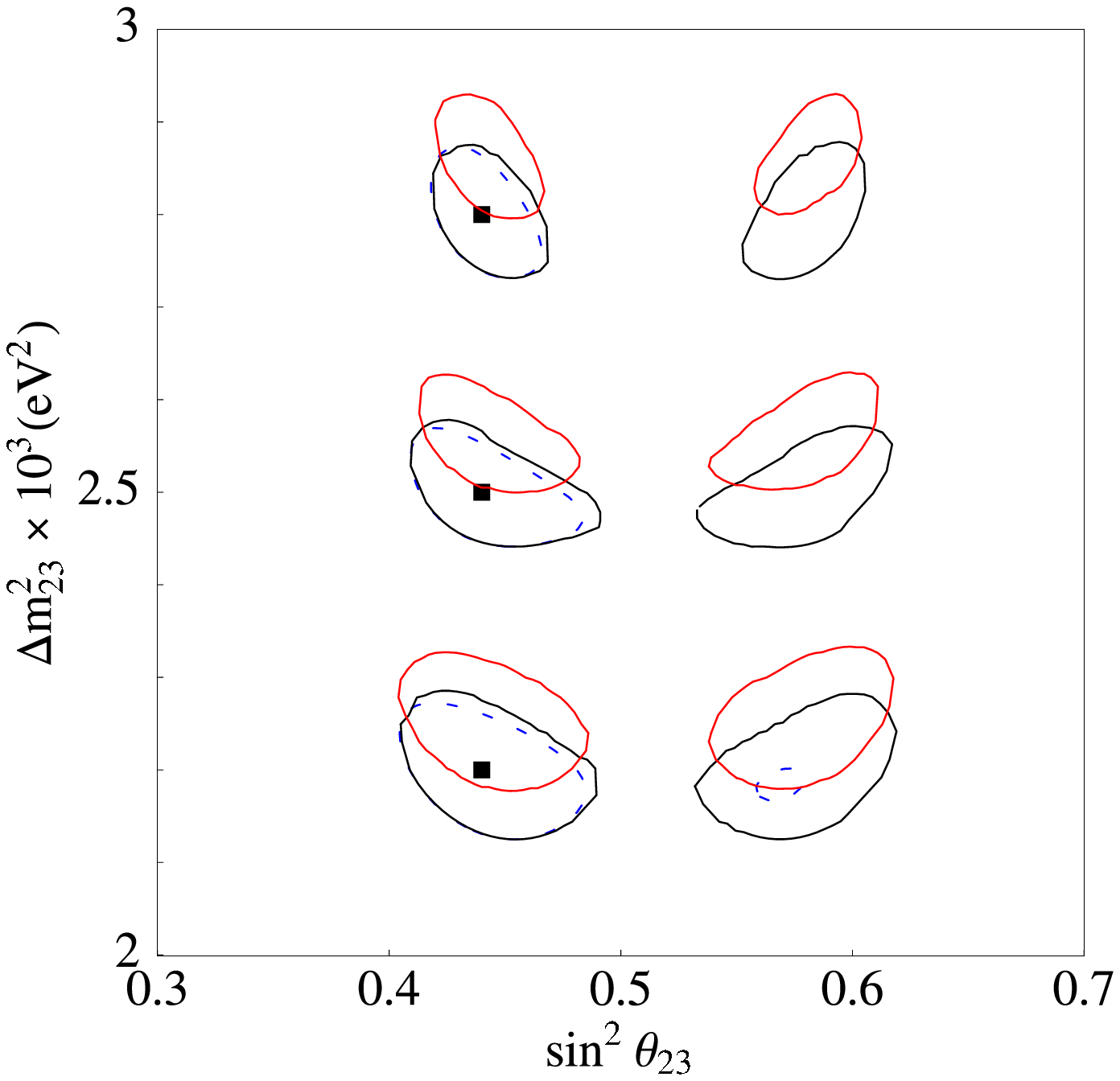}
\end{tabular}
\caption{\it Appearance and disappearance at T2K-I. Left: $\theta_{23} = 45^\circ$; right: $\theta_{23} = 41.5^\circ$;
                                                  top: $\theta_{13} = 0^\circ$; bottom: $\theta_{13} = 8^\circ$.} 
\label{fig:th13:appdis:t2k}
\end{center}
\end{figure}

\begin{figure}[t!]
\vspace{-0.5cm}
\begin{center}
\begin{tabular}{cc}
\hspace{-1.0cm} \epsfxsize8cm\epsffile{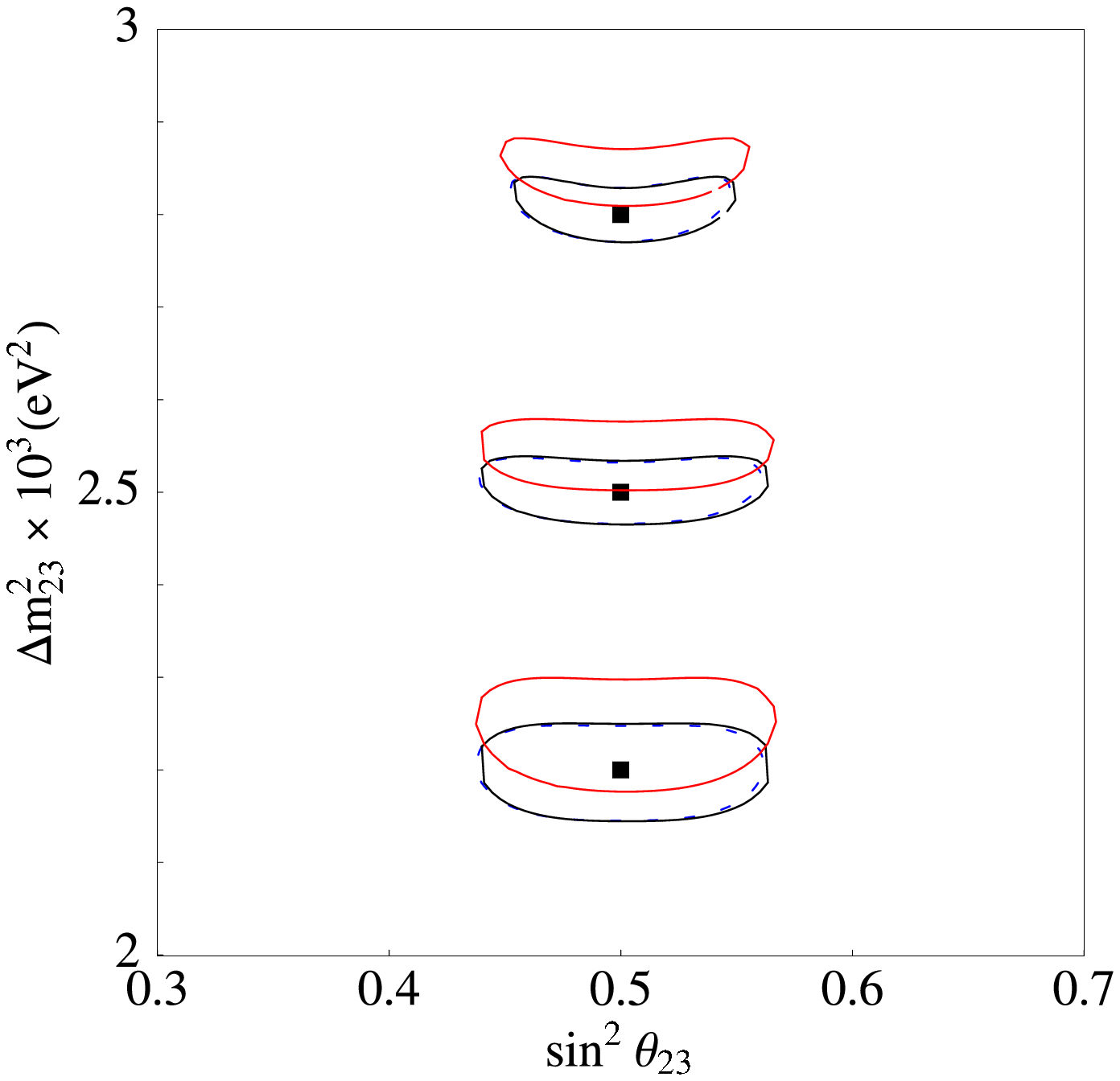} & 
\hspace{-0.5cm} \epsfxsize8cm\epsffile{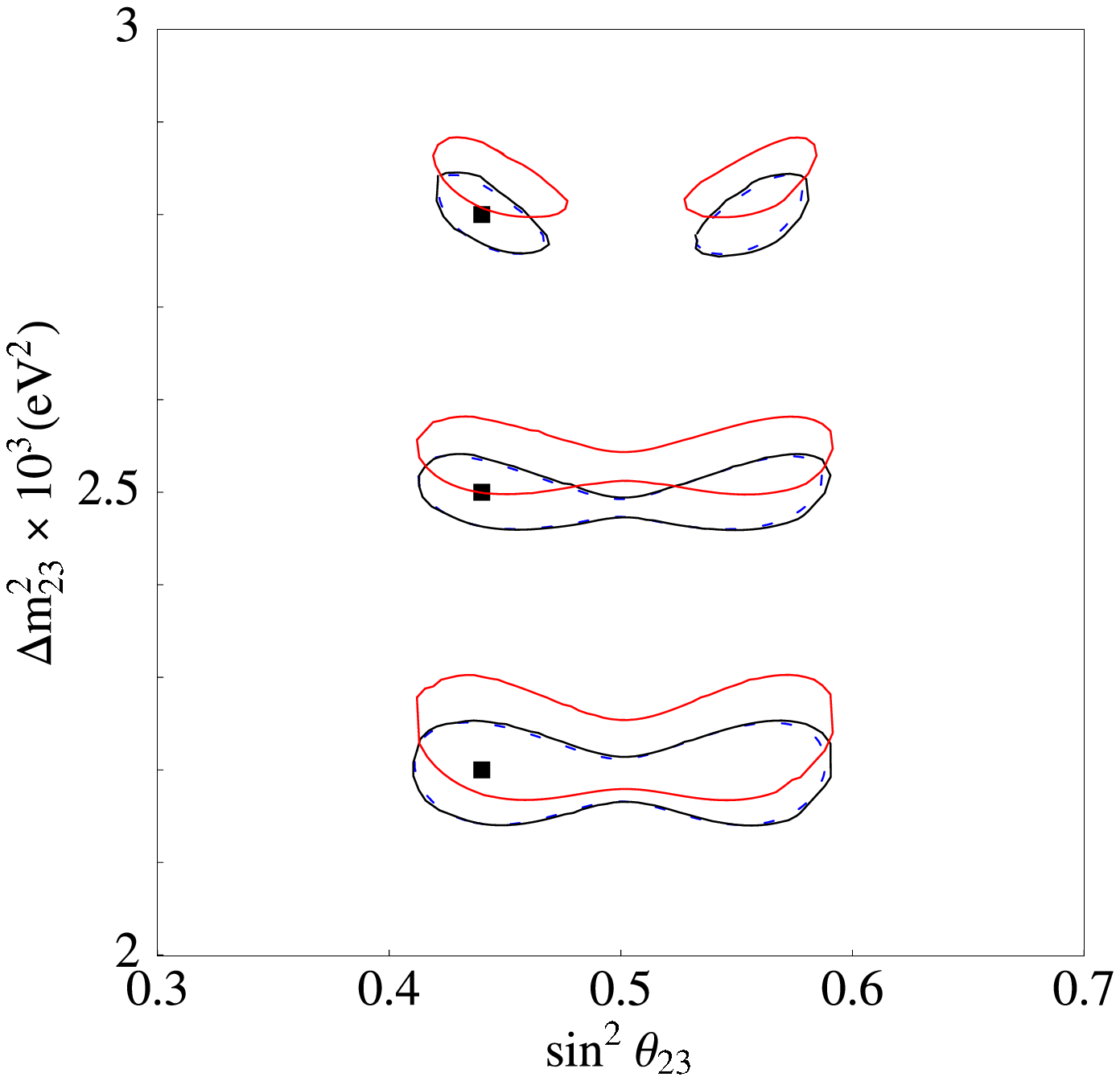} \\
\hspace{-1.0cm} \epsfxsize8cm\epsffile{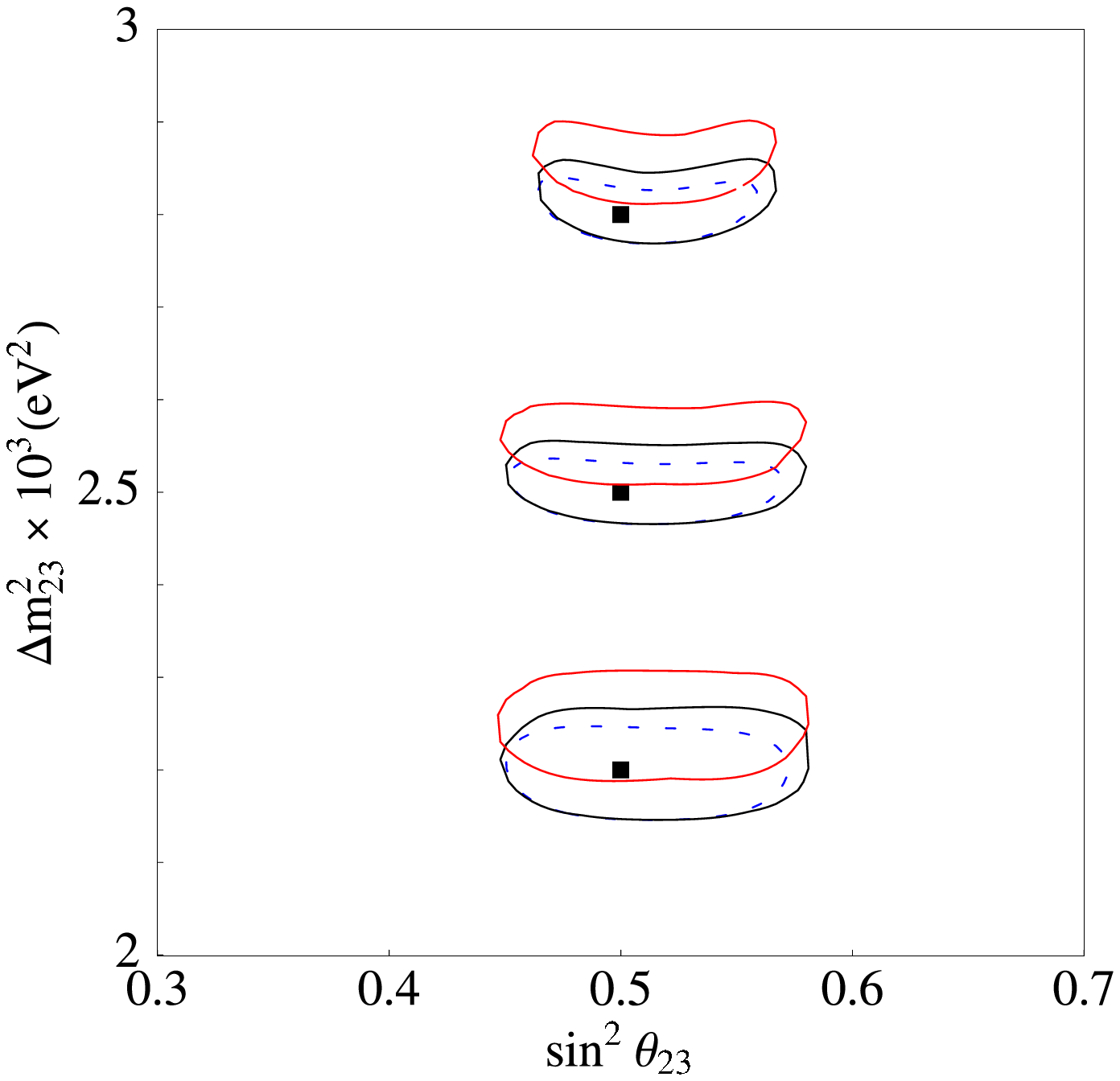} &
\hspace{-0.5cm} \epsfxsize8cm\epsffile{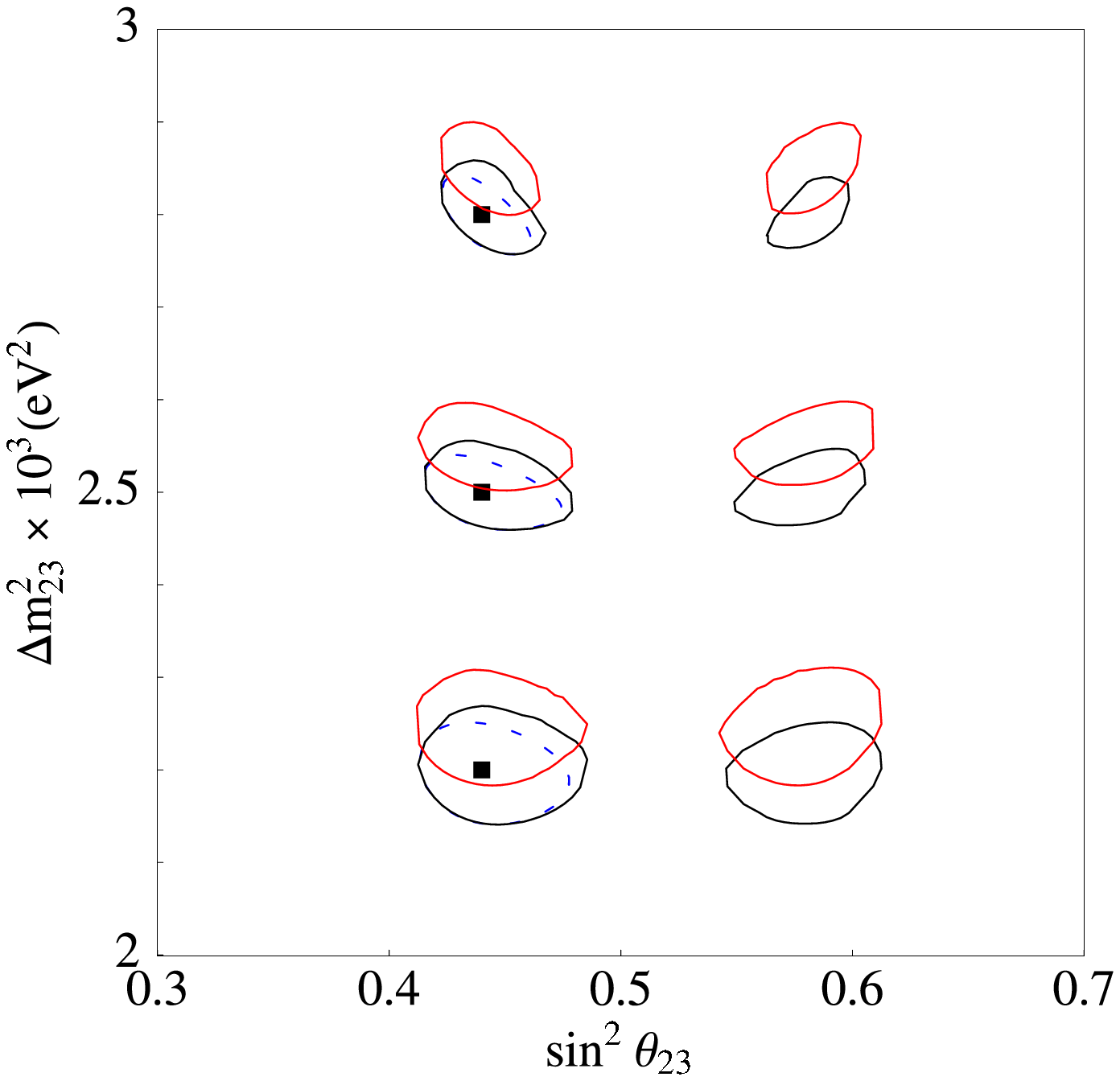}
\end{tabular}
\caption{\it Appearance and disappearance at NO$\nu$A. Left: $\theta_{23} = 45^\circ$; right: $\theta_{23} = 41.5^\circ$;
                                                  top: $\theta_{13} = 0^\circ$; bottom: $\theta_{13} = 8^\circ$.} 
\label{fig:th13:appdis:nova}
\end{center}
\end{figure}

\begin{figure}[t!]
\vspace{-0.5cm}
\begin{center}
\begin{tabular}{cc}
\hspace{-1.0cm} \epsfxsize8cm\epsffile{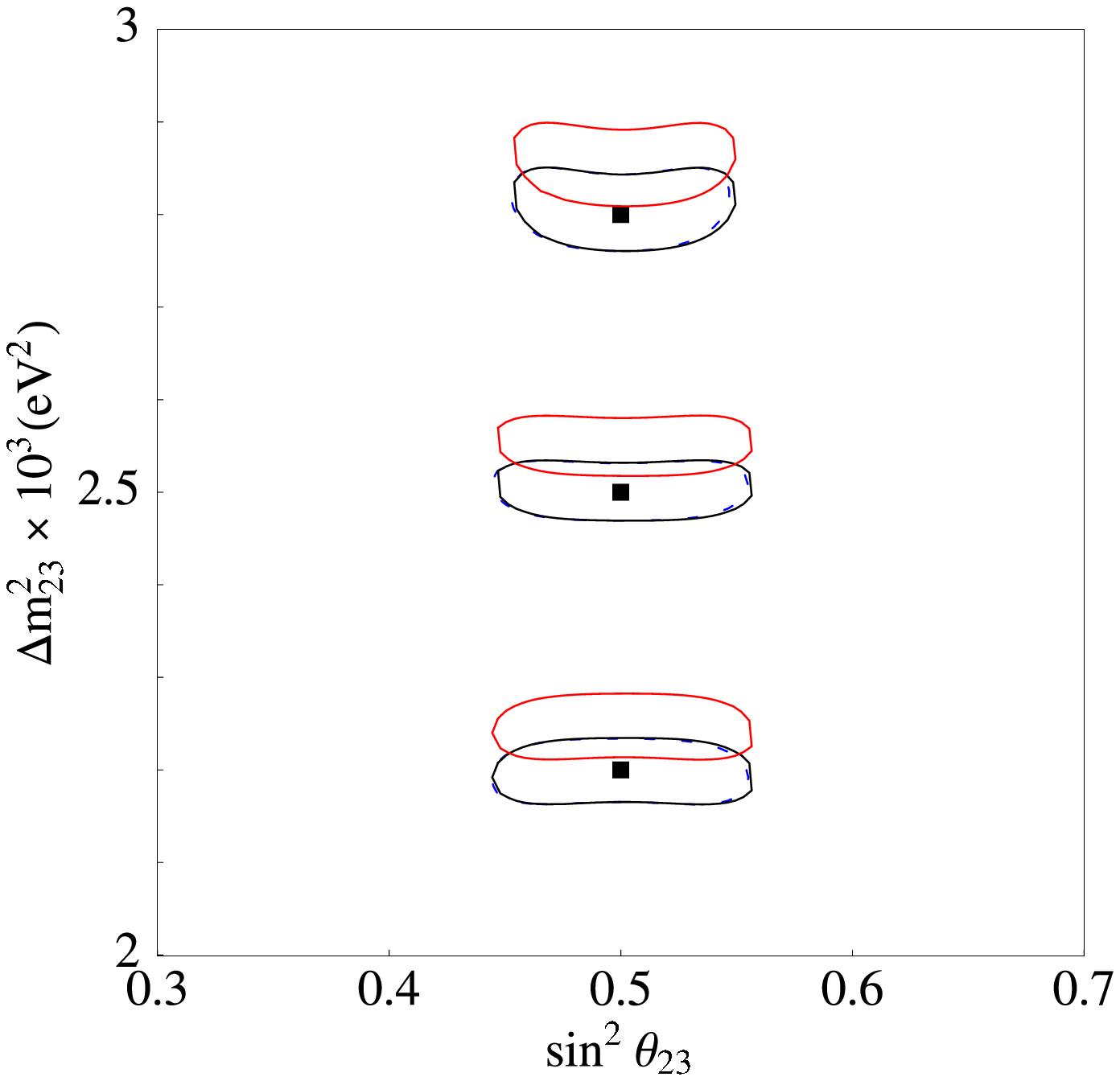} & 
\hspace{-0.5cm} \epsfxsize8cm\epsffile{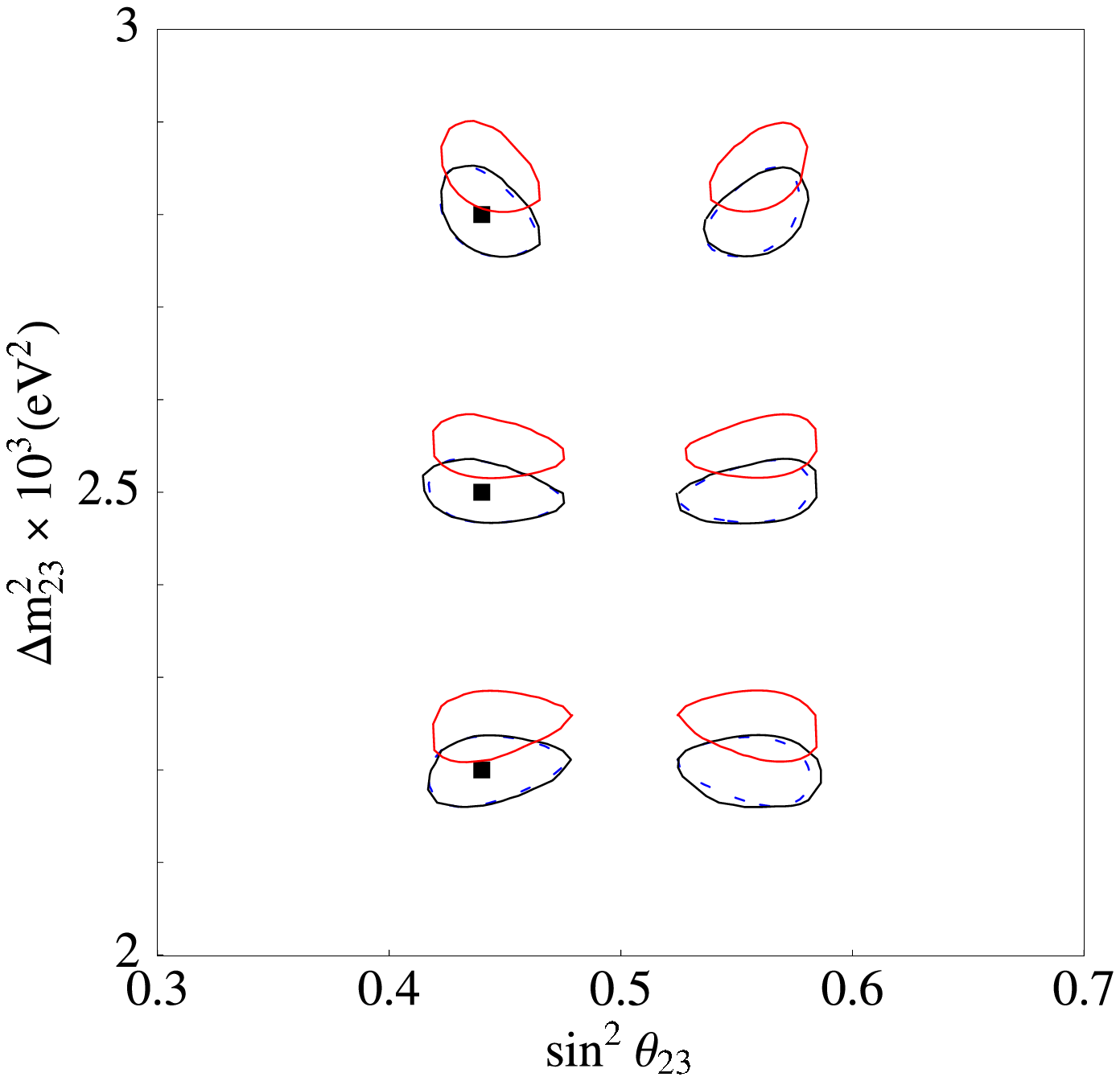} \\
\hspace{-1.0cm} \epsfxsize8cm\epsffile{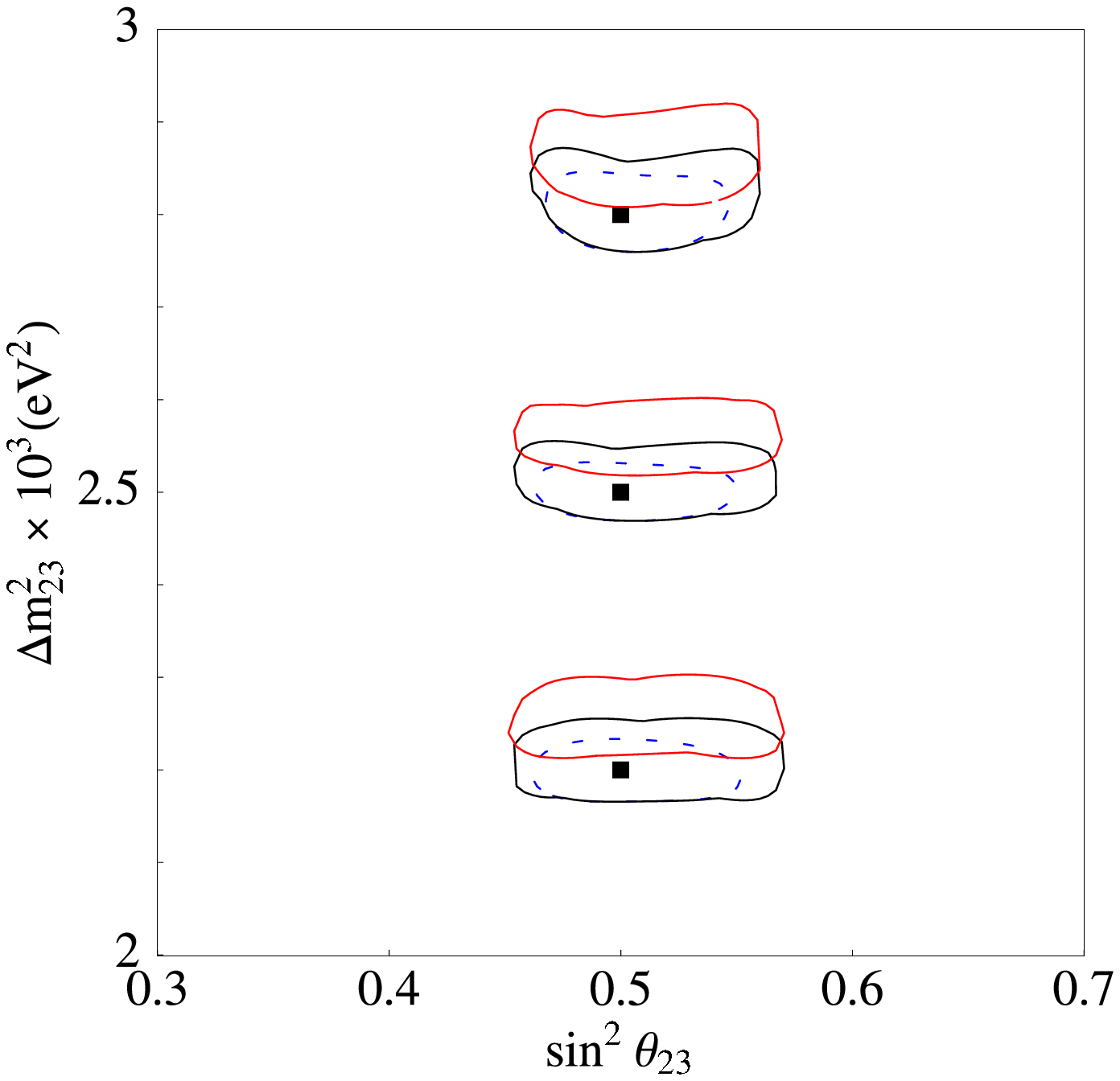} & 
\hspace{-0.5cm} \epsfxsize8cm\epsffile{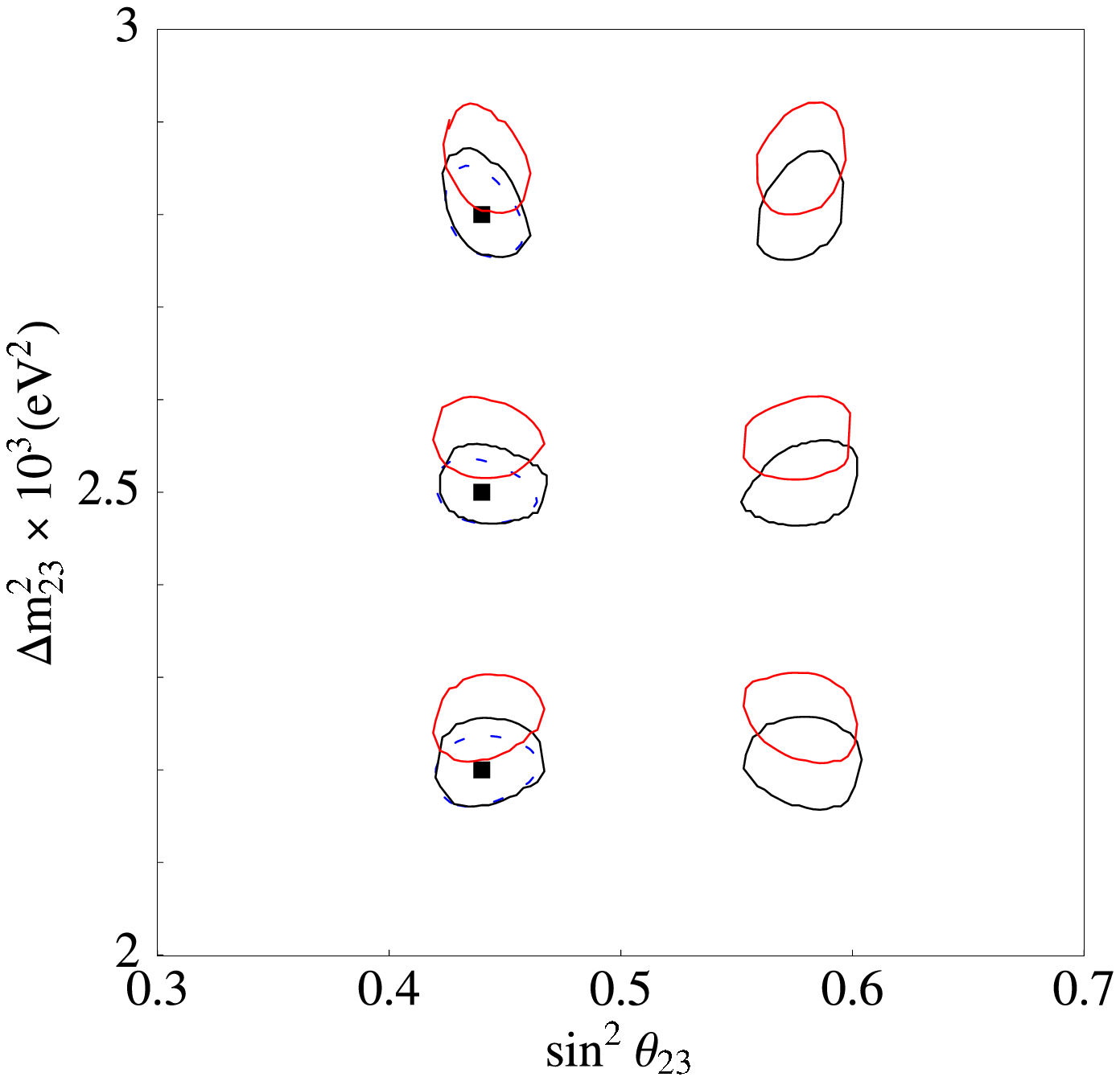}
\end{tabular}
\caption{\it  Appearance and disappearance at the standard SPL. Left: $\theta_{23} = 45^\circ$; right: $\theta_{23} = 41.5^\circ$;
                                                  top: $\theta_{13} = 0^\circ$; bottom: $\theta_{13} = 8^\circ$.} 
\label{fig:th13:appdis:spl}
\end{center}
\end{figure}

The results of a four-parameters fit in ($\theta_{23},\Delta m^2_{23},\theta_{13},\delta$) projected onto the 
($\theta_{23},\Delta m^2_{23}$) plane obtained by combining the disappearance and appearance signals at the T2K-I, 
NO$\nu$A and SPL facilities are presented in Figs.~\ref{fig:th13:appdis:t2k}-\ref{fig:th13:appdis:spl}, respectively. 
In all figures two choices of $\bar\theta_{23}$ and $\bar \theta_{13}$ are shown, $\bar \theta_{23} = 45^\circ, 41.5^\circ$ 
(left and right panels) and $\bar \theta_{13} = 0^\circ,8^\circ$ (top and bottom panels), whereas $\bar \delta = 0^\circ$. 
Solid lines represent the result of a fit with variable $\theta_{13}$ and $\delta$.
As a reference, we also present the results of a fit\footnote{Only the results for $s_{atm} = \bar s_{atm}$ are shown.}
 with $\theta_{13} = \bar \theta_{13}$ and $\delta = \bar \delta$ (dotted lines). 
It can be observed that in most of the cases the combination of disappearance with appearance signals reduces 
the spread in $\theta_{23}$ that was observed in Fig.~\ref{fig:th13}. Only for $\bar \theta_{13}$ large we can still 
see some effect. 


Eventually, the effect of a non-vanishing $\bar \delta$ can be easily understood:
the CP-conserving $\delta$-dependent term in the second row of eq.~(\ref{eq:probdismu}) is positive for 
$|\bar \delta| \in [0^\circ,90^\circ]$, negative for $|\bar \delta| \in [90^\circ,180^\circ]$
and vanishing for $|\bar \delta| = 90^\circ$. When non-vanishing, this term induces 
a shift in the fitted value of $\Delta m^2_{23}$ that is maximal (minimal) for $\bar \delta = 0^\circ$ 
($\bar \delta = 180^\circ$ ). Being, however, a $\theta_{13}$-suppressed perturbation on the subdominant 
$O(\Delta m^2_{12})$ term, its effect is rather small and it will not be shown in a separate figure. 

As a final comment, we notice that the three considered Super-Beams are indeed capable of strongly reducing the
present uncertainties on the atmospheric parameters $\theta_{23}$ and $\Delta m^2_{23}$, in particular for large
values of $\theta_{13}$ when combination of appearance and disappearance signals is most effective. 
For $\theta_{13} = 8^\circ$, for example, maximal atmospheric mixing can be excluded 
at 90 \% CL at all three facilities for $\bar \theta_{23} = 41.5^\circ$ (only the SPL, with its gigantic 1 Mton 
water \v Cerenkov, is able to exclude maximal mixing for $\bar \theta_{23} = 41.5^\circ$ for a vanishing $\theta_{13}$).
Notice, however, that while the {\it disappearance octant degeneracy} appears to be solved when combining 
disappearance and appearance signals for $\bar \theta_{13} = 8^\circ$ when $\theta_{13}$ is treated as a fixed
parameter, is indeed restored if $\theta_{13}$ is left freely varying in the presently allowed range. 

%
%%%%%%%%%%%%%%%%%%%%%%%%%%%%%%%%%%%%%%%%%%%%%%%%%%%%%%%%%%%%%%%%%%%%%%
%
\section{Appearance and disappearance channels at the Neutrino Factory.}
\label{sec:nf}
%
%%%%%%%%%%%%%%%%%%%%%%%%%%%%%%%%%%%%%%%%%%%%%%%%%%%%%%%%%%%%%%%%%%%%%%
%

The disappearance signal has not yet been deeply investigated in the framework of the Neutrino Factory 
(see, however, Ref.~\cite{Bueno:2000fg}).
The top-charted signals have been up to now the ``golden'' $\nu_e \to \nu_\mu$ \cite{Cervera:2000kp} and the 
``silver'' $\nu_e \to \nu_\tau$ \cite{Donini:2002rm} channels. However, the Neutrino Factory produces a very intense $\nu_\mu$ 
beam that can be used to measure $\nu_\mu \to \nu_\mu$ and the leading $\nu_\mu \to \nu_\tau$ appearance channel as well. 
This last transition is probably the best signal to improve significantly the atmospheric parameter uncertainties. 
However, to study it in detail we need a detector able to identify efficiently an enormous amount of $\tau$'s (roughly 1000 and 400 $\tau^-$ and 
$\tau^+$ are expected, respectively), 
such as it is expected in a Neutrino Factory beam. This detector can be in principle the same used to look for the silver
channel, but the needed scaling of the scanning power must be carefully studied and it will not be discussed here. 
We will devote this Section to a detailed study of the combination of golden, silver and disappearance channels 
at the Neutrino Factory. 

Two possible baselines are studied, $L = 3000$ km and $L = 7000$ km, i.e. the optimal distance to look for a 
CP-violating signal \cite{Cervera:2000kp} and the optimal distance to exploit matter effects through the disappearance 
channel, respectively.

\begin{figure}[t!]
\vspace{-0.5cm}
\begin{center}
\begin{tabular}{cc}
\hspace{-1.0cm} \epsfxsize8cm\epsffile{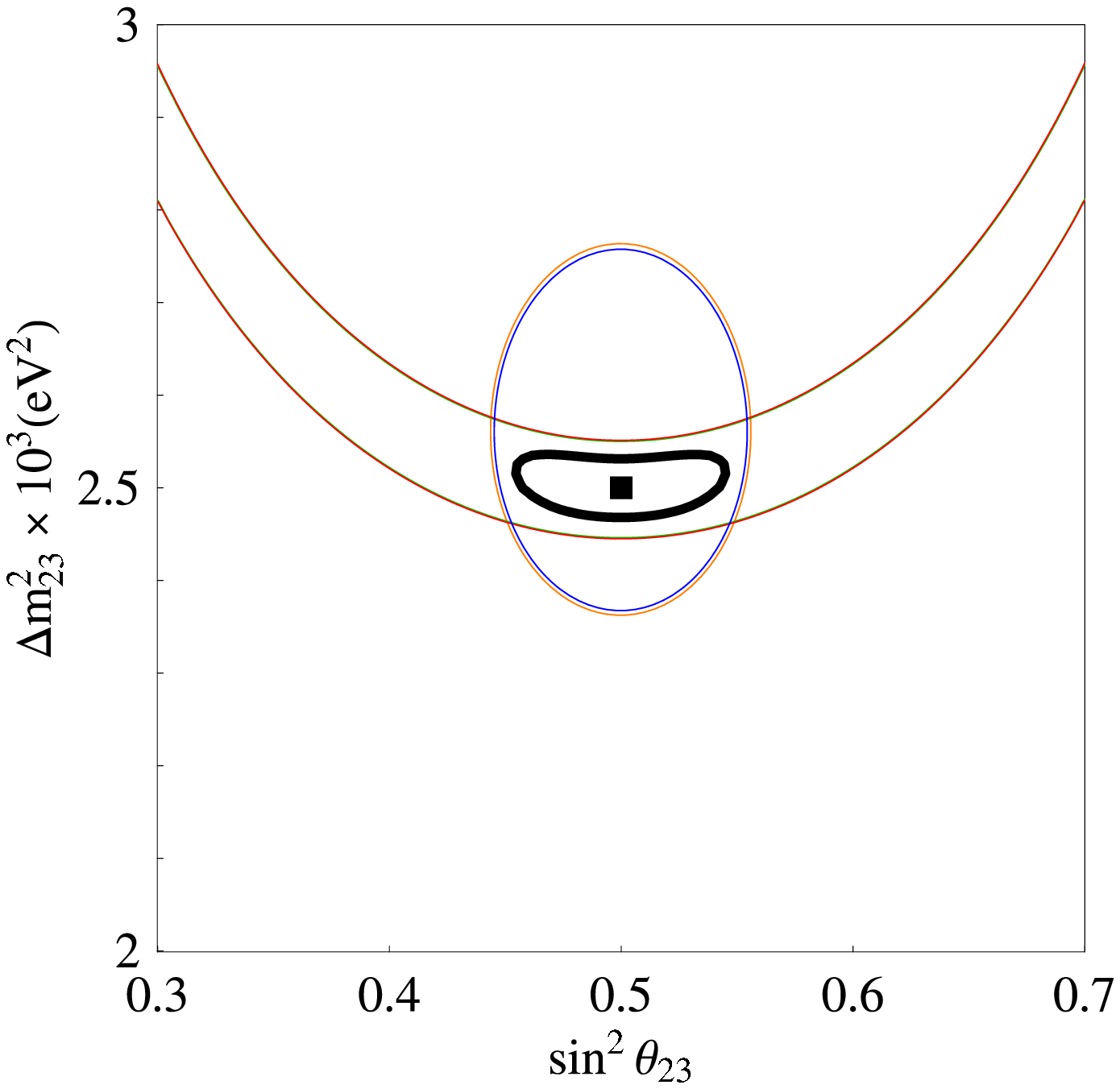} &
\hspace{-0.5cm} \epsfxsize8cm\epsffile{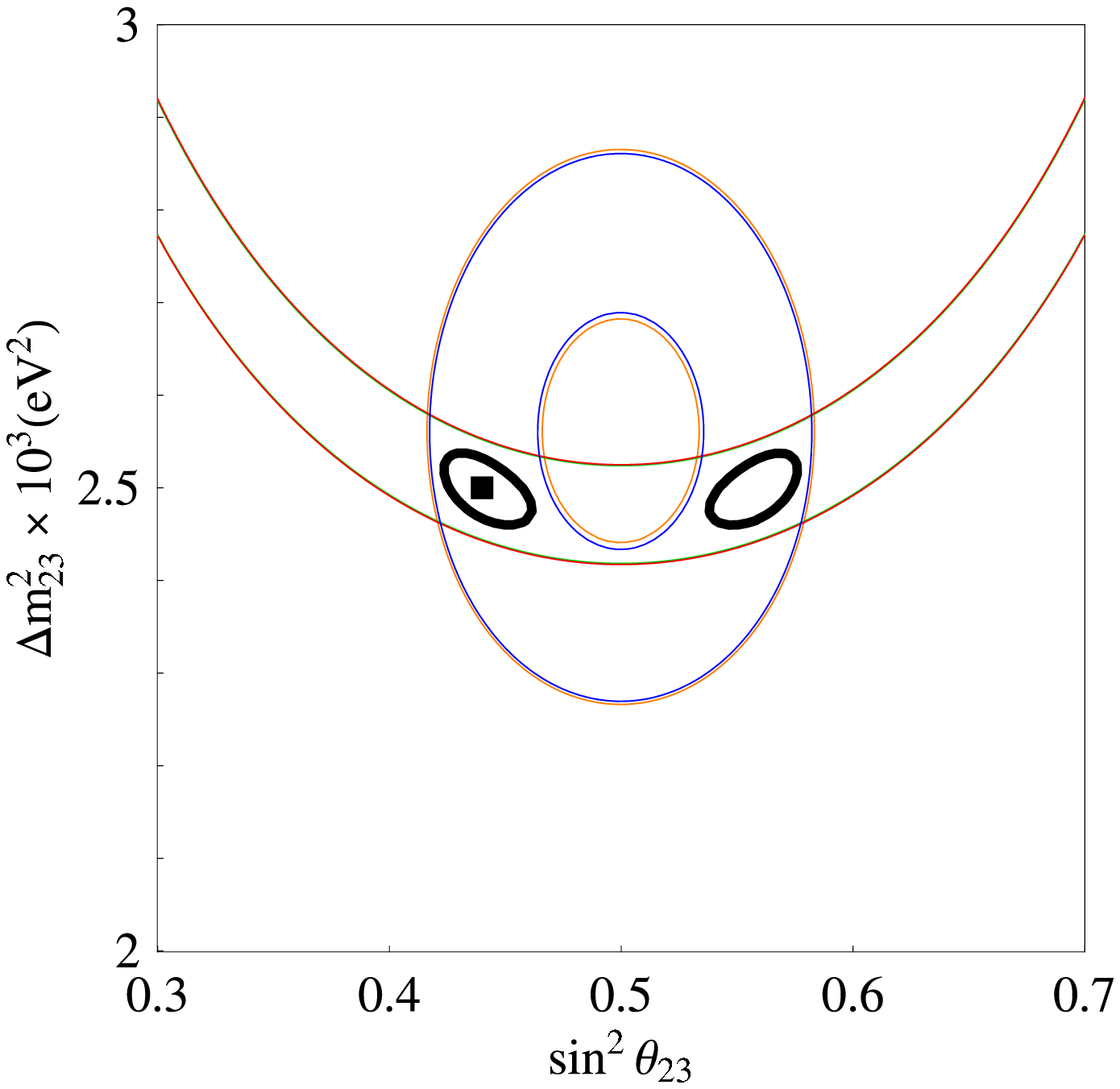} \\
\hspace{-1.0cm} \epsfxsize8cm\epsffile{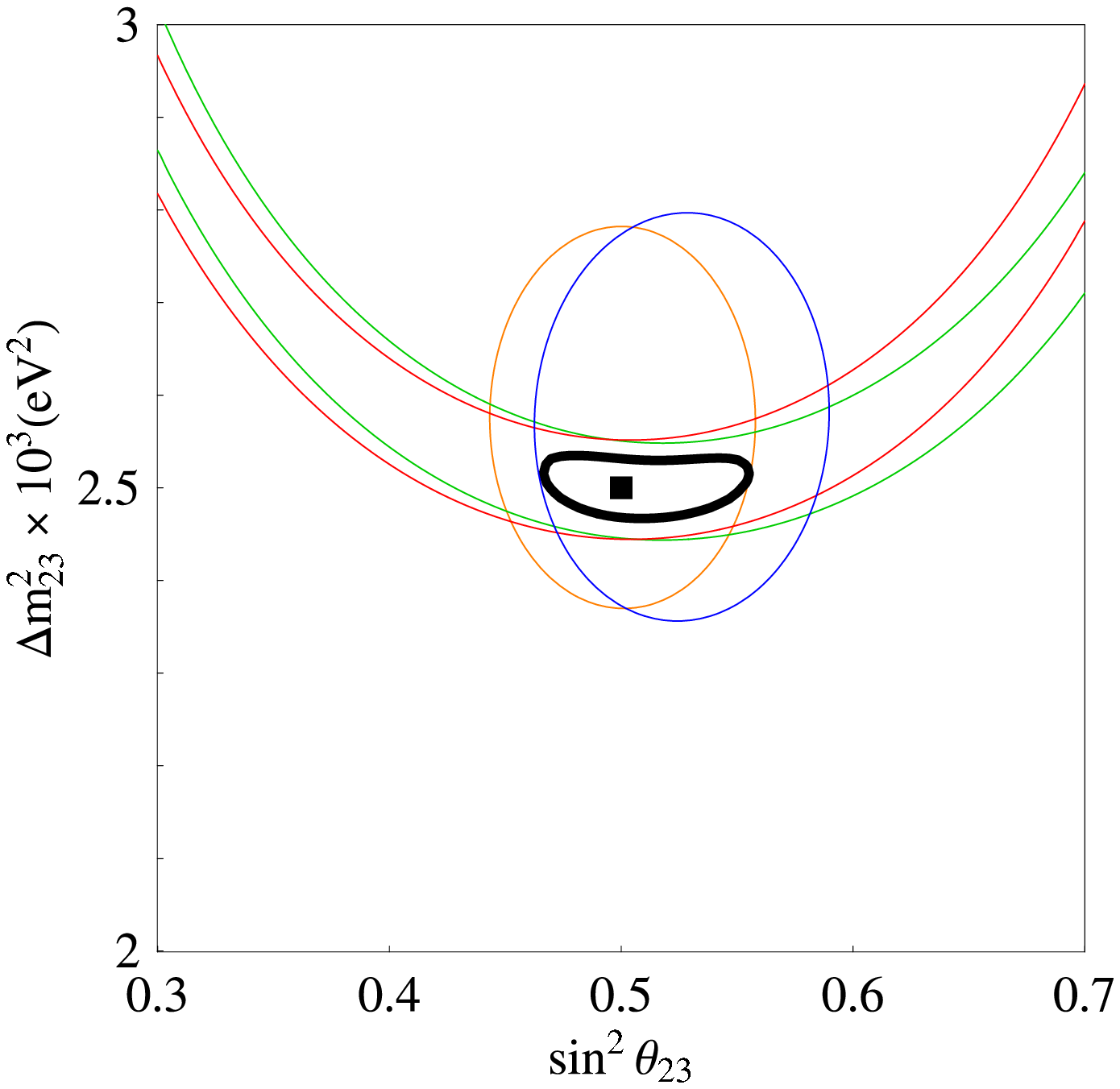} &
\hspace{-0.5cm} \epsfxsize8cm\epsffile{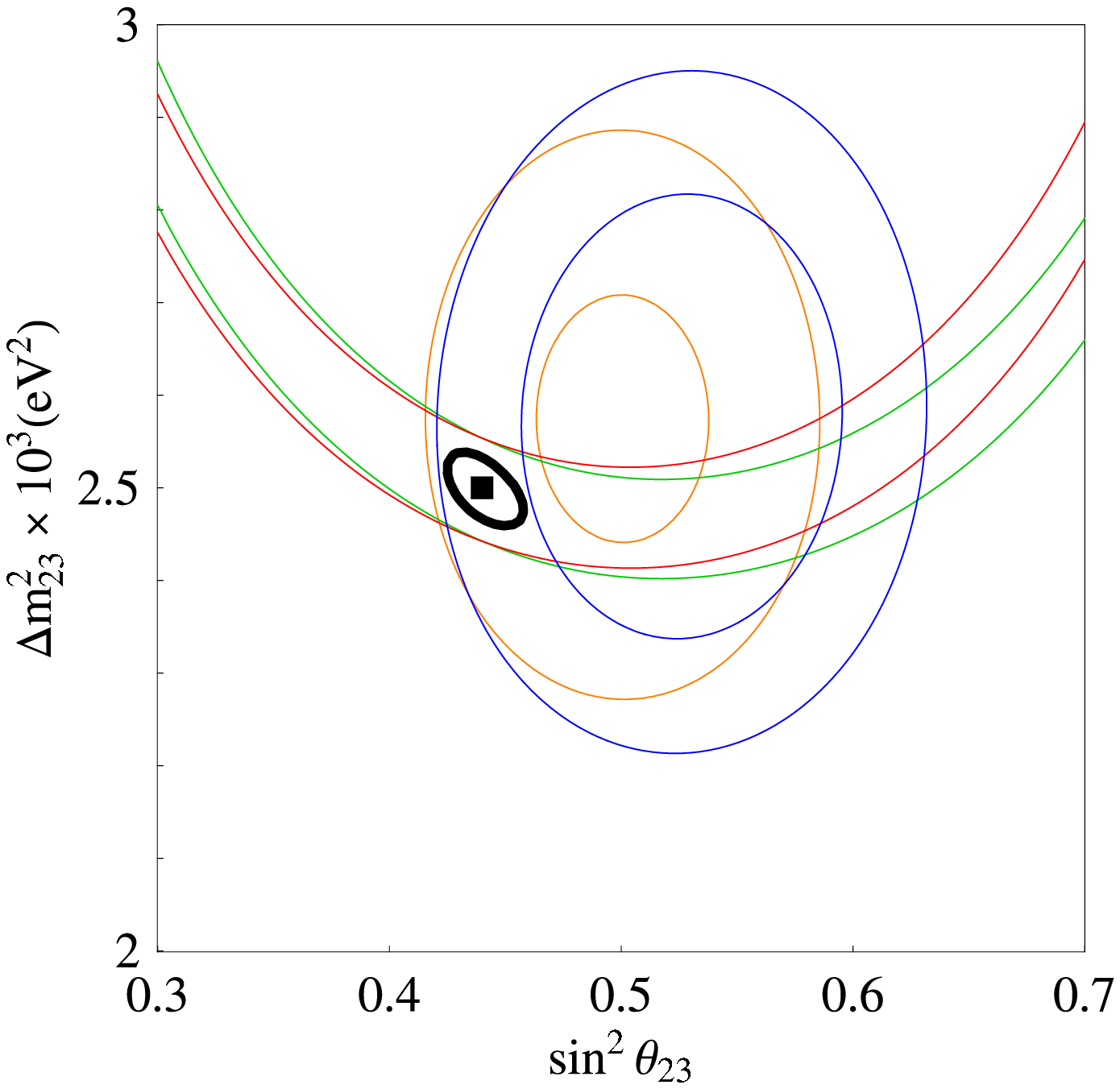} \\
\end{tabular}
\caption{\it Binning at the $L = 3000$ km Neutrino Factory.  Left: $\theta_{23} = 45^\circ$; right: $\theta_{23} = 41.5^\circ$;
                                                  top: $\theta_{13} = 0^\circ$; bottom: $\theta_{13} = 8^\circ$.} 
\label{fig:NF3000bins}
\end{center}
\end{figure}

\begin{figure}[t!]
\vspace{-0.5cm}
\begin{center}
\begin{tabular}{cc}
\hspace{-1.0cm} \epsfxsize8cm\epsffile{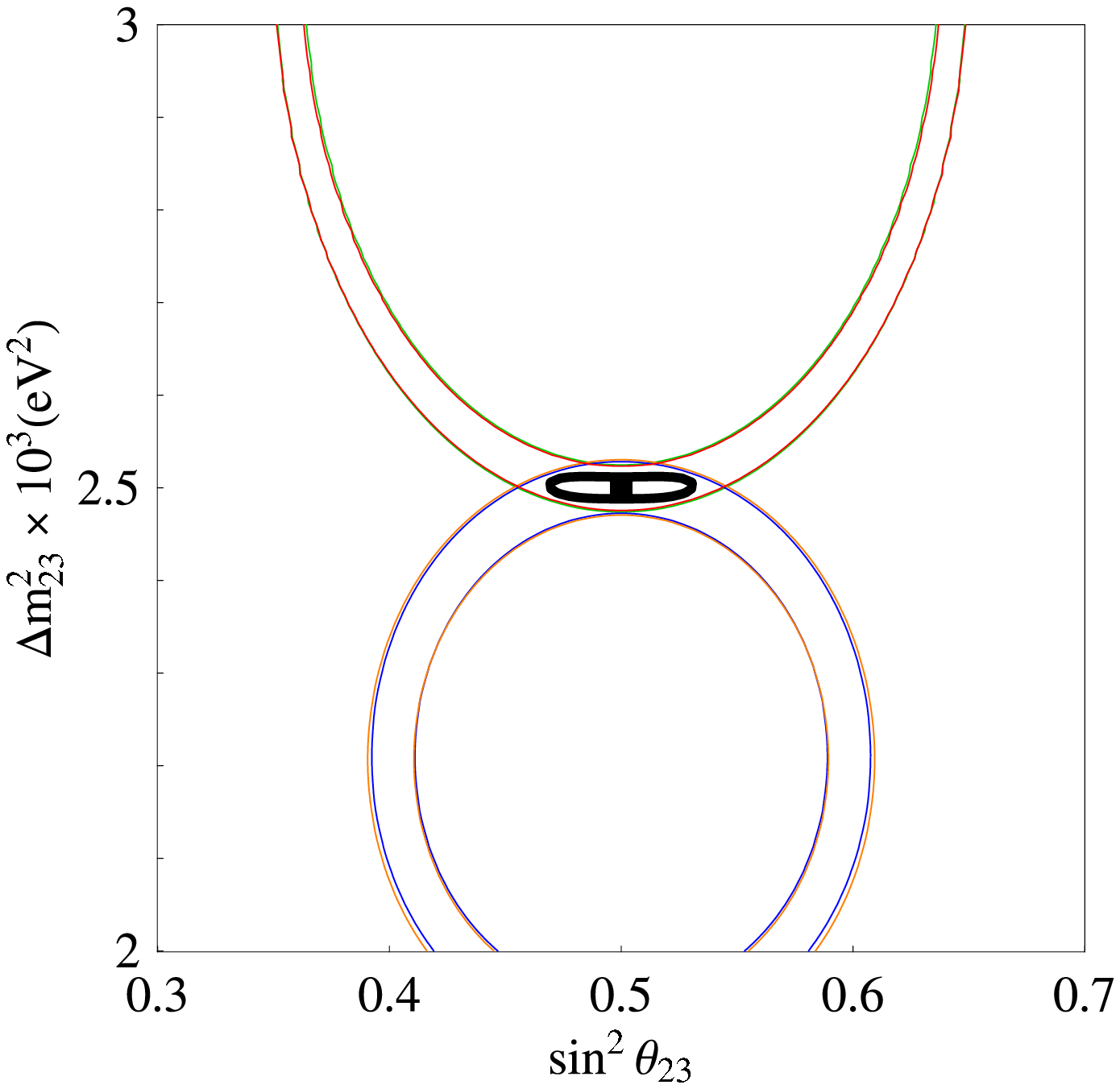} &
\hspace{-0.5cm} \epsfxsize8cm\epsffile{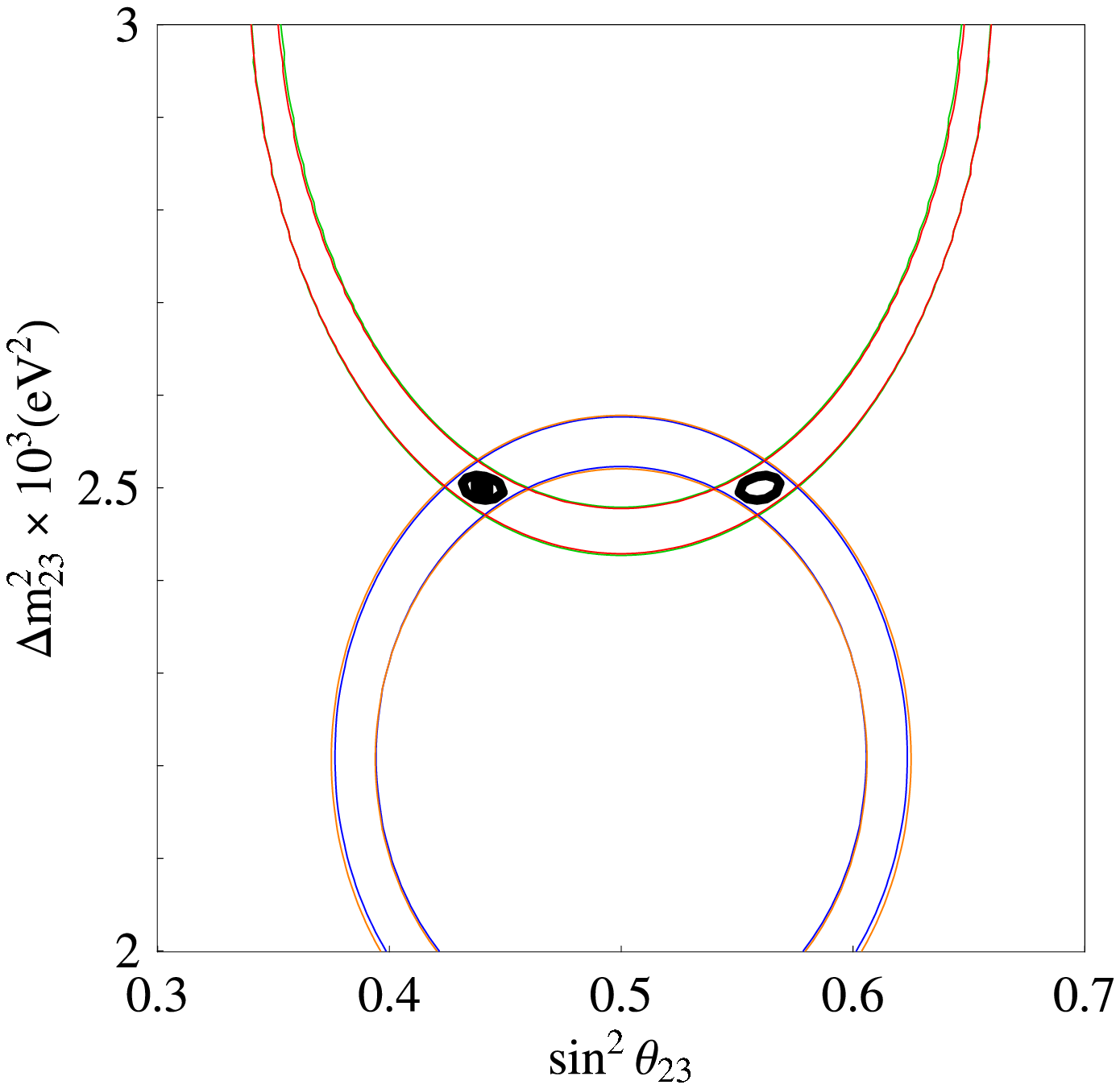} \\
\hspace{-1.0cm} \epsfxsize8cm\epsffile{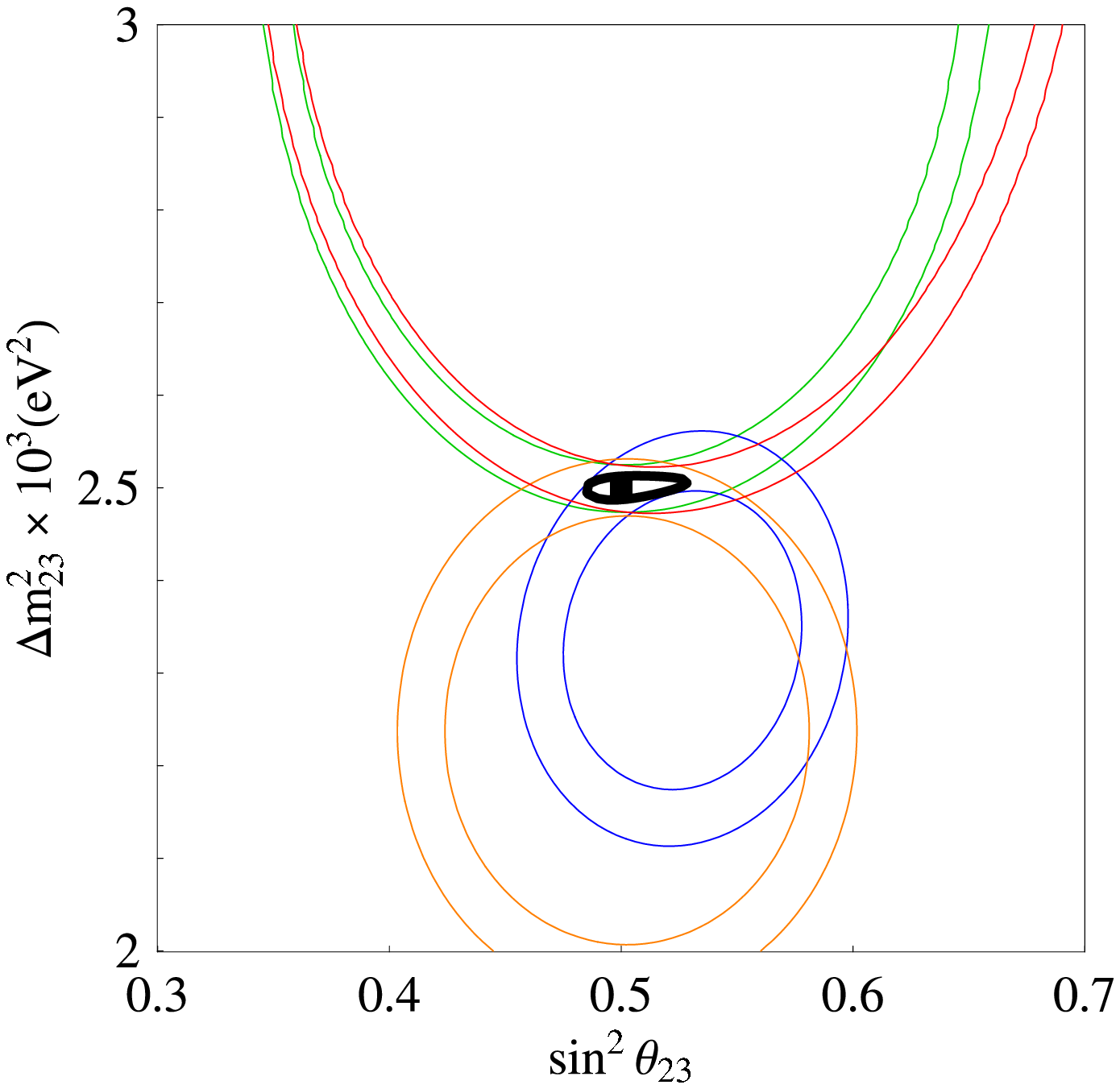} &
\hspace{-0.5cm} \epsfxsize8cm\epsffile{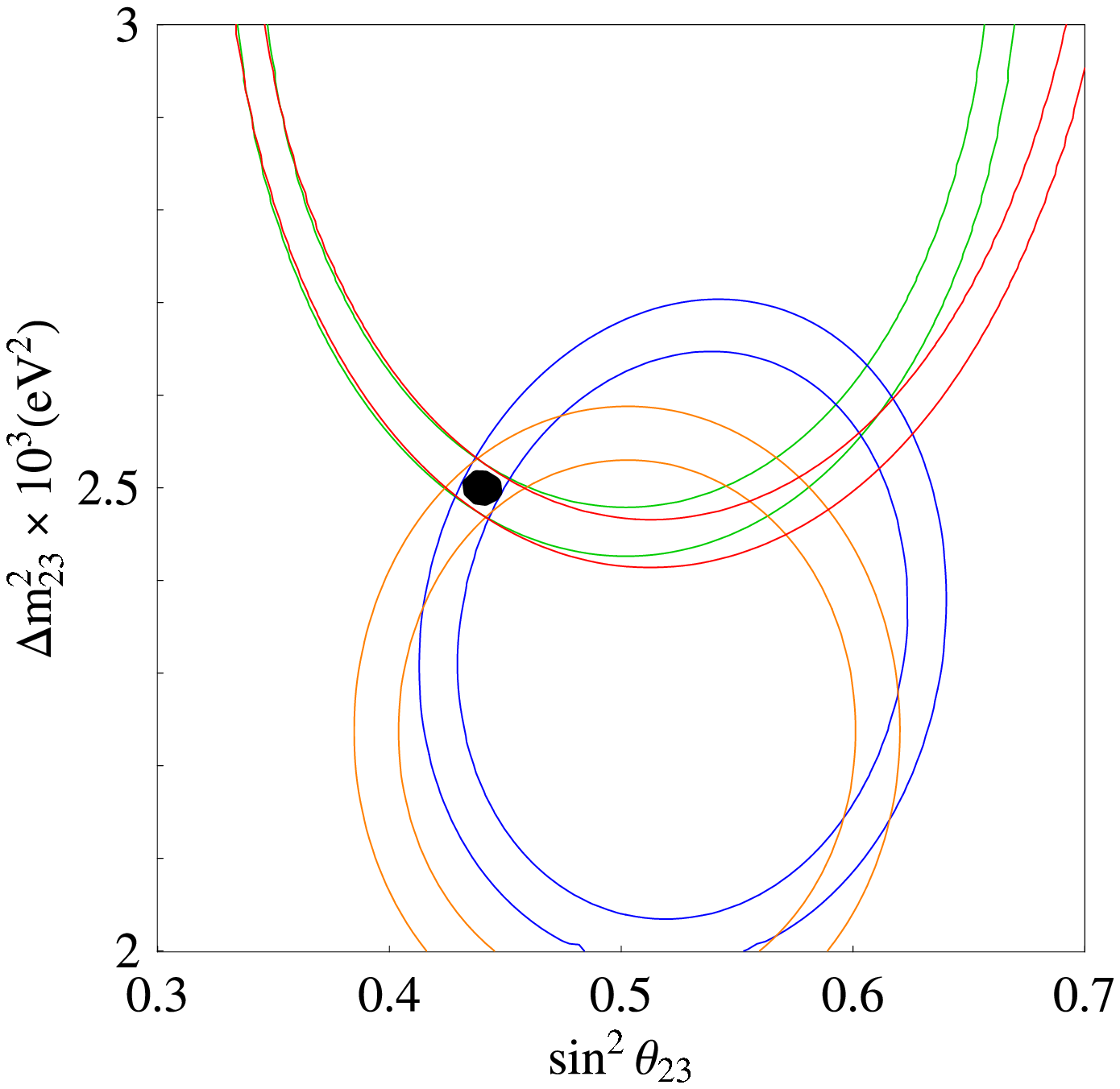} \\
\end{tabular}
\caption{\it  Binning at the $L = 7000$ km Neutrino Factory.  Left: $\theta_{23} = 45^\circ$; right: $\theta_{23} = 41.5^\circ$;
                                                  top: $\theta_{13} = 0^\circ$; bottom: $\theta_{13} = 8^\circ$.}
\label{fig:NF7000bins}
\end{center}
\end{figure}

In Figs.~\ref{fig:NF3000bins} and \ref{fig:NF7000bins} we first present the 90 \% contours of the on-peak and above-peak
disappearance channel energy bins for two values of $\theta_{23}$, $\bar \theta_{23} = 41.5^\circ,45^\circ$ 
(left and right panels, respectively), and two values of $\theta_{13}$, $\bar \theta_{13} = 0^\circ,8^\circ$
(top and bottom panels, respectively). The medium baseline results are shown in Fig.~\ref{fig:NF3000bins}, 
the long baseline results in Fig.~\ref{fig:NF7000bins}. Both neutrino and antineutrino bins are presented.
Again, solar parameters are kept fixed to their present best fit values, $\Delta m^2_{12} = 8.2 \times 10^{-5}$ eV$^2$, 
$\theta_{12} = 33^\circ$. The input value for $\delta$ is $\bar \delta = 0^\circ$.

Notice, first of all, that the resolution in $\theta_{23}$ is extremely good and that maximal mixing can be easily 
excluded for $\theta_{23} = 41.5^\circ$.
In particular, at the $L = 7000$ baseline a 4 \% error on $\theta_{23}$ for $\theta_{23} = 41.5^\circ$ is found.
This was expected, being the statistics much higher than at the Super-Beams experiments studied in Sect.~\ref{sec:sb} 
(see Tabs.~\ref{tab:SPLevents}-\ref{tab:NF7Levents}). Moreover, a new feature arises for $\theta_{13} = 8^\circ$ at both baselines.
When a rather large non-vanishing $\theta_{13}$ is switched on, matter effects become extremely important and introduce 
a strong $\theta_{23}$-asymmetry in eq.~(\ref{eq:probdismu}). The asymmetry can be clearly seen in the bottom panels of 
Figs.~\ref{fig:NF3000bins} and \ref{fig:NF7000bins}, and it is crucial in solving the {\it disappearance octant degeneracy}, 
see Sect.~\ref{sec:sb:deg}.  

We point out that the on-peak bin shows a circular shape centered around the input value, distinct from the upward(downward)-curved 
shapes of contours corresponding to above(below)-peak bins of Figs.~\ref{fig:T2Kbins} and \ref{fig:SPLbins}. This is because the 
Neutrino Factory flux is not centered around the peak energy for the chosen baselines: the peak energy for a $L = 3000$ km baseline 
would be $<E_{\nu_\mu}> \sim 6$ GeV and for a $L = 7000$ km baseline $<E_{\nu_\mu}> \sim 14$ GeV. As a consequence, we have no energy
bins below the peak energy, but only on-peak or above-peak bins. This is a major flaw of the present Neutrino Factory design, 
that could perhaps be solved with an improved detector capable to take advantage of low energy bins. 


The solving of the octant-degeneracy in the disappearance channel (for large $\theta_{13}$) is of great importance. 
It is useful to recall that, when looking for a CP-violating signal in appearance channels as 
$\nu_e \to \nu_\mu, \nu_e \to \nu_\tau$ and $\nu_\mu \to \nu_e$, we must deal in general with an eightfold-degeneracy 
\cite{Barger:2001yr} that originates from three sources: the discrete ambiguities parametrized by 
$s_{atm}$ \cite{Minakata:2001qm} and $s_{oct}$ \cite{Fogli:1996pv} and the {\it intrinsic} ambiguity due to the 
trigonometric nature of $P_{e\mu},P_{e\tau}$ and $P_{\mu e}$ in the two unknowns $\theta_{13}$ and $\delta$ 
\cite{Burguet-Castell:2001ez} (see eq.~(\ref{eq:appnue}) as an example).
As a consequence, for generic values of the input pair ($\bar \theta_{13},\bar \delta$), we get eight different solutions in
the ($\theta_{13},\delta$) plane. Different ways to solve the eightfold ambiguity have been proposed in the literature, 
such as to combine experiments with different $L/E$ \cite{Burguet-Castell:2001ez}, or the golden channel at the Neutrino
Factory with $\nu_\mu \to \nu_e$ at a SuperBeam \cite{Burguet-Castell:2002qx}, or the golden and silver channels 
at the Neutrino Factory \cite{Donini:2002rm,Autiero:2003fu}, or even to combine different SuperBeams \cite{Minakata:2003ca,Yasuda:2004gu} 
or a SuperBeam and a $\beta$-Beam \cite{Bouchez:2003fy,Donini:2004hu} (see also Refs.~\cite{Terranova:2004hu}-\cite{Donini:2005qg}).
In Ref.~\cite{Donini:2003kr} it was shown that the combination of the Neutrino Factory 
with a SPL-like SuperBeam was able to solve the eightfold-degeneracy and to give a single allowed region 
in the ($\theta_{13},\delta$) plane at the price of three specialized detectors, a 40 kton Magnetized Iron Calorimeter 
and a 4 kton Emulsion Cloud Chamber (to look for $\nu_e \to \nu_\mu$ and $\nu_e \to \nu_\tau$ oscillations, respectively) 
and a gigantic 1 Mton Water \v Cerenkov (to look for $\nu_\mu \to \nu_e$). 
The results of Figs.~\ref{fig:NF3000bins} and \ref{fig:NF7000bins} suggest, on the other hand, that some reduction of the 
eightfold ambiguity can be achieved at the Neutrino Factory combining both golden and silver appearance signal with 
the $\nu_\mu$ disappearance signal, with no need for a huge water \v Cerenkov detector to be located somewhere else
(albeit, using a 10 kton ECC). 

\begin{figure}[t!]
\vspace{-0.5cm}
\begin{center}
\begin{tabular}{cc}
\hspace{-1.0cm} \epsfxsize8cm\epsffile{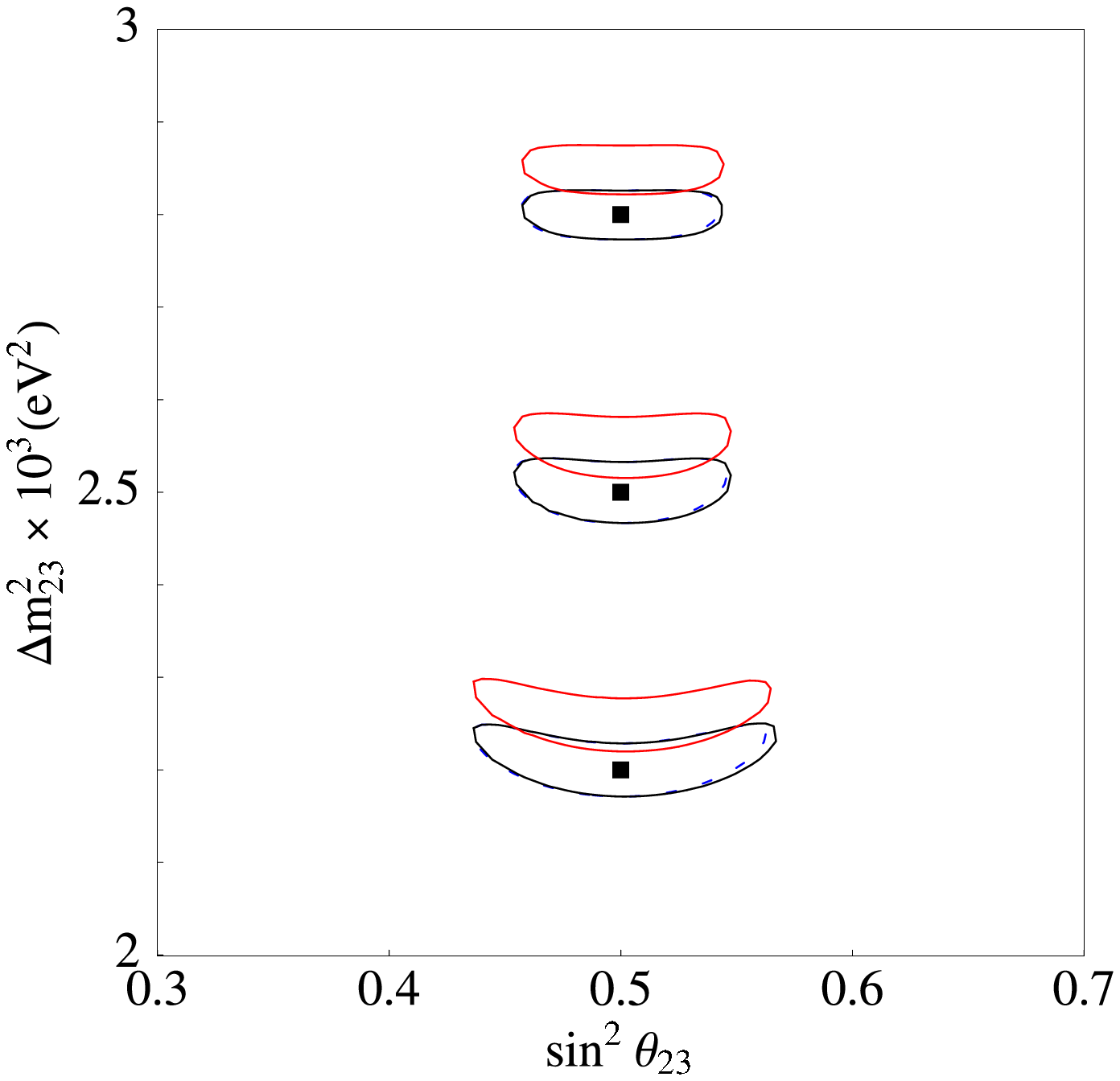} & 
\hspace{-0.5cm} \epsfxsize8cm\epsffile{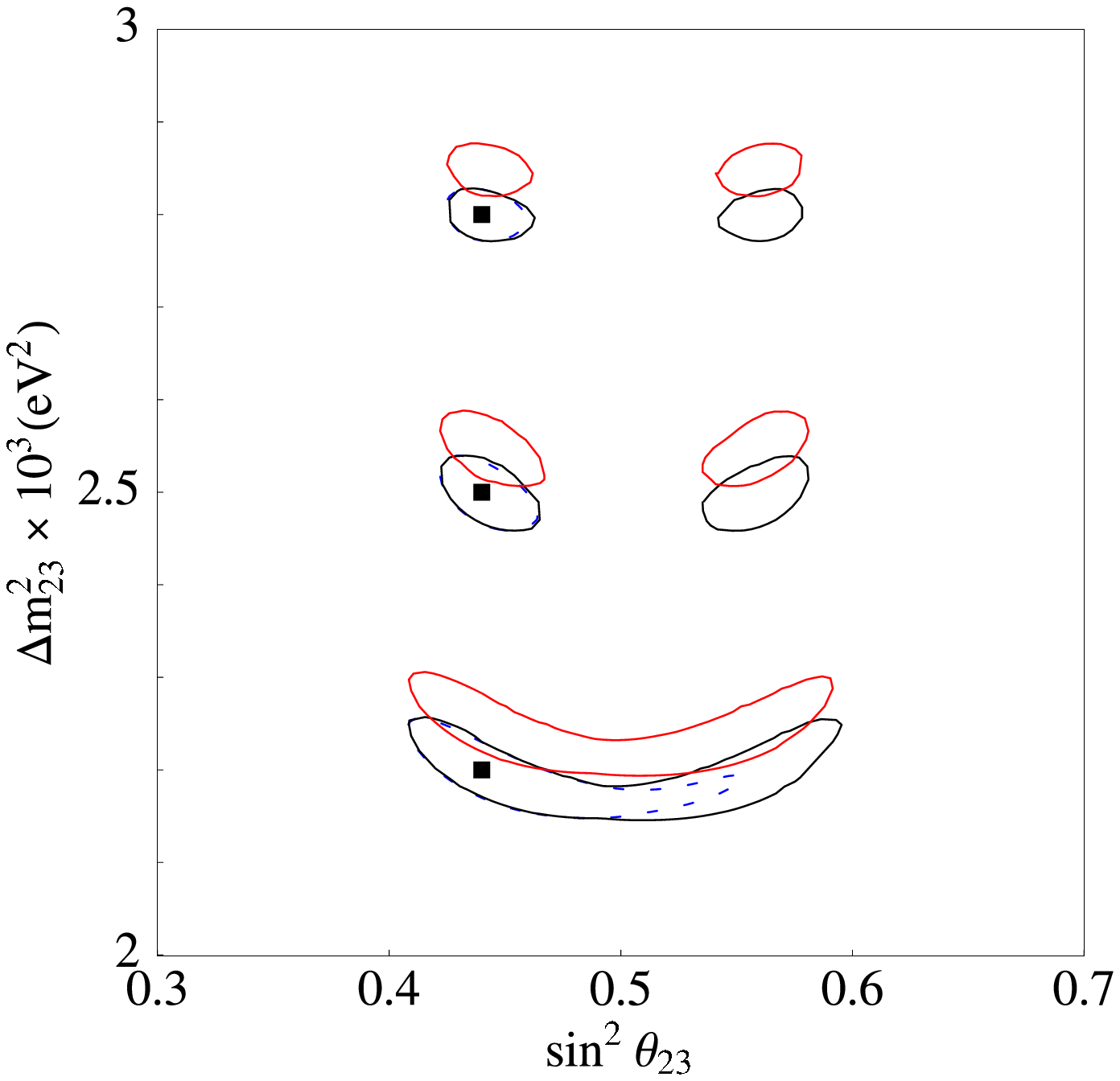} \\
\hspace{-1.0cm} \epsfxsize8cm\epsffile{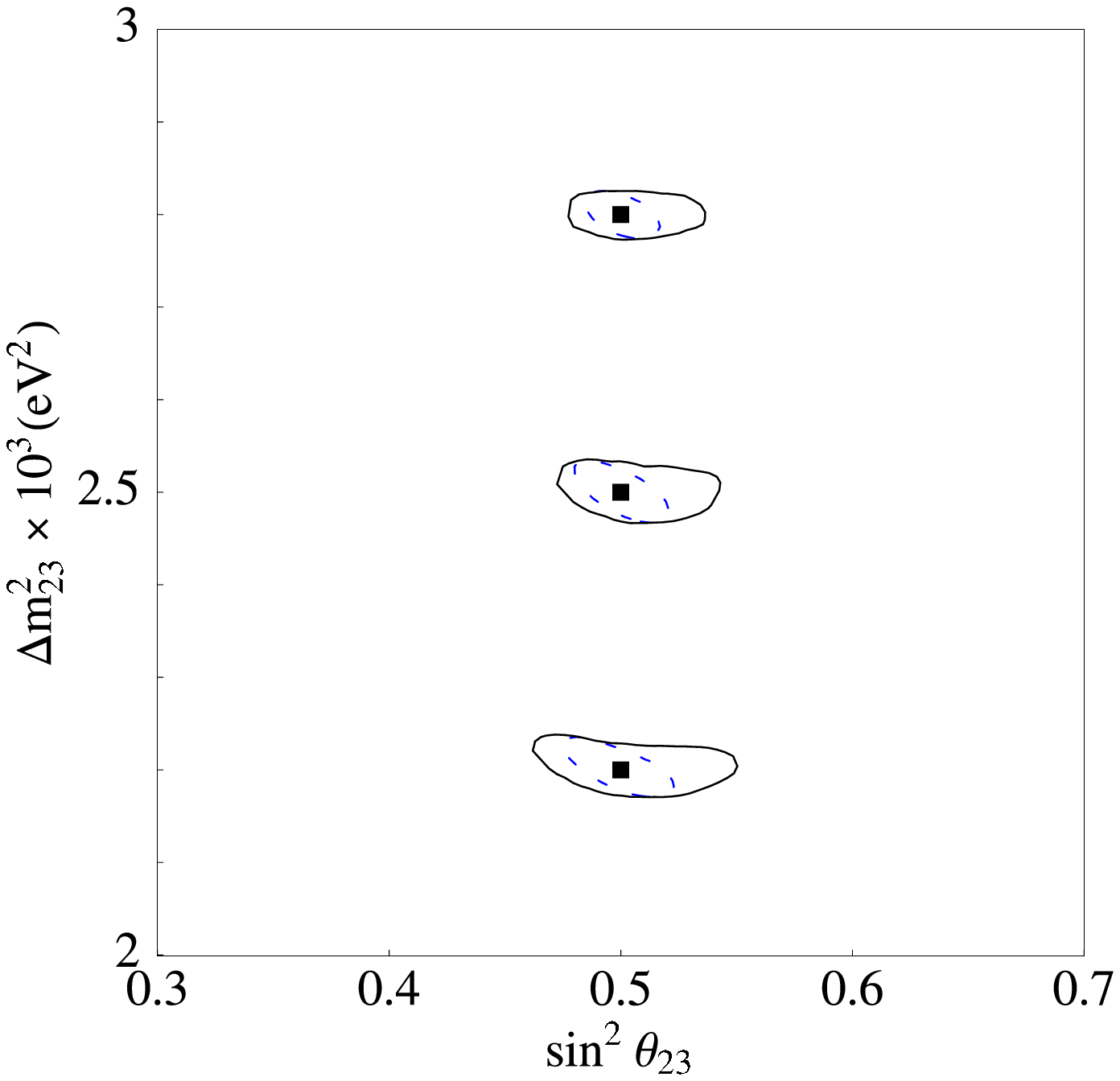} & 
\hspace{-0.5cm} \epsfxsize8cm\epsffile{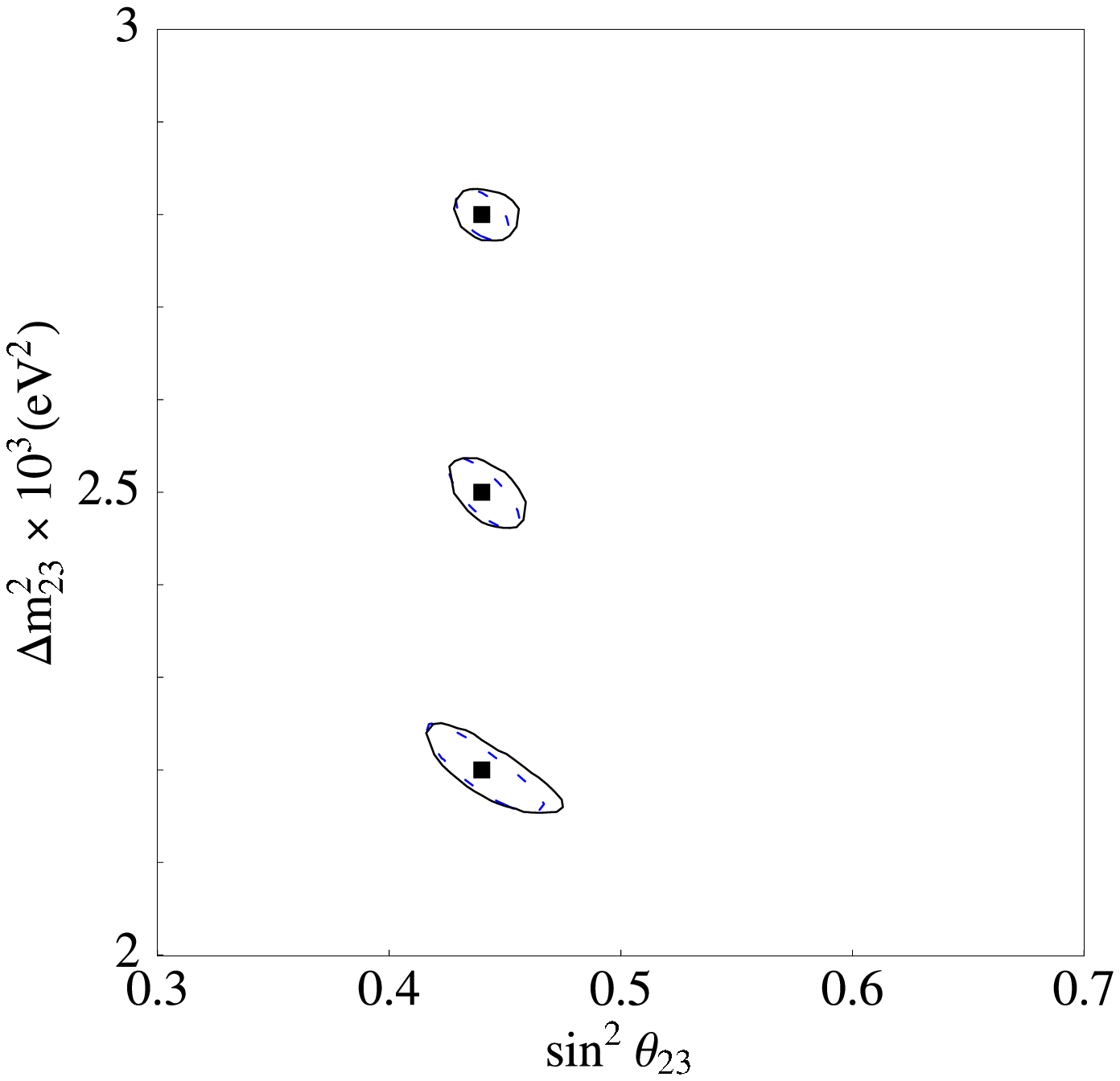}
\end{tabular}
\caption{\it Appearance and disappearance in the ($\theta_{23}, \Delta m^2_{23}$) plane at the $L = 3000$ km Neutrino Factory. 
Left: $\theta_{23} = 45^\circ$; right: $\theta_{23} = 41.5^\circ$; top: $\theta_{13} = 0^\circ$; bottom: $\theta_{13} = 8^\circ$.} 
\label{fig:th13:appdis:nf3000}
\end{center}
\end{figure}

\begin{figure}[t!]
\vspace{-0.5cm}
\begin{center}
\begin{tabular}{cc}
\hspace{-1.0cm} \epsfxsize8cm\epsffile{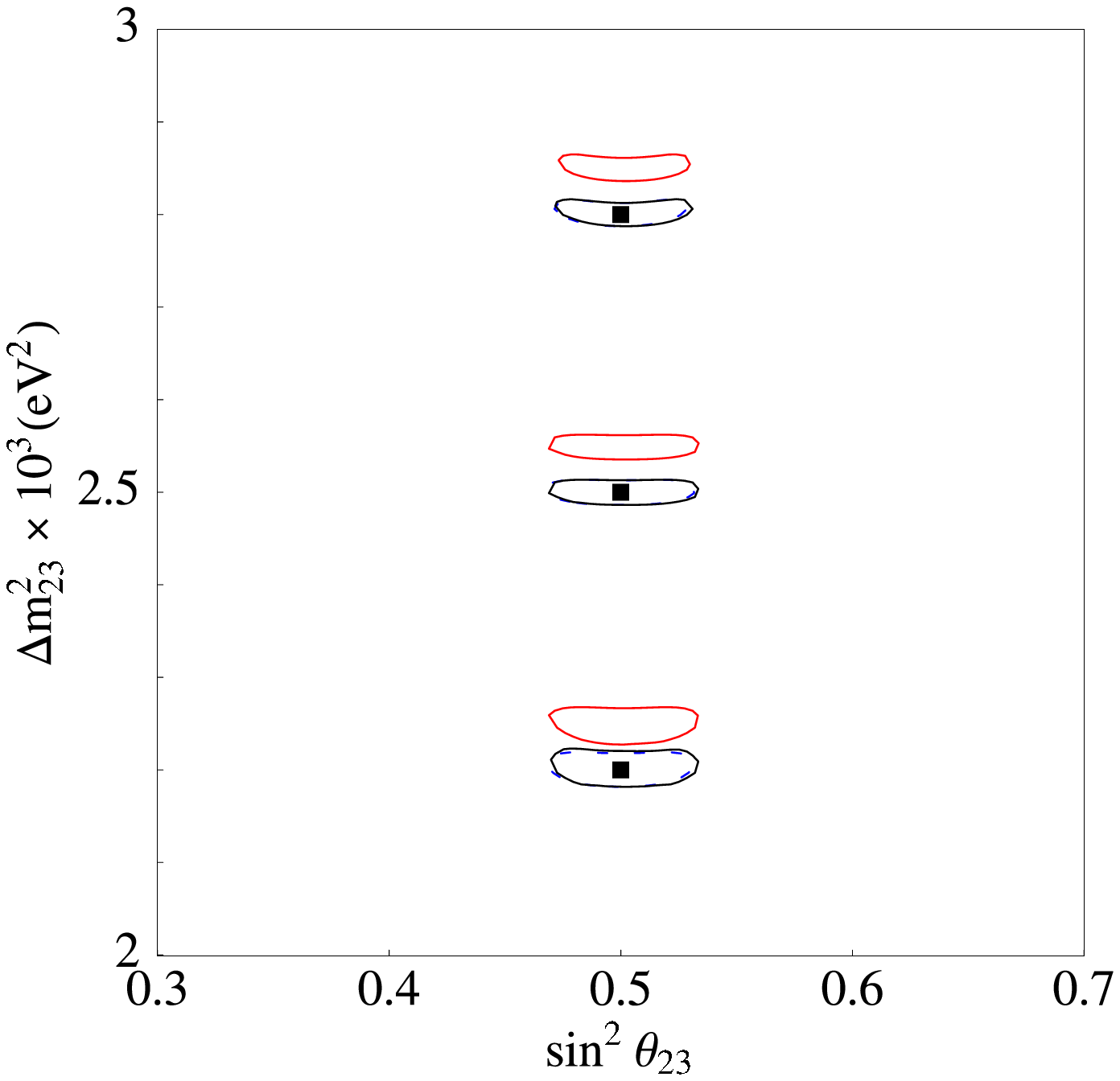} & 
\hspace{-0.5cm} \epsfxsize8cm\epsffile{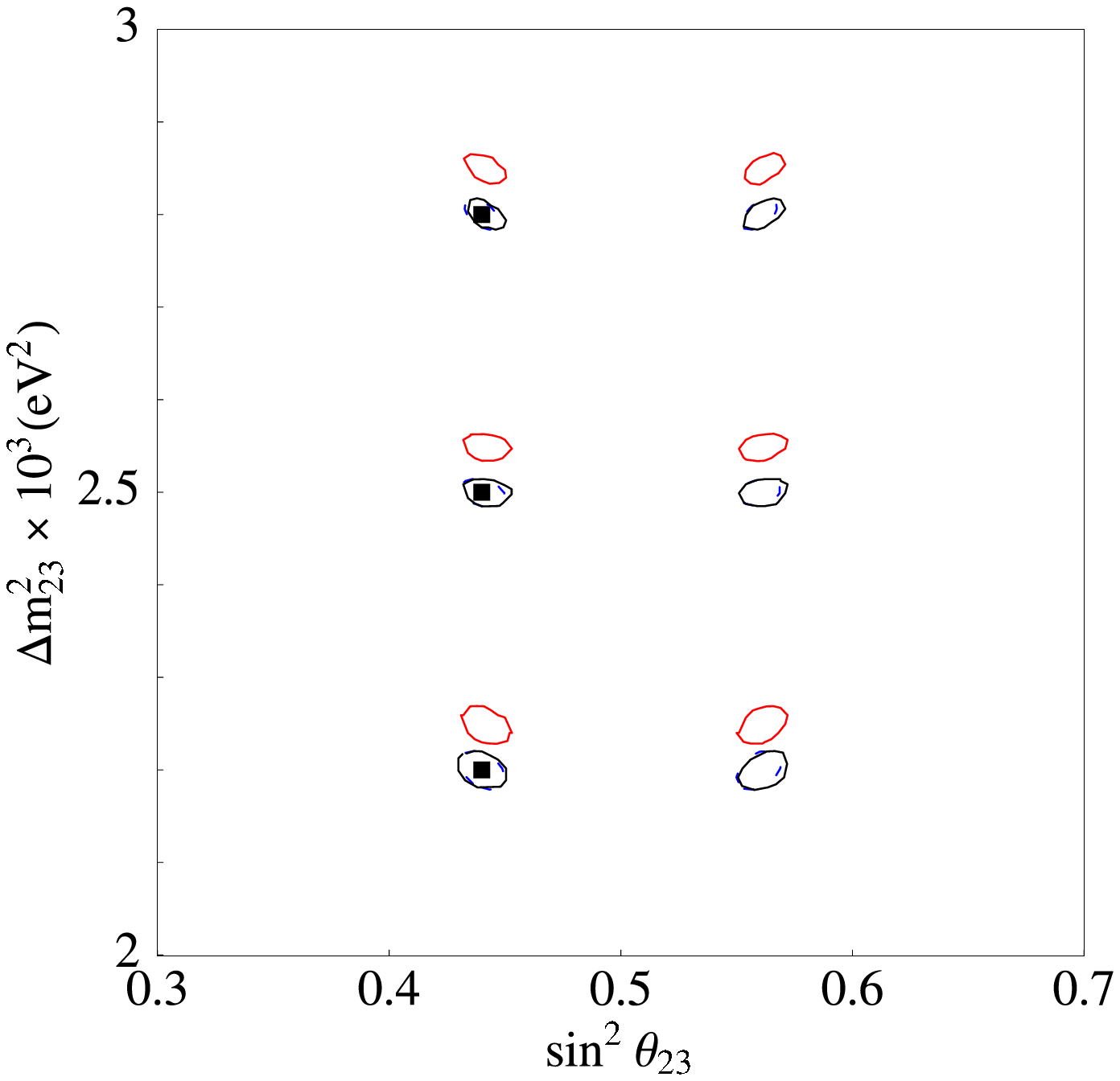} \\
\hspace{-1.0cm} \epsfxsize8cm\epsffile{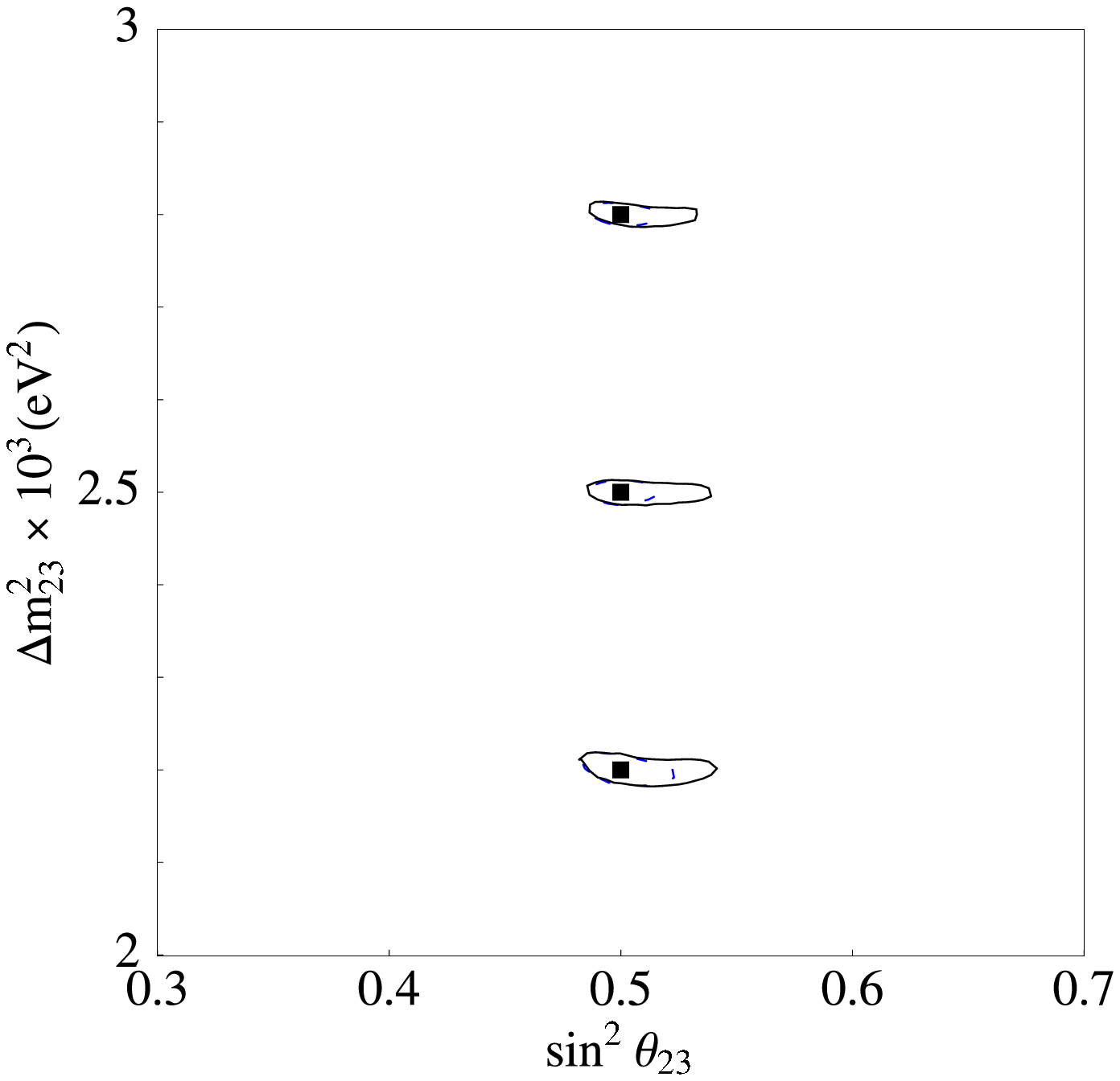} & 
\hspace{-0.5cm} \epsfxsize8cm\epsffile{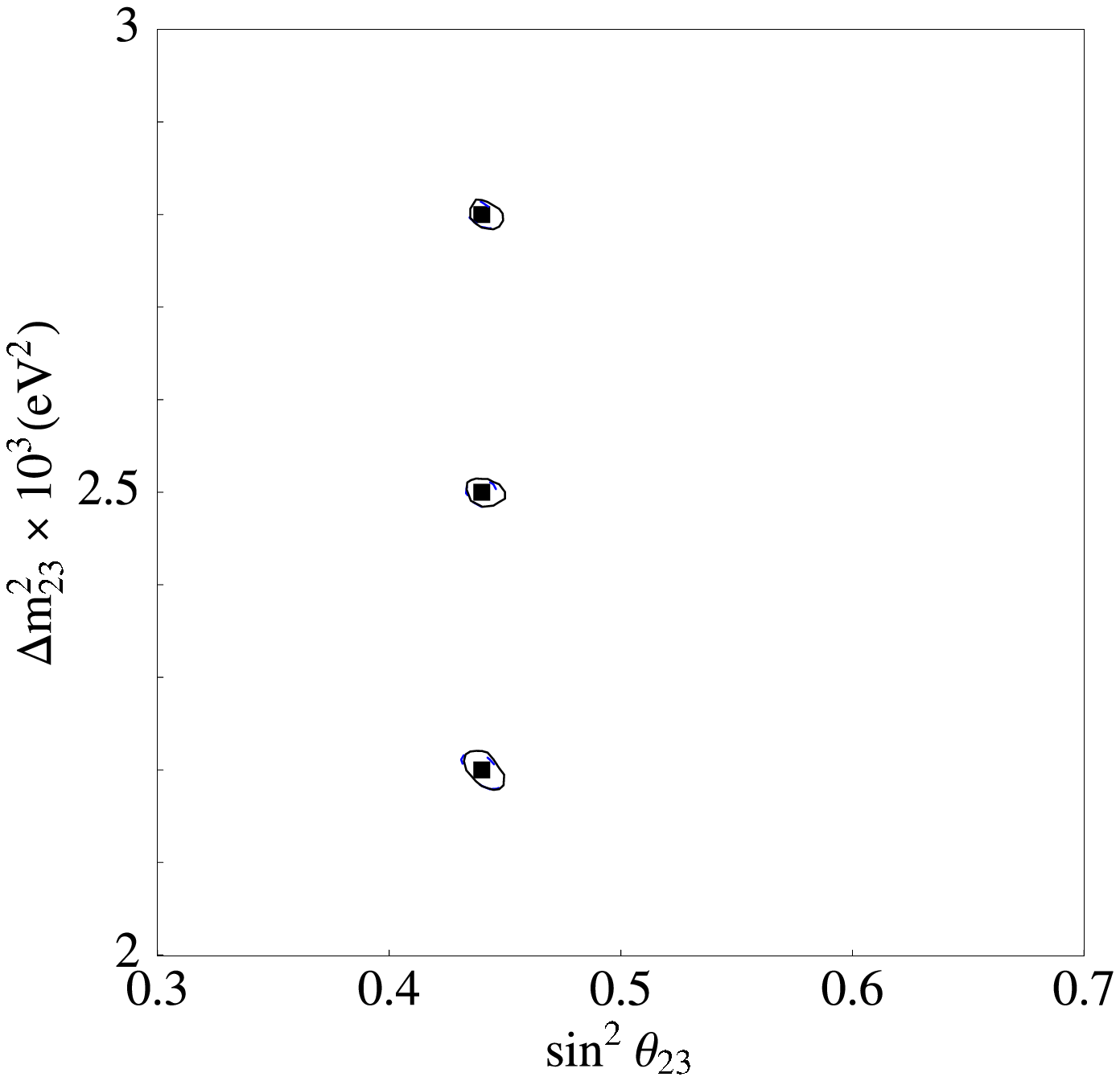}
\end{tabular}
\caption{\it Appearance and disappearance in the ($\theta_{23}, \Delta m^2_{23}$) plane at the $L = 7000$ km Neutrino Factory. 
Left: $\theta_{23} = 45^\circ$; right: $\theta_{23} = 41.5^\circ$; top: $\theta_{13} = 0^\circ$; bottom: $\theta_{13} = 8^\circ$.} 
\label{fig:th13:appdis:nf7000}
\end{center}
\end{figure}

The results of a four-parameters fit in ($\theta_{23},\Delta m^2_{23},\theta_{13},\delta$) projected onto the 
($\theta_{23},\Delta m^2_{23}$) plane\footnote{The two-parameters 90 \% CL contours are shown.} obtained by combining 
the disappearance and appearance signals at the Neutrino Factory 
are eventually presented in Fig.~\ref{fig:th13:appdis:nf3000} for the medium baseline $L = 3000$ km and in 
Fig.~\ref{fig:th13:appdis:nf7000} for the long baseline, $L = 7000$ km.
Again, two values of $\bar \theta_{23}$, $\bar \theta_{23} = 41.5^\circ,45^\circ$ (left and right panels, respectively) 
and two values of $\bar \theta_{13}$, $\bar \theta_{13} = 0^\circ,8^\circ$ (top and bottom panels, respectively) are shown.

The input value for $\delta$ is $\bar \delta = 0^\circ$.
In each figure, dotted lines refer to the fit with $\theta_{13} = \bar \theta_{13}$, $\delta = \bar \delta$
and $s_{atm} = \bar s_{atm}$. 
The solid lines, on the other hand, represent the result of a fit with variable $\theta_{13},\delta$ and $s_{atm}$.
Notice that, as a general result, the uncertainties on the atmospheric parameters are better measured at the long baseline 
than at the medium one. In particular, for $\theta_{13} = 0^\circ$, maximal mixing cannot be excluded for a low atmospheric 
mass difference, $\Delta m^2_{23} = 2.2 \times 10^{-3}$ eV$^2$ at the medium baseline: for such a low value of $\Delta m^2_{23}$, 
all the energy bins are above-peak and thus show an upward-curved shape (see Fig.~\ref{fig:NF3000bins}), reducing their complementarity.
For $\theta_{13}=8^\circ$ the octant ambiguity is always solved by the strong matter effects at both baselines.
Combining the disappearance channel with the long-studied golden and silver appearance channels at the Neutrino
Factory solves the octant-ambiguity for $\theta_{13} \geq 3^\circ$. 


\begin{figure}[t!]
\vspace{-0.5cm}
\begin{center}
\begin{tabular}{cc}
\hspace{-1.0cm} \epsfxsize8cm\epsffile{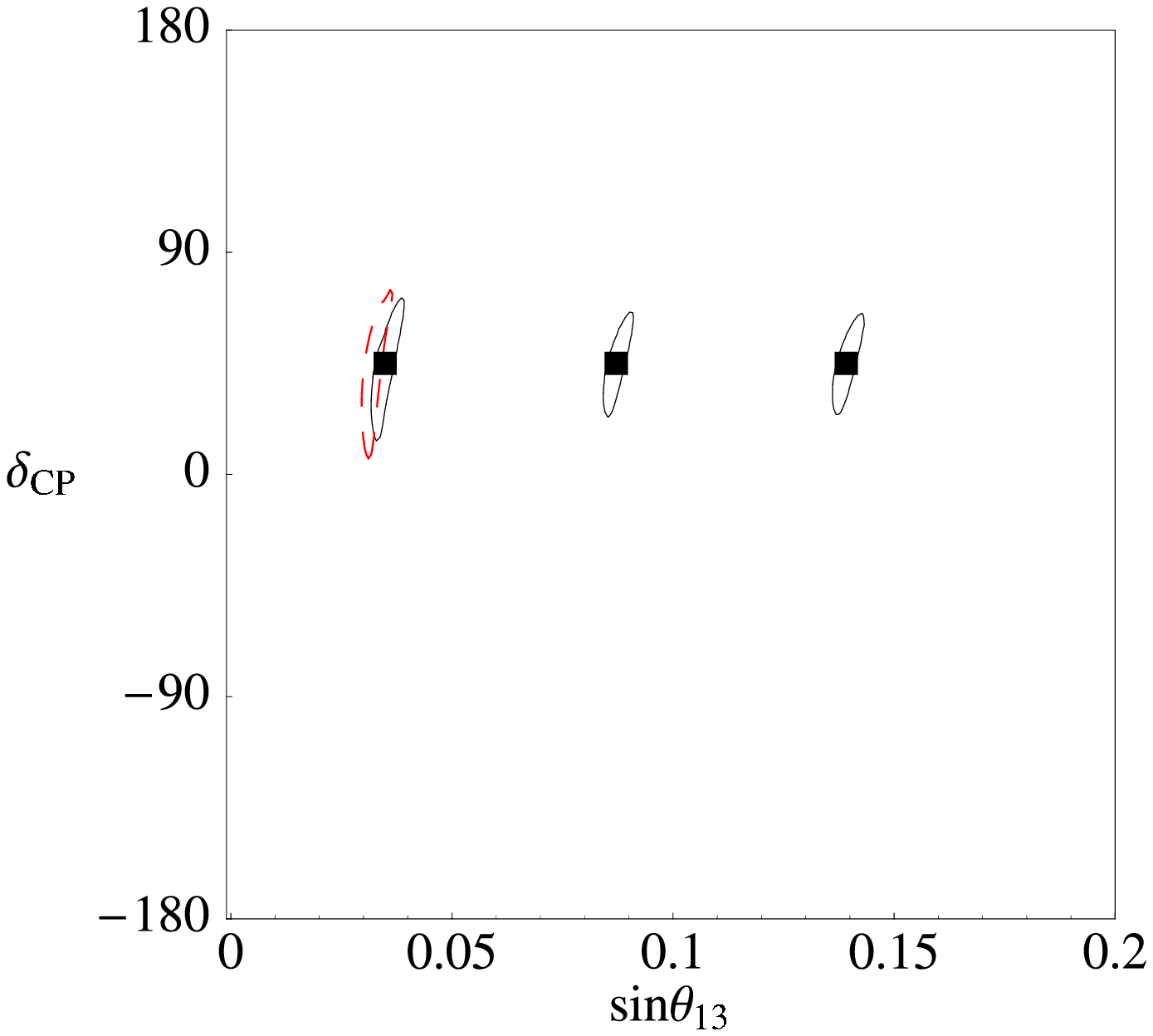} & 
\hspace{-0.5cm} \epsfxsize8cm\epsffile{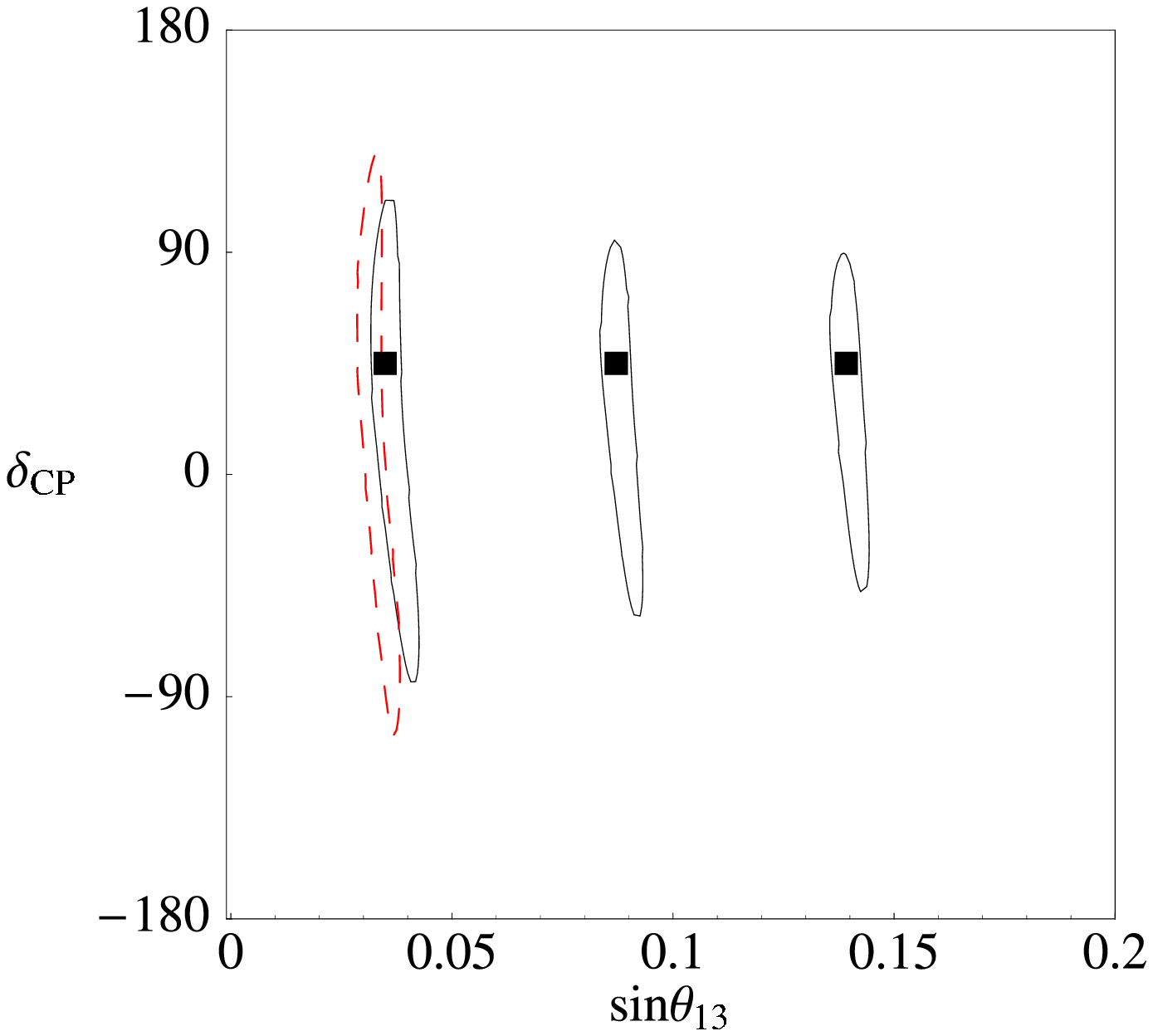} \\
\hspace{-1.0cm} \epsfxsize8cm\epsffile{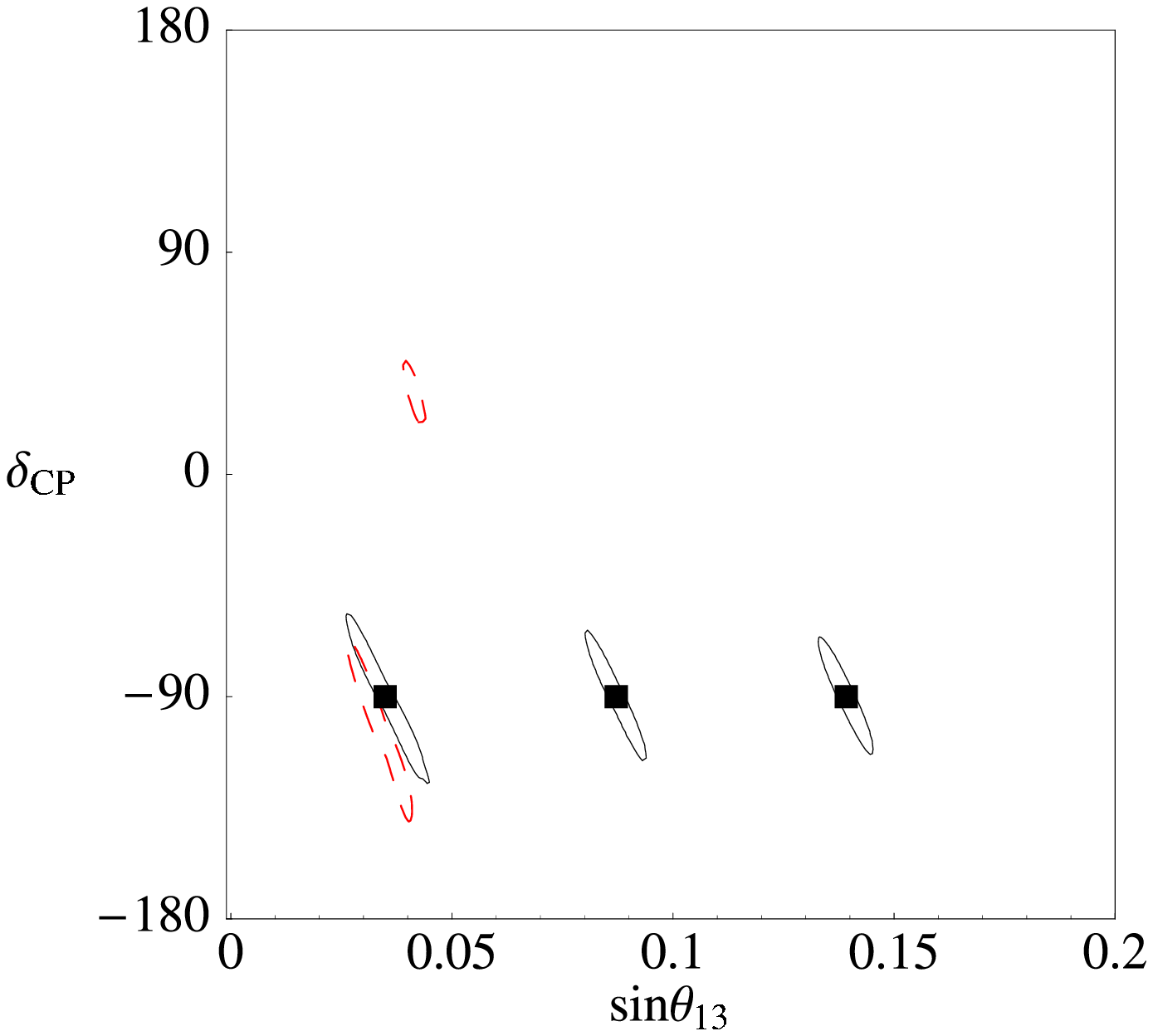} & 
\hspace{-0.5cm} \epsfxsize8cm\epsffile{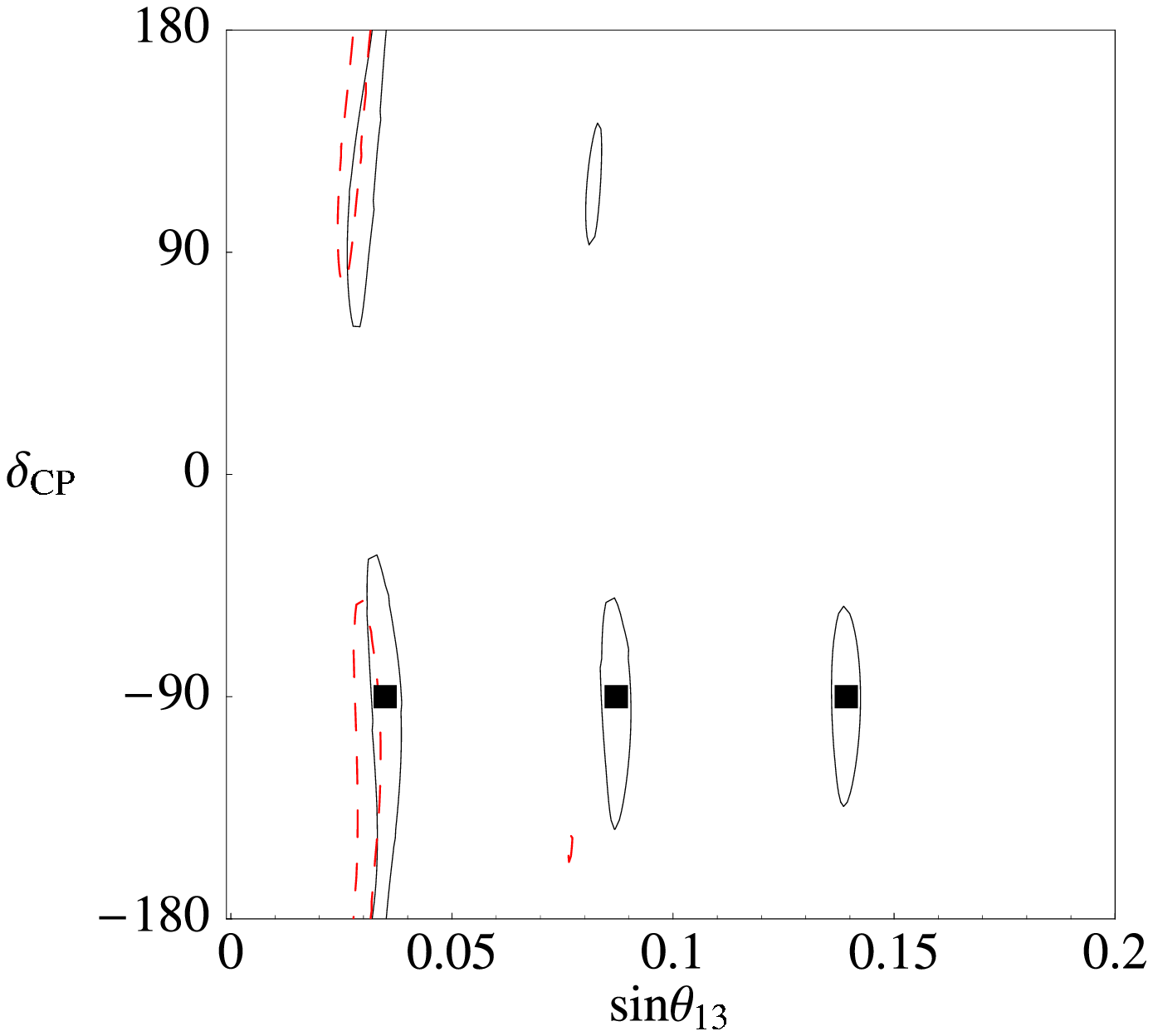}
\end{tabular}
\caption{\it Appearance and disappearance in the ($\theta_{13}, \delta$) plane at the Neutrino Factory.
Left: $L = 3000$ km baseline; right: $L = 7000$ km baseline; top: $\delta = 45^\circ$; bottom: $\delta = -90^\circ$.
The mixing angle takes the following values: $\theta_{13} = 2^\circ, 5^\circ$ and $8^\circ$.} 
\label{fig:th13:appnf}
\end{center}
\end{figure}

In Fig.~\ref{fig:th13:appnf} we eventually show the results of a four-parameters fit in
($\theta_{23},\Delta m^2_{23},\theta_{13},\delta$) projected onto the ``classic'' ($\theta_{13},\delta$) 
plane \cite{Cervera:2000kp} obtained by combining the disappearance and appearance signals 
at the Neutrino Factory, both for the $L = 3000$ km and $L = 7000$ km baselines (left and right panels, respectively). 
Two values of $\delta$ are shown, $\bar \delta = 45^\circ$ (top panels) and $\bar \delta = -90^\circ$ (bottom panels), 
and three values of $\theta_{13}$, $\bar \theta_{13} = 2^\circ, 5^\circ$ and $ 8^\circ$ 
(corresponding to $\sin \theta_{13} = 0.04, 0.09, 0.14$).
Solid lines correspond to 90\% CL contours with $(s_{atm},s_{oct}) = (\bar s_{atm},\bar s_{oct})$, 
dashed lines to $(s_{atm},s_{oct}) \neq (\bar s_{atm},\bar s_{oct})$.
As it has been known for long, the Neutrino Factory experiments have such a long baseline that the sign of the atmospheric mass 
difference is measured down to $\theta_{13} = 2^\circ$. The novelty here, is in that at 90 \% CL not only the sign clones 
but the octant and intrinsic clones are also solved down to $\theta_{13} = 2^\circ$ for most values of $\delta$.
This stands true for both baselines. However, it is manifest that for $L = 7000$ km the sensitivity to $\delta$ is lost 
(this was called the ``magic baseline'' \cite{Huber:2003ak} precisely because around this distance the effect of $\delta$ 
in the $\nu_e \to \nu_\mu$ oscillation probability vanishes, see also Refs.~\cite{Donini:1999jc} and \cite{Barger:2001yr}). 
A very good measurement of $\theta_{13}$ can be performed at this baseline, but certainly not of the leptonic CP-violating phase.
The case of $\bar \delta = -90^\circ$ is particularly bad, due to low statistics in the antineutrino sample as opposed 
to a relatively high background. In this case the combination of neutrino and antineutrino fluxes is not as effective 
as for positive $\delta$ and the allowed region increases (something that has been observed for the $\beta$-Beam also, 
see Refs.~\cite{Donini:2004hu,Donini:2004iv}).

%

%%%%%%%%%%%%%%%%%%%%%%%%%%%%%%%%%%%%%%%%%%%%%%%%%%%%%%%%%%%%%%%%%%%%%%
%
\section{Sensitivities}
\label{sec:CPdis}
%
%%%%%%%%%%%%%%%%%%%%%%%%%%%%%%%%%%%%%%%%%%%%%%%%%%%%%%%%%%%%%%%%%%%%%%

In this section we compare the sensitivities to $\theta_{13}$, the maximality of $\theta_{23}$,
the sign of the atmospheric mass difference,  the $\theta_{23}$-octant, as well as the CP-discovery potential
potential at the experimental facilities studied in this paper. Since T2K-I and NO$\nu$A will run
with neutrinos only, no sensitivity to the $\delta$ phase is expected;
we then study them separately from the other facilities in section (Sect.\ref{t2knova}). The SPL, T2K-II and 
the Neutrino Factories with $L = 3000$ km and $L = 7000$ km will be analysed in Sect.\ref{others}.
All the contours have been computed adding the
appearance and disappearance channels available at a given facility and, if not differently stated, all the parameters not
involved in the fits are fixed to their best fit values as quoted in the Introduction.

\subsection{Sensitivities at T2K-I and NO$\nu$A}
\label{t2knova}
\begin{figure}[t!]
\vspace{-0.5cm}
\begin{center}
\begin{tabular}{cc}
\hspace{-1.4cm} \epsfxsize8.75cm\epsffile{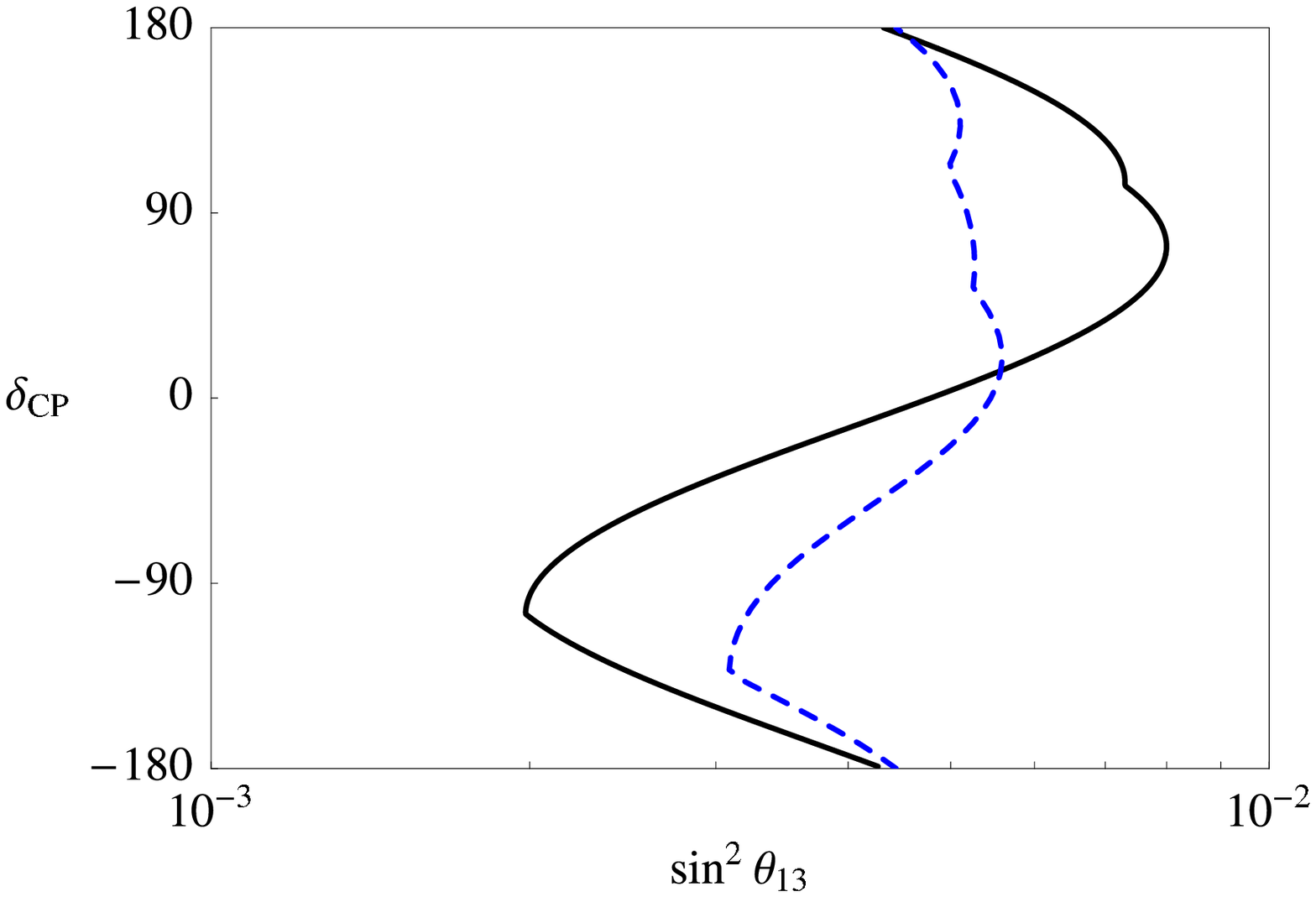} & 
\hspace{-0cm} \epsfxsize8.25cm\epsffile{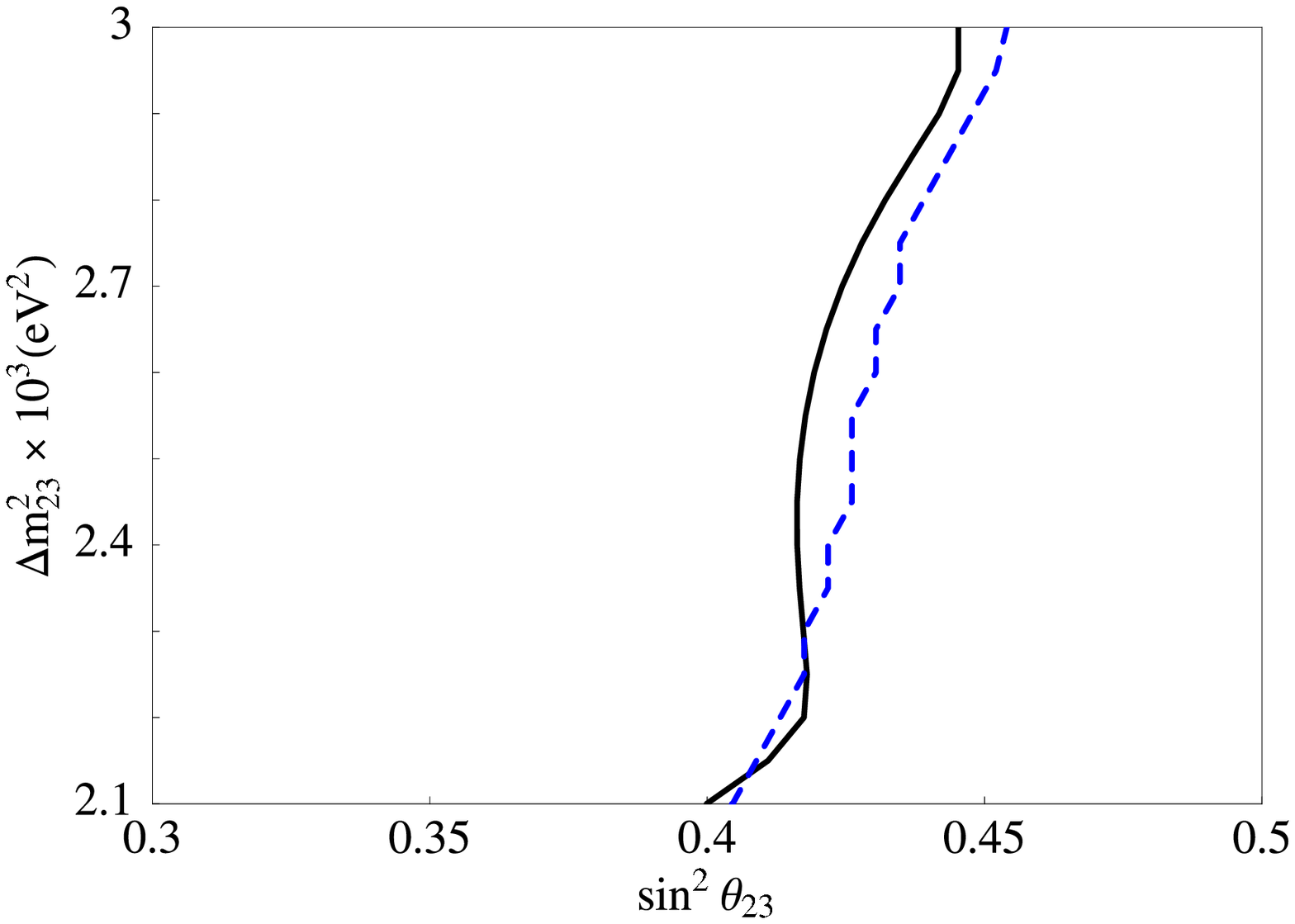} 
\end{tabular}
\caption{\it Left: 3$\sigma$ $\theta_{13}$-sensitivity; right: 3$\sigma$ sensitivity to maximal $\theta_{23}$.
Solid lines refer to T2K-I; dashed lines to NO$\nu$A.} 
\label{fig:sensit2k}
\end{center}
\end{figure}

In Fig.\ref{fig:sensit2k} we compare the $\theta_{13}$-sensitivity (left plot) 
and the sensitivity to maximal $\theta_{23}$ (right plots) at T2K-I (solid lines) and
NO$\nu$A (dashed lines). \\

\begin{itemize}

\item {\bf $\theta_{13}$-sensitivity} \\

The $\theta_{13}$-sensitivity is defined as the one-parameter $3\sigma$
excluded region as a function of $\delta$ in case of absence of signal.
The contours presented in the following figures represent the excluded valuees of $\theta_{13}$
for a given facility taking into account all posible choices of $s_{atm}$ and $s_{oct}$.
For both facilities the $\theta_{13}$-sensitivity is rather poor, with a typical excluded
region ranging from $[ \sin^2 \theta_{13} ]_{min} = [2,8] \times 10^{-3}$. Sensitivity is 
slightly better for negative $\delta$ than for positive $\delta$, with a maximal sensitivity at
$\delta = -90^\circ (-135^\circ)$ for T2K (NO$\nu$A). We have noticed that, while the T2K-I sensitivity is
basically unaffected by the $s_{atm}$, $s_{oct}$ choice, the NO$\nu$A sensitivity is
sensibly diminished in $\delta \in [-90^\circ,90^\circ]$ when choosing a wrong value for the sign of
$\Delta m_{23}^2 $ or the $\theta_{23}$-octant.

To increase both the $\theta_{13}$-sensitivity
and the CP-discovery potential it has been proposed to add a proton driver to 
the NO$\nu$A design, such as to increase
the neutrino flux from four to six times \cite{Ayres:2004js,MenaRequejo:2005hn}.
\\

\item {\bf Sensitivity to maximal $\theta_{23}$} \\

The potential to exclude maximal $\theta_{23}$ has been computed in the following way:
for a given $\Delta m^2_{23}$, we look for the largest value of
$\theta_{23}$ for which the two-parameter $3\sigma$ contours
do not touch $\theta_{23} = 45^\circ$. We have considered both octants of 
$\theta_{23}$ and we have found a behavior approximately symmetric 
of the sensitivity; we then display our results in the $\sin^2\theta_{23}$ 
variable up to the maximal mixing value of 0.5.

As we can see in Fig.\ref{fig:sensit2k} (right), the sensitivity for the two experiments is essentially the same
and strongly decreases for low values of $\Delta m^2_{23}$. At the best fit 
point
$\Delta m^2_{23}=2.5 \times 10^{-3}$ eV$^2$ deviations as small as 14\% of 
$\sin^2
\theta_{23}$ from maximal mixing could be established at both facilities.
Notice that these curves have been computed for a fixed $\theta_{13}=0$ 
(and $s_{atm}=+1$) so that
matter effects in disappearance probabilities and those from the appearance 
channel $\nu_\mu \to \nu_e$ are completely negligible. We have also
checked that for $\theta_{13}$ close to the current bound, none of the
shown results changes drastically and that everything is basically independent on
the mass hierarchy.

\end{itemize}

\subsection{Sensitivities at T2K-II, SPL and Neutrino Factories}
\label{others}
\begin{itemize}
\item {\bf $\theta_{13}$-sensitivity and CP-discovery potential} \\

\begin{figure}[t!]
\vspace{-0.5cm}
\begin{center}
\begin{tabular}{cc}
\hspace{-1cm} \epsfxsize8.25cm\epsffile{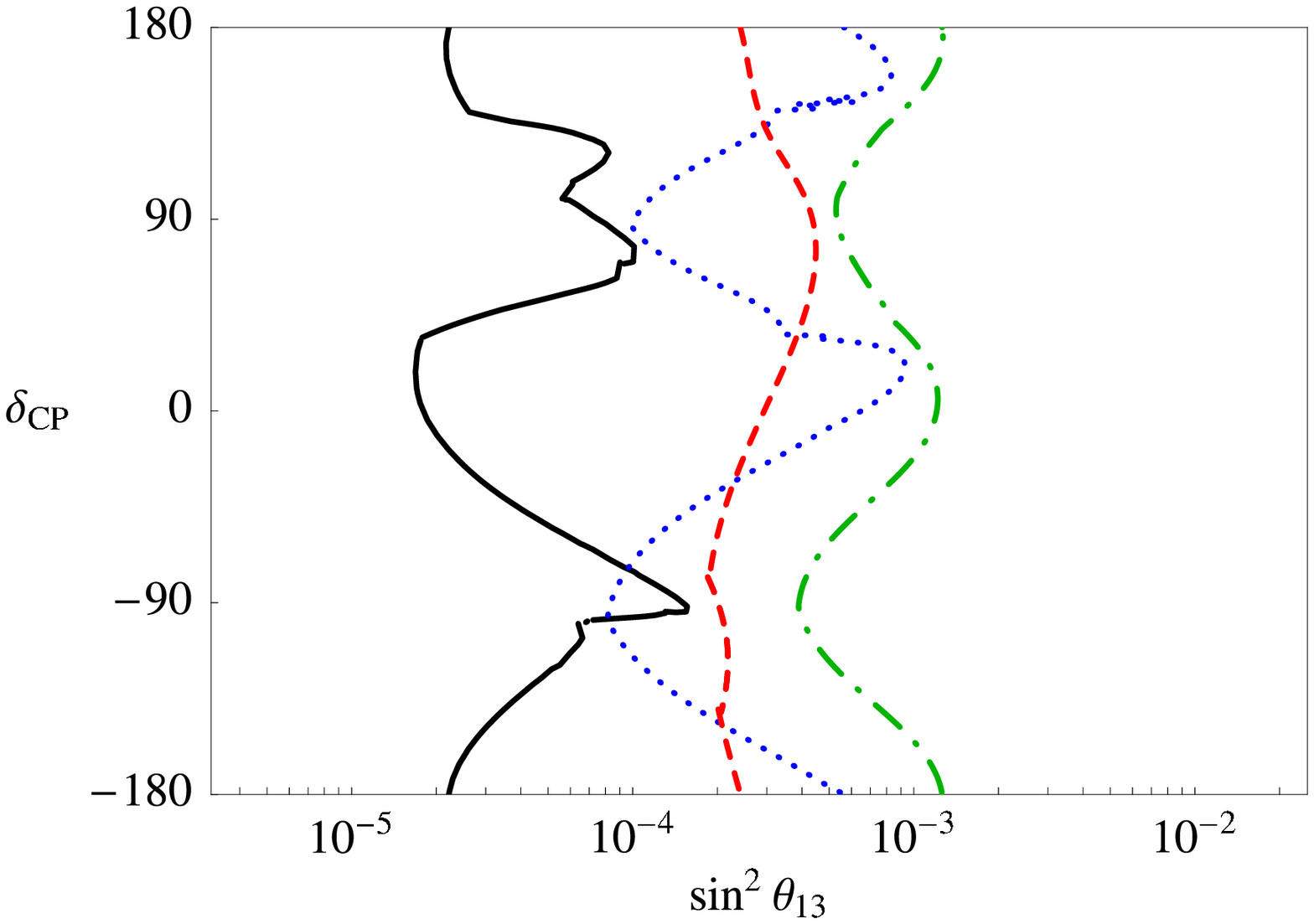} & 
\hspace{-0.5cm} \epsfxsize8.25cm\epsffile{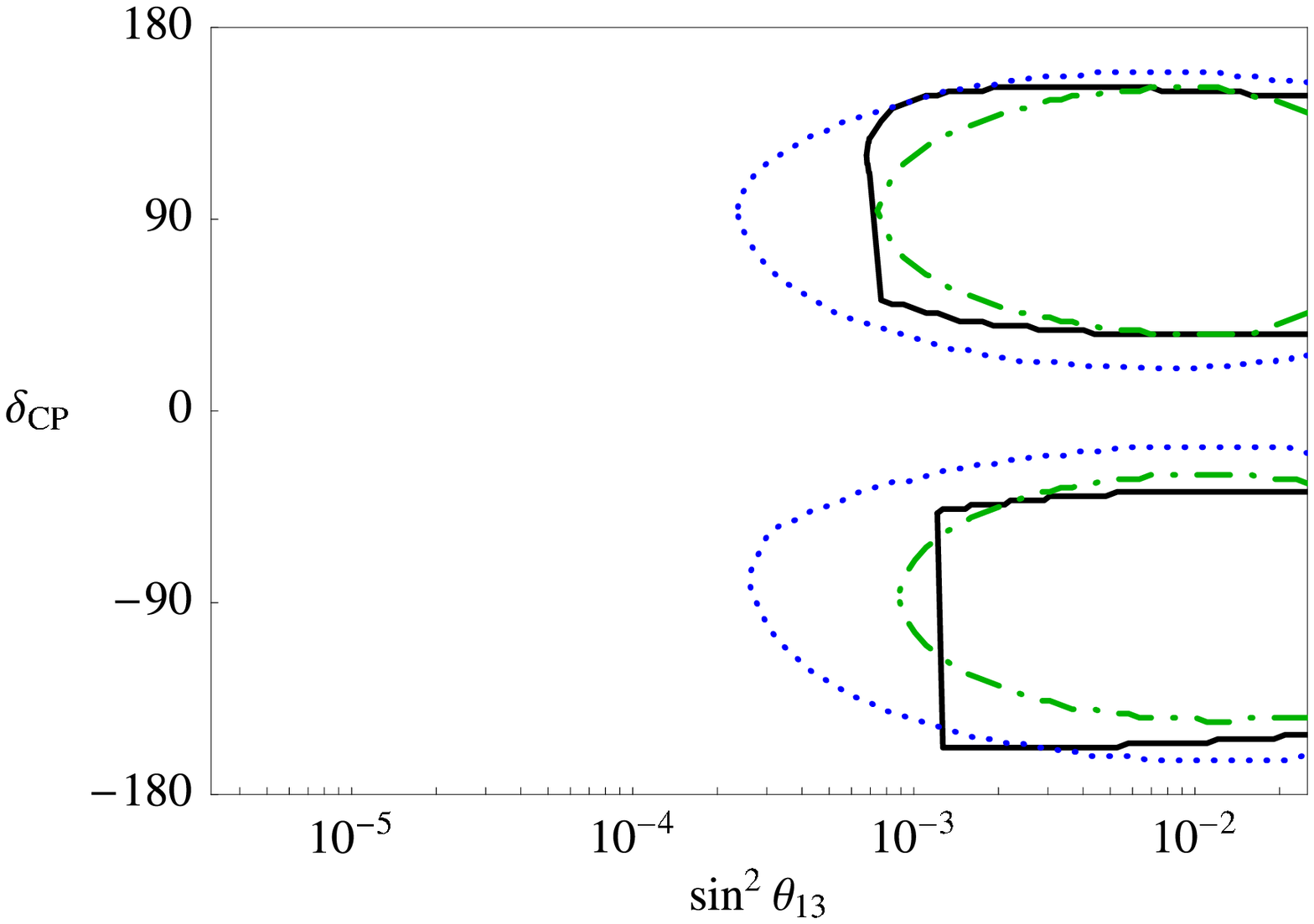} \
\end{tabular}
\caption{\it Left: 3$\sigma$ $\theta_{13}$-sensitivity; right: 3$\sigma$ CP-discovery potential.
Solid lines refer to the $L = 3000$ km Neutrino Factory; dashed lines to the $L = 7000$ km Neutrino Factory; 
dotted lines to T2K-II; dot-dashed lines to the (standard) SPL.} 
\label{fig:sensi}
\end{center}
\end{figure}

In Fig.\ref{fig:sensi} (left) we present the $\theta_{13}$-sensitivity
(computed as explained in the previous section) for 
T2K-II (dotted line), the SPL (dot-dashed line), the Neutrino Factory at 
$L = 3000$ (solid line) and $L = 7000$ (dashed line). As we can see, the NF at 3000 km shows the best
sensitivity to $\theta_{13}$ in the whole range of $\delta$ (with the best value
 of $\sin^2 \theta_{13} \sim 2 \times 10^{-5}$ for $\delta$ around
$30^\circ$) except a small 
region around $\delta=-90^\circ$ where T2K-II assures a slightly better performance (at the
level of $8 \times 10^{-5}$). 
For the Neutrino Factory at $L=7000$ km,
the loss in sensitivity is to be abscribed to the sign degeneracy.
For example, we have found that for $s_{atm} = \bar s_{atm}$ we get that $\theta_{13}$ is excluded at $3\sigma$ down to
$\sin^2 \theta_{13} \sim 7 \times 10^{-5}$ ($\theta_{13} \sim 0.5^\circ$) if $\delta = 90^\circ$. On the 
other hand, for the same value of $\delta$ (but for a wrong choice of $s_{atm}$) we can only exclude $\theta_{13}$
down to $\sin^2 \theta_{13} \sim 4 \times 10^{-4}$ 
($\theta_{13} \sim 1.1^\circ$).
As for the SPL $\theta_{13}$-sensitivity, the $3\sigma$ excluded
region varies in the range
$\sin^2 \theta_{13} \in [4 \times 10^{-4},1.5 \times 10^{-3}]$ for different
values of $\delta$, with no big differences coming from different choices of 
$s_{atm}$ and $s_{oct}$. T2K-II is significantly better than the SPL but significantly worse than the NF at 3000 km,
apart from $\delta = -90^\circ$ (where it improves a little the NF limit).

The CP-violation discovery potential has been computed as in Refs.~\cite{Donini:2004iv,Donini:2005rn}:
at a fixed $\theta_{13}$, we look for the smallest (largest) value of $|\bar \delta|$ for which the
two-parameters $3\sigma$ contours of any of the degenerate solutions
do not touch $\delta = 0^\circ$ nor $|\delta| = 180^\circ$.
Notice that, although the input $\theta_{13}$ value is fixed, the degeneracies can touch $\delta = 0^\circ, 180^\circ$
at $\theta_{13} \neq \bar \theta_{13}$, also\footnote{This is not the case of Fig.~11 in Ref.~\cite{Donini:2004hu}, 
where the excluded region in $\delta$ at fixed $\bar \theta_{13}$ in the absence of a CP-violating signal at 90\% CL 
is presented. In practice, in that figure we compare $N_\pm (\bar \theta_{13},\delta)$ with 
$N_\pm (\bar \theta_{13},0^\circ)$, thus obtaining a one-parameter sensitivity plot in $\delta$ only.}.
The outcome of this procedure is finally plotted, representing the region in the $(\theta_{13},\delta)$
parameter space for which a CP-violating signal is observed at 3$\sigma$. For
each facility we show one single curve obtained taking
into account the impact of all the degeneracies. 

First of all, notice that no CP-discovery potential has been considered for the
NF at 7000 km: due to matter effects and the choice of the baseline (close to the {\it magic
baseline}, see Ref. \cite{Huber:2003ak}), the sensitivity to
$\delta$ vanishes. The best sensitivity to the measurement of the
CP-violating phase is reached by the T2K-II experiment: 
a CP-violating signal can be 
observed at $3\sigma$ for $|\delta| \in [27^\circ,155^\circ]$ down to
$\sin^2 \theta_{13} \sim 2 \times 10^{-4}$
($\theta_{13} \sim 0.8^\circ$). At the SPL a good sensitivity can be
reached also: CP violation is clearly observed
at $3\sigma$ for $|\delta| \in [45^\circ,135^\circ]$ down to $\sin^2 \theta_{13} \sim 9 \times 10^{-4}$
($\theta_{13} \sim 1.7^\circ$), as it was shown in Ref.~\cite{Donini:2004iv}.
The NF at 3000 km is not as good as one would expect from its $\theta_{13}$-sensitivity contours,
having a CP-discovery potential very similar to the SPL;
the loss in sensitivity with respect to T2K-II is mainly due to the presence of
the sign degeneracy. We have observed that the NF starts to be insensitive to
the leptonic CP-violation just when, for $\bar \theta_{13} \sim 2^\circ$, sign degeneracies close to $\theta_{13}=0$ appear,
which do not allow to exclude the CP-conserving case any longer. This is a
consequence of the fact that, for neutrinos with $< E_\nu > \sim 30$ GeV, the effective $\theta_{13}$ in matter
\cite{Zaglauer:1988gz,Cervera:2000kp} is extremely small. As it has been shown in Ref.~\cite{Donini:2003vz},
when $\theta_{13}$ vanishes, the sign degeneracies flow to $\delta = 0^\circ$ or $\delta = 180^\circ$, as can be seen
in Figs. 2 and 3 (left) of Ref.~\cite{Donini:2003vz}. At the SPL and T2K-II, for which the vacuum parameter
$\theta_{13}$ is the relevant parameter, this happens at a much smaller value $\theta_{13} \sim 0.5^\circ$. 
Therefore, at these experiments the CP-discovery potential is statistics-dominated. \\

\item {\bf Sensitivity to the mass hierarchy} \\ 

In addition to the previous sensitivities one can ask for the sensitivity to the
sign of the atmospheric mass difference. We compute the smallest values of
$\theta_{13}$ for which the sign of $\Delta m^2_{23}$ can be measured 
in the ($\sin^2 \theta_{13}$, $\delta$) plane. The measurement
of $s_{atm}$ requires matter effects to be sizeble. For this
reason neither T2K-II nor the SPL have the capabilities to perform such a
measurement. In Fig.\ref{fig:signo} we present the results for the Neutrino
Factory only and for both posibilities of the true $s_{atm}=\pm1$.
As expected, the NF at 3000 km (solid and dotted lines for normal and inverted
hierarchy, respectively) exhibits the worst sensitivity to sign($\Delta m^2_{23}$). For
$s_{atm}=+1$ the best sensitivity is reached in a quite large region 
around $\delta=90^\circ$ 
for which $[ \sin^2 \theta_{13} ]_{min}=3.7 \times 10^{-4} \,(1.1^\circ)$ whereas for
$\delta \sim -90^\circ$  $[ \sin^2 \theta_{13}]_{min}=4 \times 10^{-3} \,(3.6^\circ)$.
The situation is completely reversed for $s_{atm}=-1$ due to the fact that in
matter a flip in the sign of $\Delta m^2_{23}$ corresponds to a change among 
neutrino and antineutrinos oscillation probabilities.
At least one order of magnitude in sensitivity can be gained at the NF at 7000 km, 
depending on the fact that around L=7000 Km the effect of $\delta$ in the $\nu_e \to
\nu_\mu$ oscillation probability vanishes and that helps to measure the sign of the
atmospheric mass difference. For the normal hierachy, a maximal sensitivity is
achieved for $0<\delta<90^\circ$ at the level of 
$[ \sin^2 \theta_{13} ]_{min}=4.4 \times 10^{-5} \,(0.4^\circ)$ whereas the largest value for
$[ \sin^2 \theta_{13} ]_{min}=1.7 \times 10^{-4} \,(0.8^\circ)$ is around
$\delta=-100^\circ$. For the inverted hierarchy the sensitivity varies in the
range $\sin^2 \theta_{13} \geq [1, 3]\;\times 10^{-4}\;
(0.6^\circ,1^\circ)$.\\

\begin{figure}[t!]
\vspace{-0.5cm}
\begin{center}
\hspace{-1cm} \epsfxsize8.25cm\epsffile{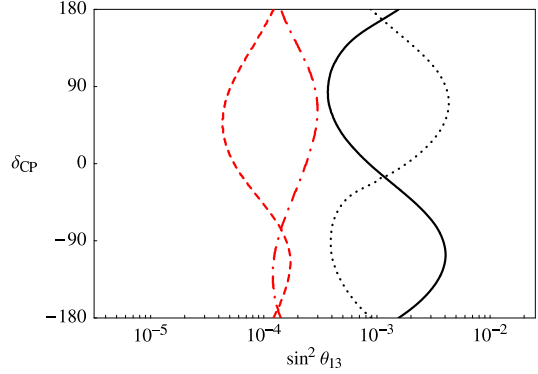} 
\caption{\it 3$\sigma$ sensitivity to the sign($\Delta m^2_{23}$).
Solid lines refer to the $L = 3000$ km Neutrino Factory with normal hierarchy; 
dashed lines to the $L = 7000$ km Neutrino Factory with normal hierarchy; 
dotted lines to the $L = 3000$ km Neutrino Factory with inverted hierarchy; 
dot-dashed lines to the $L = 7000$ km Neutrino Factory with inverted hierarchy.} 
\label{fig:signo}
\end{center}
\end{figure}

\item {\bf Sensitivity to maximal $\theta_{23}$ and the octant-discovery potential} \\

Eventually in Fig.\ref{fig:octant} we present the 
sensitivity to maximal $\theta_{23}$ (plot on the left) and the octant-discovery
potential (plot on the right). The curves have been computed for
$\theta_{13}=0$.

With respect to the sensitivity to maximal $\theta_{23}$, we observe that
the NF at 3000 km (solid line) is not as
good as one may expect since it measures far away from the oscillation maximum. 
At the value of $\Delta m^2_{23}$ in which the sensitivity is maximal, 
deviations as small as 10\% of 
$\sin^2 \theta_{23}$ from maximal mixing could be established.
Similar behaviour is expected at the SPL (dot-dashed line), except for small $\Delta m^2_{23}$ in
which it outperforms the NF at 3000 km. A big
improvement will be however achieved at a NF at 7000 km (dashed lines) and T2K-II (dotted
line); both experiments
have energies and baselines (as well as off-axis angle for T2K-II) chosen to 
match the first oscillation peak (in vacuum and matter respectively);
the reached sensitivities are at the level of
$\sin^2 \theta_{23} \in [0.45-0.48]$ almost independently on the value of
$\Delta m^2_{23}$, which means that deviations from maximal mixing of the order of 4\%
could be established.

Notice that, although this sensitivity is rather good, in general it is very 
difficult to determine the octant in which the atmospheric angle lies. 
As we can see from eq.(\ref{eq:probdismu}), it is quite difficult to break the
$\theta_{23} \to \pi/2-\theta_{23}$ symmetry induced by the leading term in the
transition probability; the subleading terms that could help in lifting this
degeneracy are very difficult to isolate. However, for values of $\theta_{13}$
different from zero, we can take full advantage of matter effects in the
disappearance of muon neutrinos, as we have
already seen in Sect.\ref{sec:nf} 
(Figs.\ref{fig:NF3000bins}-\ref{fig:NF7000bins}). Obviously the 
matter effects at SPL and T2K can never be sizeble enough to solve the octant
ambiguity; on the other hand,
the Neutrino Factory shows a (limited) capability to solve it, irrespective of the
baseline and the value of $\delta$. To illustrate this point, we fixed $\theta_{23}=41.5^\circ$ and 
$\Delta m^2_{23}= 2.5 \times 10^{-3}$ eV$^2$ and, for any value of $\delta$, we compute the minimum value of 
$\theta_{13}$ for which the octant ambiguity is solved.
As we can see in the right plot of
Fig.\ref{fig:octant}, a $3\sigma$ octant discovery is possible
for $\sin^2 \theta_{23} >  0.01 \; (6^\circ)$.

\begin{figure}[t!]
\vspace{-0.5cm}
\begin{center}
\begin{tabular}{cc}
\hspace{-1cm} \epsfxsize8.25cm\epsffile{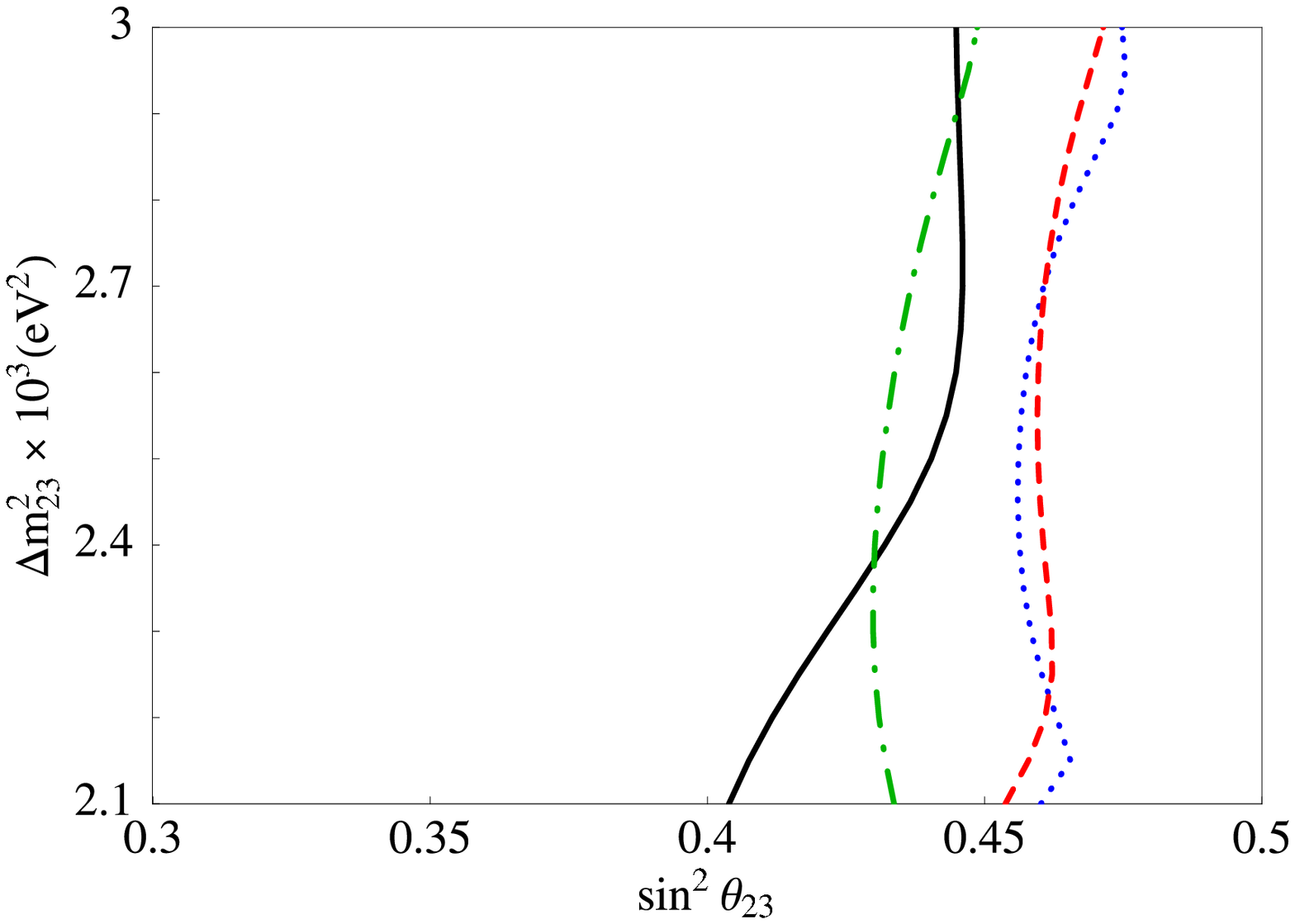} & 
\hspace{-0.5cm} \epsfxsize8.25cm\epsffile{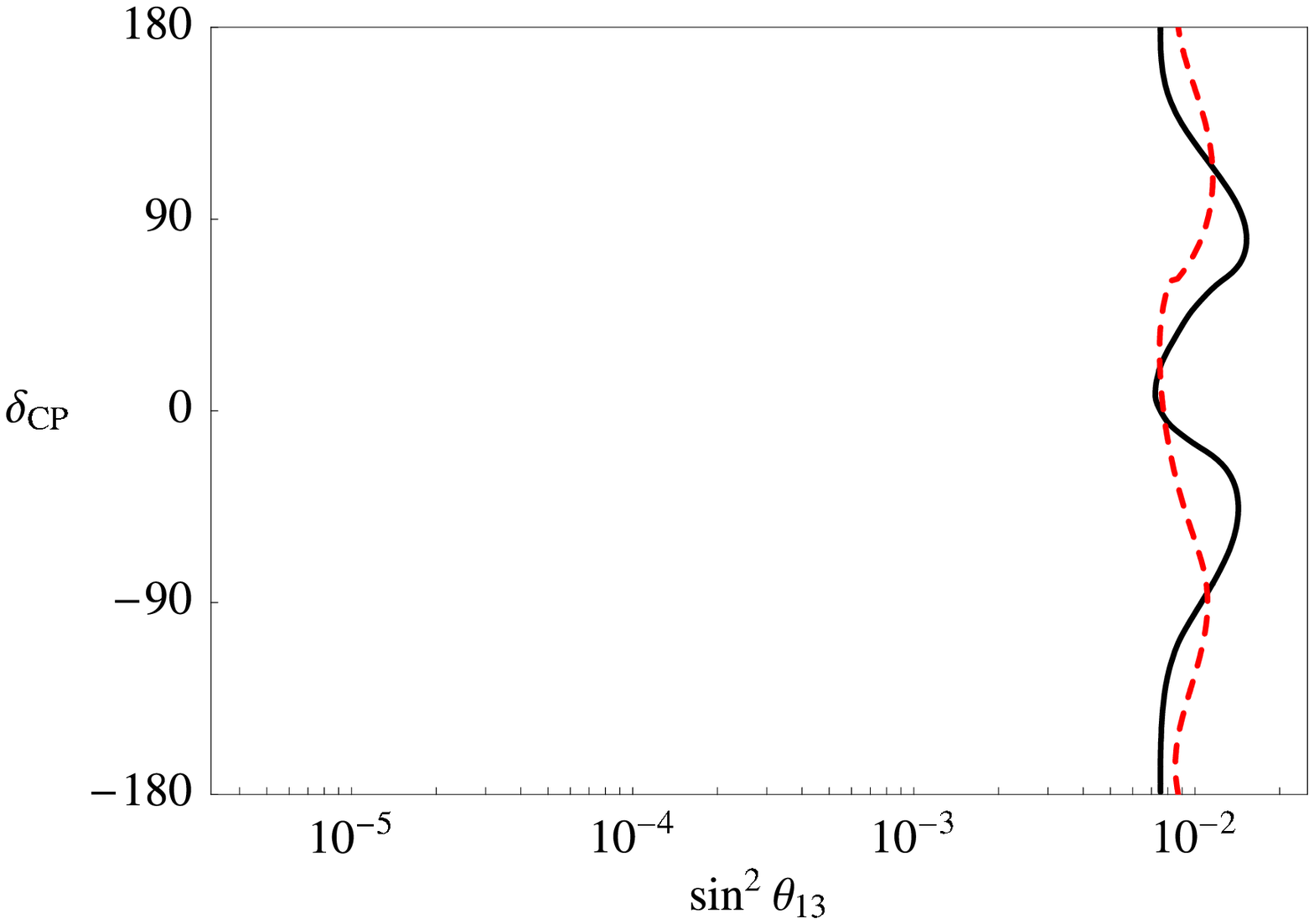} \
\end{tabular}
\caption{\it Left: 3$\sigma$ sensitivity to maximal $\theta_{23}$; right: 3$\sigma$ sensitivity to the $\theta_{23}$-octant.
Solid lines refer to the $L = 3000$ km Neutrino Factory; dashed lines to the $L = 7000$ km Neutrino Factory; 
dotted lines to T2K-II; dot-dashed lines to the (standard) SPL.} 
\label{fig:octant}
\end{center}
\end{figure}

\end{itemize}

%%%%%%%%%%%%%%%%%%%%%%%%%%%%%%%%%%%%%%%%%%%%%%%%%%%%%%%%%%%%%%%%%%%%%%
%
\section{Conclusions}
\label{sec:concl}
%
%%%%%%%%%%%%%%%%%%%%%%%%%%%%%%%%%%%%%%%%%%%%%%%%%%%%%%%%%%%%%%%%%%%%%%
%

In this paper, we studied the measurement of the atmospheric neutrino oscillation parameters, $\theta_{23}$ and $\Delta m^2_{23}$
using the $\nu_\mu$ disappearance channel at three conventional beam facilities, the SPL, T2K-phase I and NO$\nu$A, and at the Neutrino Factory. 
The precision on these two parameters will be of crucial importance in the measurement of two of the unknowns of the PMNS mixing matrix, 
$\theta_{13}$ and the leptonic CP-violating phase $\delta$. 

It has been shown that counting experiments cannot reduce significantly the uncertainties
on $\theta_{23}$ and $\Delta m^2_{23}$. Hence, we have considered detectors with (modest but non vanishing)
sensitivity to the energy of the final leptons produced via neutrino interaction. 
For T2K-I and NO$\nu$A we have indeed found that, independently of the input value of $\theta_{23}$, 
the errors on $\theta_{23}$ can be significantly reduced. Different energy bins give different allowed regions in the 
($\theta_{23}$, $\Delta m^2_{23}$)-plane and the combination of them eventually helps in reducing the uncertainties around 
the physical point. The two experiments only run with $\pi^+$, and thus are not sensitive to the CP-violating phase $\delta$.
The SPL, with a 1 Mton detector and 2 + 8 years of running time with $\pi^+$ and $\pi^-$, respectively, 
could in principle greatly improve these results. However, having neutrinos and antineutrinos at this setup 
does not help, since they have been produced with the same energy, roughly. Moreover, the baseline is too short 
for matter effects to make a difference. As a consequence, information coming from the two fluxes is not complementary
and just add statistically. Remember, however, that this is not the case in the appearance mode, $\nu_\mu \to \nu_e$, 
for which this setup was designed. In this paper, we have considered two bins of energy at the SPL: 
this already makes a big difference in the precision on $\theta_{23}$ as it can be seen comparing   
with our results of Refs.~\cite{Donini:2004hu,Donini:2004iv} and \cite{Donini:2005rn}, where only a counting experiment was considered
for this facility. 

We have then studied the impact of $\theta_{13}$ and $\delta$ in the measurement of $\theta_{23}$ and 
$\Delta m^2_{23}$. Since both parameters are unknown at present, their values should be reconstructed at
the same time with the atmospheric parameters. 
We have shown that the main effect of considering $\theta_{13}$ as a free variable is a shift of the
contours toward larger values of $\theta_{23}$ (with respect to the input point) while a free $\delta$ 
in the whole $\delta \in [-180^\circ, 180^\circ]$ range does not produce any 
significant distortion of the allowed regions. These effects can be strongly 
softened if, in addition to the disappearance channels, we introduce in the analysis the
appearance $\nu_\mu \to \nu_e$ channels, which the Super-Beam facilities
have been proposed to look for. Two important conclusions can then be drawn: if
$\theta_{13}$ is kept fixed, the combination of appearance and disappearance
channels solves the {\it disappearance octant degeneracy} for $\theta_{13}$
large enough; on the other hand, the octant degeneracy is not lifted if 
$\theta_{13}$ is free to vary in the current allowed range, 
for almost any value of $\theta_{13}$.

The situation is quite different at the Neutrino Factory. 
We have pointed out that the $\nu_\mu \to \nu_\mu$ disappearance channel is a very powerful tool to reduce the
uncertainties on the atmospheric parameters up to an unprecedented level and, in combination with appearance channels, 
to solve many of the discrete ambiguities affecting the measurement of the PMNS matrix elements.
The main feature emerging from such an analysis is that, for $\theta_{13} \geq 3^\circ - 4^\circ$, 
the sinergy between disappearance and appearance channels (the more renowned {\it golden} and {\it silver} channels) 
greatly helps in solving the octant ambiguity, at both $L = 3000$ km and $L = 7000$ km baselines, as a result of the strong matter
effects along the neutrino path and the huge statistics at hand. 
This remains true even in the case we leave $\theta_{13}$ as a free parameter, a situation which has been shown not to be true 
for the other facilities. Equally remarkable, in the same range of $\theta_{13}$, the sign {\it clones} are solved and 
they are not present in the fits, independently of the baseline and the input values of $\theta_{23}$ and $\Delta m^2_{23}$. 
This allows a precise measurement of $\theta_{13}$ at both facilities while the CP-phase $\delta$, as it is already well known, 
can only be measured at $L = 3000$ Km. For values of $\theta_{13}\le 3^\circ$, some remnant of the clones still remains and the
measurement of the two unknowns cannot be performed with huge precision.

Eventually, we have presented a comparison at the different facilities of the 3$\sigma$ sensitivity to $\theta_{13}$,
to maximal $\theta_{23}$, to the mass hierarchy, to the $\theta_{23}$-octant and their CP-discovery potential. 
T2K-I and NO$\nu$A, running with neutrinos only are not able to measure $\delta$. 
On the other hand, they have very similar performances in the $\theta_{13}$-sensitivity, with an excluded
region ranging from $[ \sin^2 \theta_{13} ]_{min} = 2 \times 10^{-3}$ to $8 \times 10^{-3}$, 
and in the sensitivity to maximal $\theta_{23}$, with the maximal $\theta_{23}$ that can be distinguished from 
$\theta_{23} = 45^\circ$ ranging from $[\sin^2 \theta_{23}]_{max} = 0.45$ to $0.40$ for $\Delta m^2_{23} \in [2.0,3.0] \times 10^{-3}$ eV$^2$.

The Neutrino Factory outperforms the considered Super-Beams for both baselines in the $\theta_{13}$-sensitivity and in the sensitivity
to the sign of the atmospheric mass difference. The longest baseline is as effective as T2K-II in the sensitivity to maximal $\theta_{23}$.
The excluded region in $\theta_{13}$ at the 3000 km Neutrino Factory ranges from $[ \sin^2 \theta_{13} ]_{min} = 2 \times 10^{-5}$ to
$2 \times 10^{-4}$ (for particular values of $\delta$). The sign of the atmospheric mass difference can be measured at the 7000 Km (3000 Km) Neutrino
Factory for $\sin^2 \theta_{13}$ as small as $10^{-4}$ ($10^{-3}$). Finally, the $\theta_{23}$-octant can be identified at both facilities
for $\sin^2 \theta_{13}$ as small as $10^{-2}$ ($\theta_{13} \geq 6^\circ$).
Notably enough, the CP-discovery potential seems to be larger for T2K-II than for Neutrino
Factory at $L = 3000$ Km (no CP-discovery potential is expected at the {\it magic baseline} $L = 7000$ Km). 
This is explained as follows: as a general rule, for small values of $\theta_{13}$ the degeneracies flow toward $\delta = 0^\circ$ 
and $|\delta| = 180^\circ$
(see Ref.~\cite{Donini:2003vz} and Ref.~\cite{Burguet-Castell:2002qx}), thus mimicking a non-CP violating phase.
Due to a {\it parametric conspiracy} between the chosen energy and baseline and the matter effects, at the Neutrino Factory
the typical value of $\theta_{13}$ for which this happens is much larger than at the SPL and T2K. 
Therefore, although from the statistical point of view the Neutrino Factory would certainly outperform both the SPL and T2K-II, 
in practice for small values of $\theta_{13}$ a CP-violating phase will be difficult to distinguish from a non--CP-violating one, 
if $s_{atm}$ and $s_{oct}$ are not measured previously.

%
%%%%%%%%%%%%%%%%%%%%%%%%%%%%%%%%%%%%%%%%%%%%%%%%%%%%%%%%%%%%%%%%%%%%%%%%%%
\section*{Acknowledgments}
%%%%%%%%%%%%%%%%%%%%%%%%%%%%%%%%%%%%%%%%%%%%%%%%%%%%%%%%%%%%%%%%%%%%%%%%%%
%
We would like to thank E.~Couce, B.~Gavela, J.~Gomez-Cadenas, P.~Hernandez, P.~Huber, O.~Mena,
P.~Migliozzi, T. Schwetz and W.~Winter for useful discussions. The authors acknowledge the financial 
support of MCYT through project FPA2003-04597 and of the European Union
through the networking activity BENE. E.F. acknowledges financial support from the UAM.
%
%%%%%%%%%%%%%%%%%%%%%%%%%%%%%%%%%%%%%%%%%%%%%%%%%%%%%%%%%%%%%%%%%%%%%%%%%%

\section*{Appendix A}
\label{app:events}

In this Appendix we recall the number of events per bin that are expected in the $\nu_\mu (\bar \nu_\mu)$ disappearance 
channel at the SPL, the T2K-I, the NO$\nu$A and the Neutrino Factory experiments.
The expected T2K-I muon and electron identification efficiencies are presented, as well.

In all tables, two different values of $\theta_{13}$ and $\delta$ are considered, as well as two choices 
of ${\rm sign} (\Delta m^2_{23})$.

In Tab.~\ref{tab:SPLevents} we show the expected event rates per bin at the SPL
after a 2 years run with $\pi^+$ and a 8 years run with $\pi^-$, using the fluxes of Ref.~\cite{gilardoni}
and a 440 kt detector at $L = 130$ km. 

\begin{table}[hbtp]
\begin{center}
\begin{tabular}{|c|c|c|c|c|} \hline
$E_\nu \in [0,250]$ MeV & No Osc. & $\theta_{13} = 0^\circ $ & $\theta_{13} = 8^\circ; \delta = 0^\circ$ &
                         $\theta_{13} = 8^\circ; \delta = 90^\circ$ \\ \hline \hline
$ N_{\mu^-}, s_{atm} = +$        & 2784 & 186 & 198 & 192 \\
\hline 
$ N_{\mu^-}, s_{atm} = -$ &       & 158 & 158 & 162 \\
\hline \hline
$E_\nu \in [250,600]$ MeV & No Osc. & $\theta_{13} = 0^\circ $ & $\theta_{13} = 8^\circ; \delta = 0^\circ$ &
                         $\theta_{13} = 8^\circ; \delta = 90^\circ$ \\ \hline \hline
$ N_{\mu^-}, s_{atm} = +$        & 21461 & 2401 & 2423 & 2455 \\
\hline 
$ N_{\mu^-}, s_{atm} = -$ &       & 2591 & 2682 & 2642 \\
\hline \hline
$E_{\bar \nu} \in [0,250]$ MeV & No Osc. & $\theta_{13} = 0^\circ $ & $\theta_{13} = 8^\circ; \delta = 0^\circ$ &
                         $\theta_{13} = 8^\circ; \delta = 90^\circ$ \\ \hline \hline
$ N_{\mu^+}, s_{atm} = +$        & 6310 & 830 & 859 & 841 \\
\hline 
$ N_{\mu^+}, s_{atm} = -$ &       & 740 & 737 & 753 \\
\hline \hline
$E_{\bar \nu} \in [250,600]$ MeV & No Osc. & $\theta_{13} = 0^\circ $ & $\theta_{13} = 8^\circ; \delta = 0^\circ$ &
                         $\theta_{13} = 8^\circ; \delta = 90^\circ$ \\ \hline \hline
$ N_{\mu^+}, s_{atm} = +$        & 19157 & 1708 & 1734 & 1757 \\
\hline 
$ N_{\mu^+}, s_{atm} = -$ &       & 1858 & 1939 & 1906 \\
\hline
\end{tabular}
\end{center}
\caption{\it Disappearance event rates for a $2+8$ years $\pi^+ + \pi^-$ run at the
standard SPL, for different values of $\theta_{13},\delta$ and of the sign of the atmospheric mass difference, $s_{atm}$.} 
\label{tab:SPLevents}
\end{table}

In Tab.~\ref{tab:T2Kevents} we show the expected event rates for the two lowest bins at T2K-I after a 5 years run with $\pi^+$, 
for a 22.5 kt detector at $L = 295$ km.

\begin{table}[hbtp]
\begin{center}
\begin{tabular}{|c|c|c|c|c|} \hline
$E_\nu \in [400,600]$ MeV & No Osc. & $\theta_{13} = 0^\circ $ & $\theta_{13} = 8^\circ; \delta = 0^\circ$ &
                         $\theta_{13} = 8^\circ; \delta = 90^\circ$ \\ \hline \hline
$ N_{\mu^-}, s_{atm} = +$        & 752 & 52 & 56 & 54 \\
\hline 
$ N_{\mu^-}, s_{atm} = -$ &       & 45 & 46 & 47 \\
\hline \hline
$E_\nu \in [600,800]$ MeV & No Osc. & $\theta_{13} = 0^\circ $ & $\theta_{13} = 8^\circ; \delta = 0^\circ$ &
                         $\theta_{13} = 8^\circ; \delta = 90^\circ$ \\ \hline \hline
$ N_{\mu^-}, s_{atm} = +$        & 2218 & 122 & 126 & 128 \\
\hline 
$ N_{\mu^-}, s_{atm} = -$ &       & 140 & 149 & 145 \\
\hline
\end{tabular}
\end{center}
\caption{\it Disappearance event rates for a $5$ years $\pi^+$ run at T2K-I, for different values of
$\theta_{13},\delta$ and of the sign of the atmospheric mass difference, $s_{atm}$.} 
\label{tab:T2Kevents}
\end{table}

In Tab.~\ref{tab:NOVAevents} we show the expected event rates per bin at NO$\nu$A after a 5 years run with $\pi^+$, 
for a 30 kt detector at $L=812$ km. 

\begin{table}[hbtp]
\begin{center}
\begin{tabular}{|c|c|c|c|c|} \hline
$E_\nu \in [1000,1666]$ MeV & No Osc. & $\theta_{13} = 0^\circ $ & $\theta_{13} = 8^\circ; \delta = 0^\circ$ &
                         $\theta_{13} = 8^\circ; \delta = 90^\circ$ \\ \hline \hline
$ N_{\mu^-}, s_{atm} = +$        & 1217 & 120 & 125 & 116 \\
\hline 
$ N_{\mu^-}, s_{atm} = -$        &      & 105 & 102 & 107 \\
\hline \hline
$E_\nu \in [1666,2333]$ MeV & No Osc. & $\theta_{13} = 0^\circ $ & $\theta_{13} = 8^\circ; \delta = 0^\circ$ &
                         $\theta_{13} = 8^\circ; \delta = 90^\circ$ \\ \hline \hline
$ N_{\mu^-}, s_{atm} = +$        & 8635 & 819 & 800 & 847 \\
\hline 
$ N_{\mu^-}, s_{atm} = -$        &      & 920 & 949 & 907 \\
\hline \hline
$E_\nu \in [2333,3000]$ MeV & No Osc. & $\theta_{13} = 0^\circ $ & $\theta_{13} = 8^\circ; \delta = 0^\circ$ &
                         $\theta_{13} = 8^\circ; \delta = 90^\circ$ \\ \hline \hline
$ N_{\mu^-}, s_{atm} = +$        & 7545 & 1909 & 1886 & 1938 \\
\hline 
$ N_{\mu^-}, s_{atm} = -$        &      & 2021 & 2048 & 2003 \\
\hline
\end{tabular}
\end{center}
\caption{\it Disappearance event rates for a $5$ years $\pi^+$ run at NO$\nu$A, for different values of
$\theta_{13},\delta$ and of the sign of the atmospheric mass difference, $s_{atm}$.} 
\label{tab:NOVAevents}
\end{table}

In Tabs.~\ref{tab:NF3Levents} and \ref{tab:NF7Levents} we show the expected event rates per bin 
after a 5 years run with $\mu^-$ and a 5 years run with $\mu^+$ at the Neutrino Factory, for a 40 kt detector 
at $L = 3000$ and $L = 7000$ km, respectively. 

\begin{table}[hbtp]
\begin{center}
\begin{tabular}{|c|c|c|c|c|} \hline
$E_\nu \in [4,8]$ GeV & No Osc. & $\theta_{13} = 0^\circ $ & $\theta_{13} = 8^\circ; \delta = 0^\circ$ &
                                               $\theta_{13} = 8^\circ; \delta = 90^\circ$ \\ \hline \hline
$N_{\mu^-}$, $ s_{atm} = +$        & 6546 & 407 & 449 & 441 \\
\hline 
$N_{\mu^-}$, $ s_{atm} = -$ &       & 418 & 430 & 426 \\
\hline \hline
$E_\nu \in [8,12]$ GeV & No Osc. & $\theta_{13} = 0^\circ $ & $\theta_{13} = 8^\circ; \delta = 0^\circ$ &
                                                $\theta_{13} = 8^\circ; \delta = 90^\circ$ \\ \hline \hline
$N_{\mu^+}$, $ s_{atm} = +$        & 26110 & 6608 & 6727 & 6776 \\
\hline 
$N_{\mu^+}$, $ s_{atm} = -$ &       & 6914 & 7034 & 7004 \\
\hline \hline
$E_\nu \in [4,8]$ GeV & No Osc. & $\theta_{13} = 0^\circ $ & $\theta_{13} = 8^\circ; \delta = 0^\circ$ &
                                               $\theta_{13} = 8^\circ; \delta = 90^\circ$ \\ \hline \hline
$N_{\mu^-}$, $ s_{atm} = +$        & 3014 & 187 & 186 & 188 \\
\hline 
$N_{\mu^-}$, $ s_{atm} = -$ &       & 192 & 210 & 210 \\
\hline \hline
$E_\nu \in [8,12]$ GeV & No Osc. & $\theta_{13} = 0^\circ $ & $\theta_{13} = 8^\circ; \delta = 0^\circ$ &
                                                $\theta_{13} = 8^\circ; \delta = 90^\circ$ \\ \hline \hline
$N_{\mu^+}$, $ s_{atm} = +$        & 12201 & 3092 & 3123 & 3136 \\
\hline 
$N_{\mu^+}$, $ s_{atm} = -$ &       & 3232 & 3328 & 3305 \\
\hline
\end{tabular}
\end{center}
\caption{\it Disappearance event rates for a $5+5$ years run at the $L = 3000$ km Neutrino Factory, for
different values of $\theta_{13},\delta$ and of the sign of the atmospheric mass difference, $s_{atm}$.} 
\label{tab:NF3Levents}
\end{table}

\begin{table}[hbtp]
\begin{center}
\begin{tabular}{|c|c|c|c|c|} \hline
$E_\nu \in [10,15]$ GeV & No Osc. & $\theta_{13} = 0^\circ $ & $\theta_{13} = 8^\circ; \delta = 0^\circ$ &
                         $\theta_{13} = 8^\circ; \delta = 90^\circ$ \\ \hline \hline
$N_{\mu^-}$, $ s_{atm} = +$        & 11129 & 639 & 438 & 413 \\
\hline 
$N_{\mu^-}$, $ s_{atm} = -$ &       & 551 & 515 & 519 \\
\hline \hline
$E_\nu \in [15,20]$ GeV & No Osc. & $\theta_{13} = 0^\circ $ & $\theta_{13} = 8^\circ; \delta = 0^\circ$ &
                         $\theta_{13} = 8^\circ; \delta = 90^\circ$ \\ \hline \hline
$N_{\mu^+}$, $ s_{atm} = +$        & 27181 & 2114 & 2582 & 2556 \\
\hline 
$N_{\mu^+}$, $ s_{atm} = -$ &       & 2374 & 2565 & 2578 \\
\hline \hline
$E_\nu \in [10,15]$ GeV & No Osc. & $\theta_{13} = 0^\circ $ & $\theta_{13} = 8^\circ; \delta = 0^\circ$ &
                         $\theta_{13} = 8^\circ; \delta = 90^\circ$ \\ \hline \hline
$N_{\mu^-}$, $ s_{atm} = +$        & 5225 & 299 & 269 & 268 \\
\hline 
$N_{\mu^-}$, $ s_{atm} = -$ &       & 259 & 206 & 216 \\
\hline \hline
$E_\nu \in [15,20]$ GeV & No Osc. & $\theta_{13} = 0^\circ $ & $\theta_{13} = 8^\circ; \delta = 0^\circ$ &
                         $\theta_{13} = 8^\circ; \delta = 90^\circ$ \\ \hline \hline
$N_{\mu^+}$, $ s_{atm} = +$        & 12799 & 997 & 1083 & 1077 \\
\hline 
$N_{\mu^+}$, $ s_{atm} = -$ &       & 1116 & 1334 & 1345 \\
\hline
\end{tabular}
\end{center}
\caption{\it Disappearance event rates for a $5+5$ years run at the $L = 7000$ km Neutrino Factory, for
different values of $\theta_{13},\delta$ and of the sign of the atmospheric mass difference, $s_{atm}$.} 
\label{tab:NF7Levents}
\end{table}

\begin{figure}[t!]
\vspace{-0.5cm}
\begin{center}
%\begin{tabular}{cc}
\hspace{-0.3cm} \epsfxsize7.5cm\epsffile{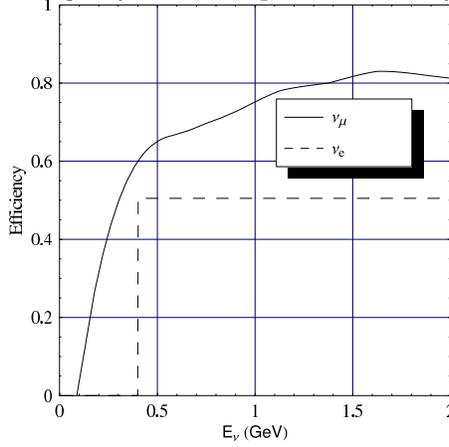}
%\hspace{-0.3cm} \epsfxsize7.5cm\epsffile{figure/NOVAfluxes.eps} \\
%\end{tabular}
\caption{\it Muon (solid) and electron (dashed) identification efficiency at T2K-I as a function of the neutrino energy.}
\label{fig:effT2K}
\end{center}
\end{figure}

Eventually, in Fig.~\ref{fig:effT2K} we present the muon (solid) and electron (dashed) identification efficiencies 
at the T2K-I experiment, as found in Ref.~\cite{Kaneyuki:2005ze} and \cite{Itow:2001ee}, respectively.

%%%%%%%%%%%%%%%%%%%%%%%%%%%%%%%%%%%%%%%%%%%%%%%%%%%%%%%%%%%%%%%%%%%%%%%%%%
%BIBLIOGRAPHY
%%%%%%%%%%%%%%%%%%%%%%%%%%%%%%%%%%%%%%%%%%%%%%%%%%%%%%%%%%%%%%%%%%%%%%%%%%

%
%%%%%%%%%%%%%%%%%%%%%%%%%%%%%%%%%%%%%%%%%%%%%%%%%%%%%%%%%%%%%%%%%%%%%%
%

\begin{thebibliography}{999}

%\cite{Fukuda:1998mi}
\bibitem{exp}
Y.~Fukuda {\it et al.}  [Super-Kamiokande Collaboration],
%``Evidence for oscillation of atmospheric neutrinos,''
Phys.\ Rev.\ Lett.\  {\bf 81} (1998) 1562 [arXiv:hep-ex/9807003];
%%CITATION = HEP-EX 9807003;%%
%\cite{Ambrosio:2001je}
M.~Ambrosio {\it et al.}  [MACRO Collaboration],
%``Matter effects in upward-going muons and sterile neutrino oscillations,''
Phys.\ Lett.\ B {\bf 517} (2001) 59 [arXiv:hep-ex/0106049]; 
%%CITATION = HEP-EX 0106049;%%
M.~H.~Ahn {\it et al.}  [K2K Collaboration],
%``Indications of neutrino oscillation in a 250-km long-baseline experiment,''
Phys.\ Rev.\ Lett.\  {\bf 90} (2003) 041801 [arXiv:hep-ex/0212007];
%%CITATION = HEP-EX 0212007;%%
B.~T.~Cleveland {\it et al.},
%``Measurement Of The Solar Electron Neutrino Flux With The Homestake  Chlorine
%Detector,''
Astrophys.\ J.\  {\bf 496} (1998) 505;
%%CITATION = ASJOA,496,505;%%
%\cite{Abdurashitov:1999zd}
J.~N.~Abdurashitov {\it et al.}  [SAGE Collaboration],
%``Measurement of the solar neutrino capture rate with gallium metal,''
Phys.\ Rev.\ C {\bf 60} (1999) 055801 [arXiv:astro-ph/9907113];
%%CITATION = ASTRO-PH 9907113;%%
%\cite{Hampel:1998xg}
W.~Hampel {\it et al.}  [GALLEX Collaboration],
%``GALLEX solar neutrino observations: Results for GALLEX IV,''
Phys.\ Lett.\ B {\bf 447} (1999) 127;
%%CITATION = PHLTA,B447,127;%%
%\cite{Fukuda:2001nj}
S.~Fukuda {\it et al.}  [Super-Kamiokande Collaboration],
%``Solar B-8 and he p neutrino measurements from 1258 days of  Super-Kamiokande
%data,''
Phys.\ Rev.\ Lett.\  {\bf 86} (2001) 5651 [arXiv:hep-ex/0103032];
%%CITATION = HEP-EX 0103032;%%
%\cite{Ahmad:2001an}
Q.~R.~Ahmad {\it et al.}  [SNO Collaboration],
%``Measurement of the charged current interactions produced by B-8  solar
%neutrinos at the Sudbury Neutrino Observatory,''
Phys.\ Rev.\ Lett.\  {\bf 87} (2001) 071301 [arXiv:nucl-ex/0106015]; 
%%CITATION = NUCL-EX 0106015;%%
K.~Eguchi {\it et al.}  [KamLAND Collaboration],
%``First results from KamLAND: Evidence for reactor anti-neutrino
%disappearance,''
Phys.\ Rev.\ Lett.\  {\bf 90} (2003) 021802 [arXiv:hep-ex/0212021].
%%CITATION = HEP-EX 0212021;%%

%\cite{Athanassopoulos:1997pv}
\bibitem{lsnd}
C.~Athanassopoulos {\it et al.}  [LSND Collaboration],
%``Evidence for nu/mu $\to$ nu/e neutrino oscillations from LSND,''
Phys.\ Rev.\ Lett.\  {\bf 81} (1998) 1774 [arXiv:nucl-ex/9709006];
%%CITATION = NUCL-EX 9709006;%%
%\cite{Aguilar:2001ty}
A.~Aguilar {\it et al.}  [LSND Collaboration],
%``Evidence for neutrino oscillations from the observation of anti-nu/e
%appearance in a anti-nu/mu beam,''
Phys.\ Rev.\ D {\bf 64} (2001) 112007
[arXiv:hep-ex/0104049].
%%CITATION = HEP-EX 0104049;%%

%\cite{Stancu:14, dr}
\bibitem{boone}
I.~Stancu {\it et al.}  [MiniBooNE collaboration],
%``The Miniboone Detector Technical Design Report,''
FERMILAB-TM-2207.
%\href{http://www.slac.stanford.edu/spires/find/hep/www?r=fermilab-tm-2207}{SPIRES entry}
%\cite{Bahcall:2004ut}

%\cite{Pontecorvo:cp}
\bibitem{neutrino_osc}
B.~Pontecorvo,
%``Mesonium And Antimesonium,''
Sov.\ Phys.\ JETP {\bf 6} (1957) 429 [Zh.\ Eksp.\ Teor.\ Fiz.\  {\bf 33} (1957) 549];
%%CITATION = SPHJA,6,429;%%
%\cite{Maki:mu}
Z.~Maki, M.~Nakagawa and S.~Sakata,
%``Remarks On The Unified Model Of Elementary Particles,''
Prog.\ Theor.\ Phys.\  {\bf 28} (1962) 870;
%%CITATION = PTPKA,28,870;%%
%\cite{Pontecorvo:fh}
B.~Pontecorvo,
%``Neutrino Experiments And The Question Of Leptonic-Charge  Conservation,''
Sov.\ Phys.\ JETP {\bf 26} (1968) 984 [Zh.\ Eksp.\ Teor.\ Fiz.\  {\bf 53} (1967) 1717];
%%CITATION = SPHJA,26,984;%%
%\cite{Gribov:1968kq}
V.~N.~Gribov and B.~Pontecorvo,
%``Neutrino Astronomy And Lepton Charge,''
Phys.\ Lett.\ B {\bf 28} (1969) 493.
%%CITATION = PHLTA,B28,493;%%

%\cite{Apollonio:1999ae}
\bibitem{chooz}
M.~Apollonio {\it et al.}  [CHOOZ Collaboration],
%``Limits on neutrino oscillations from the CHOOZ experiment,''
Phys.\ Lett.\ B {\bf 466} (1999) 415 [arXiv:hep-ex/9907037];
%%CITATION = HEP-EX 9907037;%%
%``Search for neutrino oscillations on a long base-line at the CHOOZ  nuclear power station,''
Eur.\ Phys.\ J.\ C {\bf 27} (2003) 331 [arXiv:hep-ex/0301017].
%%CITATION = HEP-EX 0301017;%%

%\cite{Fogli:2005cq}
\bibitem{Fogli:2005cq}
  G.~L.~Fogli, E.~Lisi, A.~Marrone and A.~Palazzo,
  %``Global analysis of three-flavor neutrino masses and mixings,''
  arXiv:hep-ph/0506083;
  %%CITATION = HEP-PH 0506083;%%
  G.~L.~Fogli, E.~Lisi, A.~Marrone, A.~Palazzo and A.~M.~Rotunno,
  %``Neutrino mass and mixing parameters: A short review,''
  arXiv:hep-ph/0506307.
  %%CITATION = HEP-PH 0506307;%%

\bibitem{OPERA} 
H.~Pessard  [OPERA Collaboration], arXiv:hep-ex/0504033;
%%CITATION = HEP-EX 0504033;%%
 M.~Guler {\it et al.}  [OPERA Collaboration], 
``OPERA: An appearance experiment to search for $\nu_\mu \rightarrow \nu_\tau$ oscillations in the CNGS beam. Experimental proposal,''
CERN-SPSC-2000-028.

\bibitem{ICARUS}
  F.~Arneodo {\it et al.},
  %``ICARUS: An innovative detector for underground physics,''
  Nucl.\ Instrum.\ Meth.\ A {\bf 461} (2001) 324.
  %%CITATION = NUIMA,A461,324;%%
  P.~Aprili {\it et al.}  [ICARUS Collaboration],
  %``The ICARUS experiment: A second-generation proton decay experiment and
  %neutrino observatory at the Gran Sasso laboratory. Cloning of T600 modules to
  %reach the design sensitive mass. (Addendum),''
  CERN-SPSC-2002-027.

\bibitem{SNO}
S.~N.~Ahmed {\it et al.}  [SNO Collaboration],
%``Measurement of the total active B-8 solar neutrino flux at the Sudbury
%Neutrino Observatory with enhanced neutral current sensitivity,''
Phys.\ Rev.\ Lett.\  {\bf 92} (2004) 181301
[arXiv:nucl-ex/0309004].

%\cite{Cervera:2000kp}
\bibitem{Cervera:2000kp}
A.~Cervera, A.~Donini, M.~B.~Gavela, J.~J.~Gomez Cadenas, P.~Hernandez, O.~Mena and S.~Rigolin,
%``Golden measurements at a neutrino factory,''
Nucl.\ Phys.\ B {\bf 579} (2000) 17 [Erratum-ibid.\ B {\bf 593} (2001) 731] [arXiv:hep-ph/0002108].
%%CITATION = HEP-PH 0002108;%%

\bibitem{Burguet-Castell:2001ez}
J.~Burguet-Castell, M.~B.~Gavela, J.~J.~Gomez-Cadenas, P.~Hernandez and O.~Mena,
%``On the measurement of leptonic CP violation,''
Nucl.\ Phys.\ B {\bf 608} (2001) 301 [arXiv:hep-ph/0103258].
%%CITATION = HEP-PH 0103258;%%

%\cite{Minakata:2001qm}
\bibitem{Minakata:2001qm}
H.~Minakata and H.~Nunokawa,
%``Exploring neutrino mixing with low energy superbeams,''
JHEP {\bf 0110} (2001) 001 [arXiv:hep-ph/0108085].
%%CITATION = HEP-PH 0108085;%%

%\cite{Fogli:1996pv}
\bibitem{Fogli:1996pv}
G.~L.~Fogli and E.~Lisi,
%``Tests of three-flavor mixing in long-baseline neutrino oscillation experiments,''
Phys.\ Rev.\ D {\bf 54} (1996) 3667 [arXiv:hep-ph/9604415].
%%CITATION = HEP-PH 9604415;%%

%\cite{Barger:2001yr}
\bibitem{Barger:2001yr}
V.~Barger, D.~Marfatia and K.~Whisnant,
%``Breaking eight-fold degeneracies in neutrino CP violation, mixing, and  mass hierarchy,''
Phys.\ Rev.\ D {\bf 65} (2002) 073023 [arXiv:hep-ph/0112119].
%%CITATION = HEP-PH 0112119;%%

%\cite{Donini:2005rn}
\bibitem{Donini:2005rn}
  A.~Donini, D.~Meloni and S.~Rigolin,
  %``The impact of solar and atmospheric parameter uncertainties on the
  %measurement of theta(13) and delta,''
  arXiv:hep-ph/0506100;
  %%CITATION = HEP-PH 0506100;%%
  D.~Meloni,
  %``Neutrinos as a probe of physics beyond the standard model,''
  Acta Phys.\ Polon.\ B {\bf 35} (2004) 2781;
  %%CITATION = APPOA,B35,2781;%%
  %``The measurement of theta(13) and delta: The role of the uncertainties on
  %the solar and atmospheric parameters,''
  arXiv:hep-ph/0509370.
  %%CITATION = HEP-PH 0509370;%%

%\cite{Itow:2001ee}
\bibitem{Itow:2001ee}
Y.~Itow {\it et al.},
%``The JHF-Kamioka neutrino project,''
arXiv:hep-ex/0106019.
%%CITATION = HEP-EX 0106019;%%

%\cite{Zucchelli:sa}
\bibitem{Zucchelli:sa}
P.~Zucchelli,
%``A Novel Concept For A Anti-Nu/E / Nu/E Neutrino Factory: The Beta Beam,''
Phys.\ Lett.\ B {\bf 532} (2002) 166.
%%CITATION = PHLTA,B532,166;%%

%\cite{Geer:1997iz}
\bibitem{Geer:1997iz}
S.~Geer,
%``Neutrino beams from muon storage rings: Characteristics and physics potential,''
Phys.\ Rev.\ D {\bf 57} (1998) 6989 [Erratum-ibid.\ D {\bf 59} (1999) 039903] [arXiv:hep-ph/9712290];
%%CITATION = HEP-PH 9712290;%%
A.~De Rujula, M.~B.~Gavela and P.~Hernandez,
%``Neutrino oscillation physics with a neutrino factory,''
Nucl.\ Phys.\ B {\bf 547} (1999) 21 [arXiv:hep-ph/9811390].
%%CITATION = HEP-PH 9811390;%%

%\cite{Apollonio:2002en}
\bibitem{Apollonio:2002en}
C.~Albright {\it et al.},
%``Physics at a neutrino factory,''
arXiv:hep-ex/0008064;
%%CITATION = HEP-EX 0008064;%%
M.~Apollonio {\it et al.},
%``Oscillation physics with a neutrino factory. ((G)) ((U)),''
arXiv:hep-ph/0210192.
%%CITATION = HEP-PH 0210192;%%

%\cite{Donini:2004iv}
\bibitem{Donini:2004iv}
  A.~Donini, E.~Fern\'andez-Mart\'{\i}nez and S.~Rigolin,
  %``Appearance and disappearance signals at a beta-beam and a super-beam facility,''
  Phys.\ Lett.\ B {\bf 621}, 276 (2005) [arXiv:hep-ph/0411402];
  %%CITATION = HEP-PH 0411402;%%
  S.~Rigolin,
  %``Degeneracies at a beta-beam and a super-beam facility,''
  arXiv:hep-ph/0411403.
  %%CITATION = HEP-PH 0411403;%%

%\cite{Huber:2005ep}
\bibitem{Huber:2005ep}
  P.~Huber, M.~Maltoni and T.~Schwetz,
  %``Resolving parameter degeneracies in long-baseline experiments by atmospheric neutrino data,''
  Phys.\ Rev.\ D {\bf 71} (2005) 053006 [arXiv:hep-ph/0501037].
  %%CITATION = HEP-PH 0501037;%%

%\cite{Gandhi:2005wa}
\bibitem{Gandhi:2005wa}
  R.~Gandhi, P.~Ghoshal, S.~Goswami, P.~Mehta and S.~Uma Sankar,
  %``Probing the nu mass hierarchy via atmospheric nu/mu + anti-nu/mu survival
  %rates in Megaton water Cerenkov detectors,''
  arXiv:hep-ph/0506145.
  %%CITATION = HEP-PH 0506145;%%

%\cite{Gandhi:2004bj}
\bibitem{Gandhi:2004bj}
  R.~Gandhi, P.~Ghoshal, S.~Goswami, P.~Mehta and S.~Uma Sankar,
  %``Earth matter effects at very long baselines and the neutrino mass
  %hierarchy,''
  arXiv:hep-ph/0411252.
  %%CITATION = HEP-PH 0411252;%%

%\cite{CHOUBEY}
\bibitem{Choubey:2005zy}
S.~Choubey and P.~Roy,
%``Probing the deviation from maximal mixing of atmospheric neutrinos,''
arXiv:hep-ph/0509197.
%%CITATION = HEP-PH 0509197;%%

%\cite{Donini:2005gy}
\bibitem{Donini:2005gy}
  A.~Donini, E.~Fern\'andez-Mart\'{\i}nez and S.~Rigolin,
  %``nu/mu disappearance at the SPL, T2K-I and the neutrino factory,''
  arXiv:hep-ph/0509349.
  %%CITATION = HEP-PH 0509349;%%

%\cite{Ayres:2004js}
\bibitem{Ayres:2004js}
  D.~S.~Ayres {\it et al.}  [NOvA Collaboration],
  %``NOvA proposal to build a 30-kiloton off-axis detector to study neutrino
  %oscillations in the Fermilab NuMI beamline,''
  arXiv:hep-ex/0503053.
  %%CITATION = HEP-EX 0503053;%%

%\cite{Gomez-Cadenas:2001eu}
\bibitem{Gomez-Cadenas:2001eu}
J.~J.~Gomez-Cadenas {\it et al.}  [CERN working group on Super Beams Collaboration],
%``Physics potential of very intense conventional neutrino beams,''
arXiv:hep-ph/0105297.
%%CITATION = HEP-PH 0105297;%%

%\cite{Barger:1999jj}
\bibitem{Barger:1999jj}
  V.~D.~Barger, S.~Geer, R.~Raja and K.~Whisnant,
  %``Long-baseline study of the leading neutrino oscillation at a neutrino factory,''
  Phys.\ Rev.\ D {\bf 62} (2000) 013004 [arXiv:hep-ph/9911524].
  %%CITATION = HEP-PH 9911524;%%

%\cite{Bueno:2000fg}
\bibitem{Bueno:2000fg}
  A.~Bueno, M.~Campanelli and A.~Rubbia,
  %``Physics potential at a neutrino factory: Can we benefit from more than just detecting muons?,''
  Nucl.\ Phys.\ B {\bf 589} (2000) 577 [arXiv:hep-ph/0005007].
  %%CITATION = HEP-PH 0005007;%%

%\cite{Donini:2002rm}
\bibitem{Donini:2002rm}
A.~Donini, D.~Meloni and P.~Migliozzi,
%``The silver channel at the neutrino factory,''
Nucl.\ Phys.\ B {\bf 646} (2002) 321 [arXiv:hep-ph/0206034];
%%CITATION = HEP-PH 0206034;%%
%``The nu/e $\to$ nu/tau channel as a tool to solve ambiguities,''
J.\ Phys.\ G {\bf 29} (2003) 1865 [arXiv:hep-ph/0209240].
%%CITATION = HEP-PH 0209240;%%

%\cite{Donini:1999jc}
\bibitem{Donini:1999jc}
  A.~Donini, M.~B.~Gavela, P.~Hernandez and S.~Rigolin,
  %``Neutrino mixing and CP-violation,''
  Nucl.\ Phys.\ B {\bf 574} (2000) 23 [arXiv:hep-ph/9909254].
  %%CITATION = HEP-PH 9909254;%%

%\cite{Huber:2003ak}
\bibitem{Huber:2003ak}
  P.~Huber and W.~Winter,
  %``Neutrino factories and the 'magic' baseline,''
  Phys.\ Rev.\ D {\bf 68} (2003) 037301 [arXiv:hep-ph/0301257].
  %%CITATION = HEP-PH 0301257;%%

%\cite{Burguet-Castell:2002qx}
\bibitem{Burguet-Castell:2002qx}
  J.~Burguet-Castell, M.~B.~Gavela, J.~J.~Gomez-Cadenas, P.~Hernandez and O.~Mena,
  %``Superbeams plus neutrino factory: The golden path to leptonic CP violation,''
  Nucl.\ Phys.\ B {\bf 646} (2002) 301 [arXiv:hep-ph/0207080].
  %%CITATION = HEP-PH 0207080;%%

%\cite{Donini:2003vz}
\bibitem{Donini:2003vz}
  A.~Donini, D.~Meloni and S.~Rigolin,
  %``Clone flow analysis for a theory inspired neutrino experiment planning,''
  JHEP {\bf 0406}, 011 (2004) [arXiv:hep-ph/0312072].
  %%CITATION = HEP-PH 0312072;%%

%\cite{Nakaya:2005gz}
\bibitem{Nakaya:2005gz}
  T.~Nakaya,
  %``K2K results,''
  Nucl.\ Phys.\ Proc.\ Suppl.\  {\bf 143} (2005) 96.
  %%CITATION = NUPHZ,143,96;%%

%\cite{Kopp:2004sc}
\bibitem{Kopp:2004sc}
  S.~E.~Kopp,
  %``The NuMI neutrino beam at Fermilab,''
  AIP Conf.\ Proc.\  {\bf 773}, 276 (2005) [arXiv:hep-ex/0412052].
  %%CITATION = HEP-EX 0412052;%%

%\cite{Kodama:2004db}
\bibitem{Kodama:2004db}
  K.~Kodama  [OPERA Collaboration],
  %``CNGS, OPERA and ICARUS status,''
  AIP Conf.\ Proc.\  {\bf 721} (2004) 231.
  %%CITATION = APCPC,721,231;%%

%\cite{Huber:2004ka}
\bibitem{Huber:2004ka}
  P.~Huber, M.~Lindner and W.~Winter,
  %``Simulation of long-baseline neutrino oscillation experiments with GLoBES,''
  Comput.\ Phys.\ Commun.\  {\bf 167} (2005) 195 [arXiv:hep-ph/0407333].
  %%CITATION = HEP-PH 0407333;%%

%\cite{Albrow:2005kw}
\bibitem{Albrow:2005kw}
  M.~G.~Albrow {\it et al.},
  %``Physics at a Fermilab Proton Driver,''
  arXiv:hep-ex/0509019.
  %%CITATION = HEP-EX 0509019;%%
  
%\cite{MenaRequejo:2005hn}
\bibitem{MenaRequejo:2005hn}
  O.~Mena Requejo, S.~Palomares-Ruiz and S.~Pascoli,
  %``Super-NOvA: A long-baseline neutrino experiment with two off-axis
  %detectors,''
  Phys.\ Rev.\ D {\bf 72}, 053002 (2005) [arXiv:hep-ph/0504015];
  %%CITATION = HEP-PH 0504015;%%
  O.~Mena and S.~Parke,
  %``Physics potential of the Fermilab NuMI beamline,''
  Phys.\ Rev.\ D {\bf 72} (2005) 053003 [arXiv:hep-ph/0505202];
  %%CITATION = HEP-PH 0505202;%%
  O.~Mena, S.~Palomares-Ruiz and S.~Pascoli,
  %``Determining the neutrino mass hierarchy and CP violation in NOnuA with a
  %second off-axis detector,''
  arXiv:hep-ph/0510182.
  %%CITATION = HEP-PH 0510182;%%

%\cite{Huber:2002rs}
\bibitem{Huber:2002rs}
  P.~Huber, M.~Lindner and W.~Winter,
  %``Synergies between the first-generation JHF-SK and NuMI superbeam experiments,''
  Nucl.\ Phys.\ B {\bf 654} (2003) 3 [arXiv:hep-ph/0211300].
  %%CITATION = HEP-PH 0211300;%%

\bibitem{Budd:2005ga}
  H.~Budd,
  %``The design and performance of the MINERvA detector,''
  arXiv:hep-ex/0503024.
  %%CITATION = HEP-EX 0503024;%%

\bibitem{gilardoni}
S.~Gilardoni, CERN Thesis;
S.~Gilardoni, G.~Grawer, G.~Maire, J.~M.~Maugain, S.~Rangod and F.~Voelker,
%``Status Of A Magnetic Horn For A Neutrino Factory,''
J.\ Phys.\ G {\bf 29} (2003) 1801.
%%CITATION = JPHGB,G29,1801;%%

%\cite{Campagne:2004wt}
\bibitem{Campagne:2004wt}
  J.~E.~Campagne and A.~Cazes,
  %``The theta(13) and delta(CP) sensitivities of the SPL-Frejus project revisited,''
  arXiv:hep-ex/0411062.
  %%CITATION = HEP-EX 0411062;%%

%\cite{Jung:1999jq}
\bibitem{Jung:1999jq}
C.~K.~Jung,
%``Feasibility of a next generation underground water Cherenkov detector: UNO,''
arXiv:hep-ex/0005046.
%%CITATION = HEP-EX 0005046;%%

%\cite{Burguet-Castell:2003vv}
\bibitem{Burguet-Castell:2003vv}
  J.~Burguet-Castell, D.~Casper, J.~J.~Gomez-Cadenas, P.~Hernandez and F.~Sanchez,
  %``Neutrino oscillation physics with a higher gamma beta-beam,''
  Nucl.\ Phys.\ B {\bf 695}, 217 (2004) [arXiv:hep-ph/0312068].
  %%CITATION = HEP-PH 0312068;%%

%\cite{Blondel:2004cx}
\bibitem{Blondel:2004cx}
  A.~Blondel, M.~Campanelli and M.~Fechner,
  %``Energy reconstruction in quasi-elastic events unfolding physics and
  %detector effects,''
  Nucl.\ Instrum.\ Meth.\ A {\bf 535} (2004) 665.
  %%CITATION = NUIMA,A535,665;%%

%\cite{Casper:2002sd}
\bibitem{Casper:2002sd}
D.~Casper,
%``The nuance neutrino physics simulation, and the future,''
Nucl.\ Phys.\ Proc.\ Suppl.\  {\bf 112} (2002) 161 [arXiv:hep-ph/0208030].
%%CITATION = HEP-PH 0208030;%%

%\cite{Blondel:2000gj}
\bibitem{Blondel:2000gj}
A.~Blondel {\it et al.},
%``The neutrino factory: Beam and experiments,''
Nucl.\ Instrum.\ Meth.\ A {\bf 451} (2000) 102.
%%CITATION = NUIMA,A451,102;%%

%\cite{Cervera:2000vy}
\bibitem{Cervera:2000vy}
A.~Cervera, F.~Dydak and J.~Gomez Cadenas,
%``A large magnetic detector for the neutrino factory,''
Nucl.\ Instrum.\ Meth.\ A {\bf 451} (2000) 123.
%%CITATION = NUIMA,A451,123;%%

%\cite{Autiero:2003fu}
\bibitem{Autiero:2003fu}
D.~Autiero {\it et al.},
%``The synergy of the golden and silver channels at the Neutrino Factory,''
Eur.\ Phys.\ J.\ C {\bf 33} (2004) 243 [arXiv:hep-ph/0305185].
%%CITATION = HEP-PH 0305185;%%

%\cite{Zeller:2003ey}
\bibitem{Zeller:2003ey}
G.~P.~Zeller,
%``Low energy neutrino cross sections: Comparison of various Monte Carlo
%predictions to experimental data,''
arXiv:hep-ex/0312061.
%%CITATION = HEP-EX 0312061;%%

%\cite{Lipari:1994pz}
\bibitem{lipari}
P.~Lipari, private communication;
P.~Lipari, M.~Lusignoli and F.~Sartogo,
%``The Neutrino cross-section and upward going muons,''
Phys.\ Rev.\ Lett.\  {\bf 74}, 4384 (1995) [arXiv:hep-ph/9411341].
%%CITATION = HEP-PH 9411341;%%

%\cite{Serreau:2004kx}
\bibitem{Serreau:2004kx}
  J.~Serreau and C.~Volpe,
  %``Neutrino nucleus interaction rates at a low-energy beta-beam facility,''
  Phys.\ Rev.\ C {\bf 70}, 055502 (2004) [arXiv:hep-ph/0403293].
  %%CITATION = HEP-PH 0403293;%%

%\cite{Petyt:1998zd}
\bibitem{Petyt:1998zd}
  D.~A.~Petyt,
  %``A Study of parameter measurement in a long baseline neutrino oscillation
  %experiment,''
FERMILAB-THESIS-1998-66
%\href{http://www.slac.stanford.edu/spires/find/hep/www?r=fermilab-thesis-1998-66}{SPIRES entry}

%\cite{Minakata:2004pg}
\bibitem{Minakata:2004pg}
  H.~Minakata, M.~Sonoyama and H.~Sugiyama,
  %``Determination of theta(23) in long-baseline neutrino oscillation
  %experiments with three-flavor mixing effects,''
  Phys.\ Rev.\ D {\bf 70}, 113012 (2004) [arXiv:hep-ph/0406073].
  %%CITATION = HEP-PH 0406073;%%

%\cite{Akhmedov:2004ny}
\bibitem{Akhmedov:2004ny}
E.~K.~Akhmedov {\em et al.},
%``Series expansions for three-flavor neutrino oscillation probabilities in
%matter,''
JHEP {\bf 0404} (2004) 078 [arXiv:hep-ph/0402175].
%%CITATION = HEP-PH 0402175;%%

%\cite{Fogli:2001wi}
\bibitem{Fogli:2001wi}
  G.~L.~Fogli, E.~Lisi and A.~Palazzo,
  %``Quasiaveraged solar neutrino oscillations,''
  Phys.\ Rev.\ D {\bf 65} (2002) 073019 [arXiv:hep-ph/0105080].
  %%CITATION = HEP-PH 0105080;%%

%\cite{Donini:2004hu}
\bibitem{Donini:2004hu}
  A.~Donini, E.~Fern\'andez-Mart\'{\i}nez, P.~Migliozzi, S.~Rigolin and L.~Scotto Lavina,
  %``Study of the eightfold degeneracy with a standard beta-beam and a
  %super-beam facility,''
  Nucl.\ Phys.\ B {\bf 710} (2005) 402 [arXiv:hep-ph/0406132];
  %%CITATION = HEP-PH 0406132;%%
  S.~Rigolin,
  %``Why care about (theta(13), delta) degeneracy at future neutrino
  %experiments,''
  arXiv:hep-ph/0407009.
  %%CITATION = HEP-PH 0407009;%%

%\cite{Minakata:2003ca}
\bibitem{Minakata:2003ca}
H.~Minakata, H.~Nunokawa and S.~J.~Parke,
%``The complementarity of eastern and western hemisphere long-baseline
%neutrino oscillation experiments,''
Phys.\ Rev.\ D {\bf 68} (2003) 013010 [arXiv:hep-ph/0301210];
%%CITATION = HEP-PH 0301210;%%
  M.~Ishitsuka, T.~Kajita, H.~Minakata and H.~Nunokawa,
  %``Resolving neutrino mass hierarchy and CP degeneracy by two identical
  %detectors with different baselines,''
  Phys.\ Rev.\ D {\bf 72} (2005) 033003 [arXiv:hep-ph/0504026];
  %%CITATION = HEP-PH 0504026;%%
  %``Establishing neutrino mass hierarchy and CP violation by two identical
  %detectors with different baselines using the J-PARC nu beam,''
  arXiv:hep-ph/0510166.
  %%CITATION = HEP-PH 0510166;%%

%\cite{Yasuda:2004gu}
\bibitem{Yasuda:2004gu}
  O.~Yasuda,
  %``New plots and parameter degeneracies in neutrino oscillations,''
  New J.\ Phys.\  {\bf 6} (2004) 83 [arXiv:hep-ph/0405005];
  %%CITATION = HEP-PH 0405005;%%
  %``Degeneracy and strategies of long baseline and reactor experiments,''
  Nucl.\ Phys.\ Proc.\ Suppl.\  {\bf 149} (2005) 170 [arXiv:hep-ph/0412405].
  %%CITATION = HEP-PH 0412405;%%

%\cite{Bouchez:2003fy}
\bibitem{Bouchez:2003fy}
  J.~Bouchez, M.~Lindroos and M.~Mezzetto,
  %``Beta beams: Present design and expected performances,''
  AIP Conf.\ Proc.\  {\bf 721} (2004) 37 [arXiv:hep-ex/0310059];
  %%CITATION = HEP-EX 0310059;%%
  M.~Mezzetto,
  %``Beta beams,''
  Nucl.\ Phys.\ Proc.\ Suppl.\  {\bf 143} (2005) 309 [arXiv:hep-ex/0410083].
  %%CITATION = HEP-EX 0410083;%%

%\cite{Terranova:2004hu}
\bibitem{Terranova:2004hu}
  F.~Terranova, A.~Marotta, P.~Migliozzi and M.~Spinetti,
  %``High energy beta beams without massive detectors,''
  Eur.\ Phys.\ J.\ C {\bf 38}, 69 (2004) [arXiv:hep-ph/0405081].
  %%CITATION = HEP-PH 0405081;%%

%\cite{Burguet-Castell:2005pa}
\bibitem{Burguet-Castell:2005pa}
  J.~Burguet-Castell, D.~Casper, E.~Couce, J.~J.~Gomez-Cadenas and P.~Hernandez,
  %``Optimal beta-beam at the CERN-SPS,''
  Nucl.\ Phys.\ B {\bf 725}, 306 (2005)
  [arXiv:hep-ph/0503021].
  %%CITATION = HEP-PH 0503021;%%

%\cite{Huber:2005jk}
\bibitem{Huber:2005jk}
  P.~Huber, M.~Lindner, M.~Rolinec and W.~Winter,
  %``Physics and optimization of beta-beams: From low to very high gamma,''
  arXiv:hep-ph/0506237.
  %%CITATION = HEP-PH 0506237;%%

%\cite{Rigolin:2005mr}
\bibitem{Rigolin:2005mr}
  S.~Rigolin,
  %``Physics reach of beta-beams and nu-factories: The problem of degeneracies,''
  arXiv:hep-ph/0509366.
  %%CITATION = HEP-PH 0509366;%%

%\cite{Donini:2005qg}
\bibitem{Donini:2005qg}
  A.~Donini, E.~Fern\'andez-Mart\'{\i}nez, P.~Migliozzi, S.~Rigolin, L. Scotto Lavina, T.~T.~de Fatis and F.~Terranova,
  %``Perspectives for a neutrino program based on the upgrades of the CERN
  %accelerator complex,''
  arXiv:hep-ph/0511134.
  %%CITATION = HEP-PH 0511134;%%

%\cite{Donini:2003kr}
\bibitem{Donini:2003kr}
  A.~Donini,
  %``NUFACT'03: The fate of the clones,''
  AIP Conf.\ Proc.\  {\bf 721}, 219 (2004) [arXiv:hep-ph/0310014].
  %%CITATION = HEP-PH 0310014;%%

  %\cite{Zaglauer:1988gz}
\bibitem{Zaglauer:1988gz}
  H.~W.~Zaglauer and K.~H.~Schwarzer,
  %``The Mixing Angles In Matter For Three Generations Of Neutrinos And The Msw Mechanism,''
  Z.\ Phys.\ C {\bf 40} (1988) 273.
  %%CITATION = ZEPYA,C40,273;%%

%\cite{Kaneyuki:2005ze}
\bibitem{Kaneyuki:2005ze}
  K.~Kaneyuki  [T2K Collaboration],
  %``T2K experiment,''
  Nucl.\ Phys.\ Proc.\ Suppl.\  {\bf 145} (2005) 178.
  %%CITATION = NUPHZ,145,178;%%

%%%%------------------------------------------------------------

\end{thebibliography}
\end{document}